\documentclass[preprint]{aa}
\usepackage[varg]{txfonts}
\usepackage{amssymb}
\usepackage{subcaption}
\usepackage{natbib}
\usepackage{tabu}
\usepackage{placeins}
\usepackage{float}

\bibpunct{(}{)}{;}{a}{}{,}

\begin{document}

\title{Spectral and orbital survey of medium-sized meteoroids\thanks{Tables 5 and 6 are only available in electronic form
at the CDS via anonymous ftp to cdsarc.u-strasbg.fr (130.79.128.5) or via http://cdsweb.u-strasbg.fr/cgi-bin/qcat?J/A+A/}}

\author{Pavol Matlovič\inst{1}
  \and Juraj Tóth\inst{1}
  \and Regina Rudawska\inst{2}   
  \and Leonard Kornoš\inst{1}
  \and Adriana Pisarčíková\inst{1}}

\institute{Faculty of Mathematics, Physics and Informatics,
  Comenius University, Bratislava, Slovakia\\
  \email{matlovic@fmph.uniba.sk}
  \and ESTEC/ESA, Keplerlaan 1, 2201 AZ Noordwijk, The Netherlands}

\date{Received 2019}

\abstract {} {We investigate the spectra, material properties, and orbital distribution of millimeter- to decimeter-sized meteoroids. Our study aims to distinguish the characteristics of populations of  differently sized meteoroids and reveal the heterogeneity of identified meteoroid streams. We verify the surprisingly large ratio of pure iron meteoroids on asteroidal orbits detected among mm-sized bodies.} {Emission spectra and multi-station meteor trajectories were collected within the AMOS network observations. The sample is based on 202 meteors of -1 to -14 magnitude, corresponding to meteoroids of mm to dm sizes. Meteoroid composition is studied by spectral classification based on relative intensity ratios of Na, Mg, and Fe and corresponding monochromatic light curves. Heliocentric orbits, trajectory parameters, and material strengths inferred from empirical $K_B$ and $P_E$ parameters were determined for 146 meteoroids.} {An overall increase of Na content compared to the population of mm-sized meteoroids was detected, reflecting weaker effects of space weathering processes on larger meteoroids. The preservation of volatiles in larger meteoroids is directly observed. We report a very low ratio of pure iron meteoroids and the discovery of a new spectral group of Fe-rich meteors. The majority of meteoroids on asteroidal orbits were found to be chondritic. Thermal processes causing Na depletion and physical processes resulting in Na-rich spectra are described and linked to characteristically increased material strengths. Numerous major and minor shower meteors were identified in our sample, revealing various degrees of heterogeneity within Halley-type, ecliptical, and sungrazing meteoroid streams. Our results imply a scattered composition of the fragments of comet 2P/Encke and 109P/Swift-Tuttle. The largest disparities were detected within $\alpha$-Capricornids of the inactive comet 169P/NEAT and $\delta$-Aquarids of the sungrazing 96P/Machholz. We also find a spectral similarity between $\kappa$-Cygnids and Taurids, which could imply a similar composition of the parent objects of the two streams.} {}

\keywords{Meteorites, meteors, meteoroids; Minor planets, asteroids: general; Comets: general; techniques: spectroscopic}
\maketitle

%
%________________________________________________________________

\section{Introduction}  \label{introduction}

One of the ultimate goals of meteor spectroscopy is to reveal meteoroid composition from ground-based atmospheric observations. This task is however quite difficult; the composition of the radiating plasma of a meteor does not fully reflect the chemical composition of the original meteoroid. This is particularly apparent for the refractory elements of Al, Ca, and Ti, which are lacking in the radiating plasma. Similarly, Si is a major component of most meteorites but is underrepresented in meteor spectra. There have been successful models of meteor radiation providing relevant results of relative abundances of the main elements. These models have been mostly applied to individual bright fireball spectra captured in high resolution \citep{1993A&A...279..627B, 2003M&PS...38.1283T, 2018A&A...610A..73F, 2018A&A...613A..54D}. In order to survey larger samples of meteor spectra, a simplified method of spectral classification \citep{2005Icar..174...15B} has been introduced, which is also applicable to more effective low-resolution video observations.

The spectral classification method has since been applied by several authors. Within wider surveys, these studies were mainly focused on fainter meteors \citep{2005Icar..174...15B, 2015A&A...580A..67V, 2019A&A...621A..68V}. Other spectral analyses focused on individual meteor showers \citep{2010pim7.conf...42B, 2014EM&P..112...45R, 2014Icar..231..356M, 2017P&SS..143..104M} or on linking specific spectral features with ablation properties \citep{2017P&SS..143...28B, 2019A&A...621A..68V}. 

The present work represents the widest spectral study of mid-sized meteoroids and also includes a large sample of orbital and structural properties, enabling a more complex view of the analyzed meteoroid population. The presented results complement previous works focused on fainter meteors corresponding to approximately mm-sized bodies \citep{2001ESASP.495..203B, 2005Icar..174...15B, 2014EM&P..112...45R, 2014me13.conf..117J}, and bright fireballs often caused by  meteoroids of over one meter in diameter that potentially break up and give rise to meteorites \citep{1993A&A...279..615C, 2003M&PS...38.1283T, 2013MNRAS.433..571M}. We aim to define the spectral properties of mm--dm sized meteoroids and reveal the differences compared to mm-sized bodies studied by 
\citet{2005Icar..174...15B} and \citet{2015A&A...580A..67V}.

From the fragments of asteroid 2008 TC\textsubscript{3} we know that meter-scale meteoroids can be compositionally heterogeneous \citep{2009Natur.458..485J, 2011Icar..212..697K}. Here we try to find out how the material heterogeneities are reflected in low-resolution meteor spectra and to what degree they are exhibited within individual meteoroids from one parent object. Furthermore, we focus on defining the characteristic spectral properties of meteoroids from different orbital sources (main belt, Jupiter-family, and Halley-type).

It was suggested that while detailed composition (e.g., specific types of chondrites) cannot be easily distinguished from low-resolution meteor spectra \citep{2015aste.book..257B}, the rough composition (chondritic, achondritic, metallic) can be identified. We attempt to verify this hypothesis and try to look for additional information about the meteoroid nature and structure that can be revealed from determined orbital and atmospheric parameters. 

One of the most surprising results of the study by \citet{2005Icar..174...15B} was the large number of the detected pure-iron meteoroids, suggesting that iron meteoroids prevail among mm-sized meteoroids on asteroidal orbits. These results are not consistent with the expectations based on what we know about the transport of large main-belt asteroids (MBA) to near-Earth orbit (NEO) space \citep{2017A&A...598A..52G}. Our work provides more detail into the spectral properties of mm- to dm-sized asteroidal meteoroids, which could help to clarify this inconsistency.

\section{Methods and observations}

\begin{figure}
\centerline{\includegraphics[width=\columnwidth,angle=0]{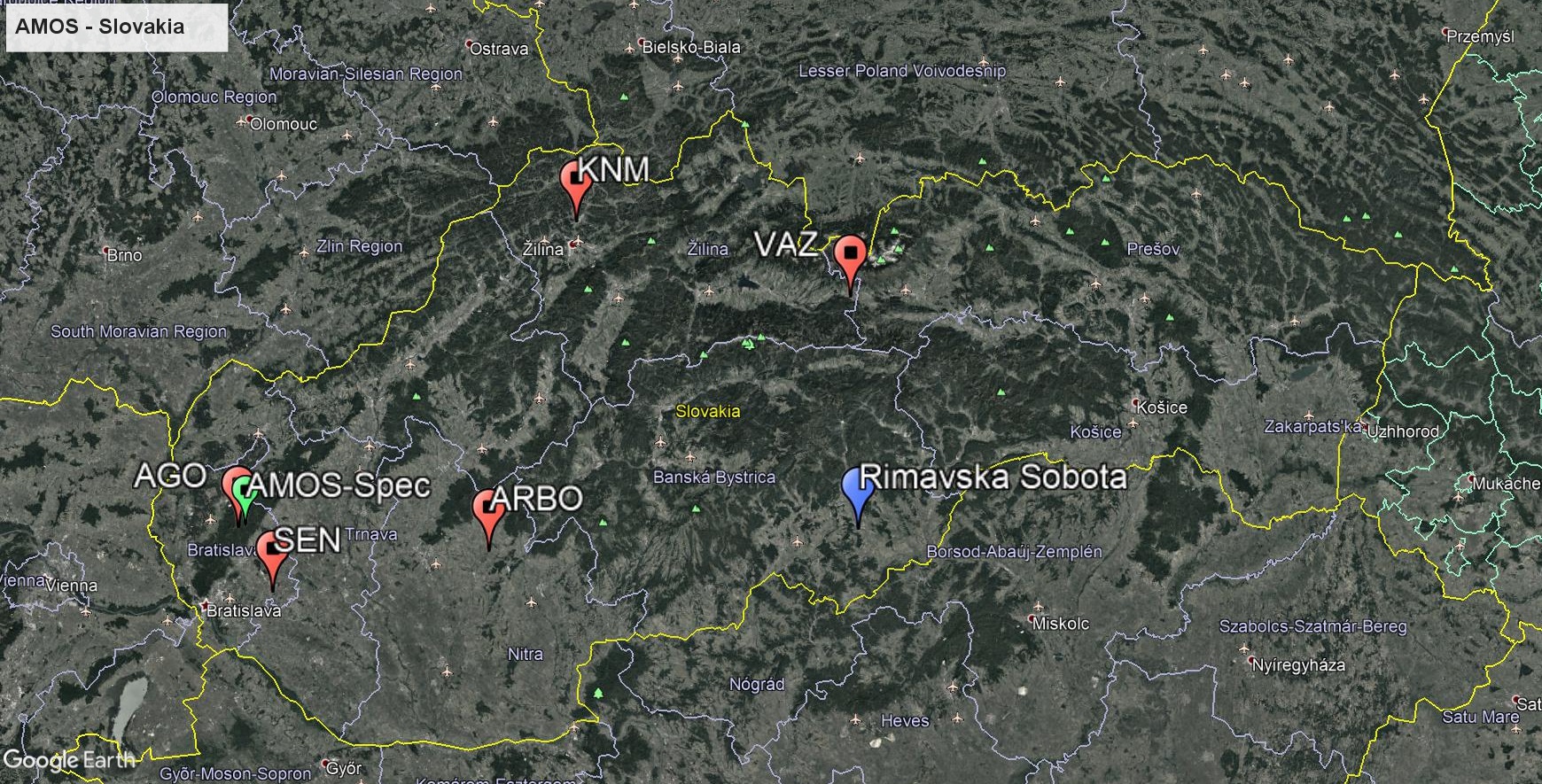}} \caption[f1]{Slovak part of the global AMOS network. Observations of the AMOS-Spec and AMOS stations by the Astronomical and Geophysical Observatory in Modra (AGO), Arborétum Tesárske Mlyňany (ARBO), Kysucké Nové Mesto (KNM), and Važec (VAZ) were used in this work. Red, green, and blue labels designate operating standard AMOS cameras,  spectral stations, and  planned stations.} 
\label{AMOS_SK}
\end{figure}

The meteor data analyzed in this work come from observations of the All-sky Meteor Orbit System (AMOS) network developed and operated by the Comenius University in Bratislava. The network consists globally of 11 standard AMOS systems and 5 spectral AMOS-Spec systems. The stations are positioned at different locations in Slovakia, the Canary Islands, Chile, and Hawaii. Standard AMOS systems provide continuous all-sky detection of meteors and are used to determine the atmospheric trajectory and original heliocentric orbit of observed meteoroids. In this work, we use the observations of the Slovak part of the AMOS network (Fig. \ref{AMOS_SK}). All of the spectra presented here were observed by the AMOS-Spec station at the Modra Observatory. More information about the AMOS network can be found in \citet{2015P&SS..118..102T, TothNEOSST19} and \citet{2013pimo.conf...18Z}.

The presented spectra span the entire year and include sporadic meteors as well as meteors from known meteoroid streams. All spectra with S/N $>$ 4  per pixel were selected for our analysis. In some cases, poor atmospheric conditions (clouds, strong Moon illumination) and acute entry angle hampered the quality of collected spectra. Approximately 10\% of collected spectra were excluded from this analysis due to poor quality. Furthermore, in cases of bright meteor flares, only frames without significant saturation were used for processing.

\subsection{AMOS instrumentation}

The standard AMOS system consists of four major components: a fish-eye lens, an image intensifier, a projection lens, and a digital video camera. The resulting field of view (FOV) of AMOS is 180$^{\circ}$ x 140$^{\circ}$ with an image resolution of 8.4 arcmin/pixel, and a video frame rate of 15 fps. Limiting magnitude for stars is about +5 mag for a single frame, and the detection efficiency is lower for moving objects, approximately +4 mag at typical meteor speeds due to the trailing loss.

\begin{figure}
\centerline{\includegraphics[width=\columnwidth,angle=0]{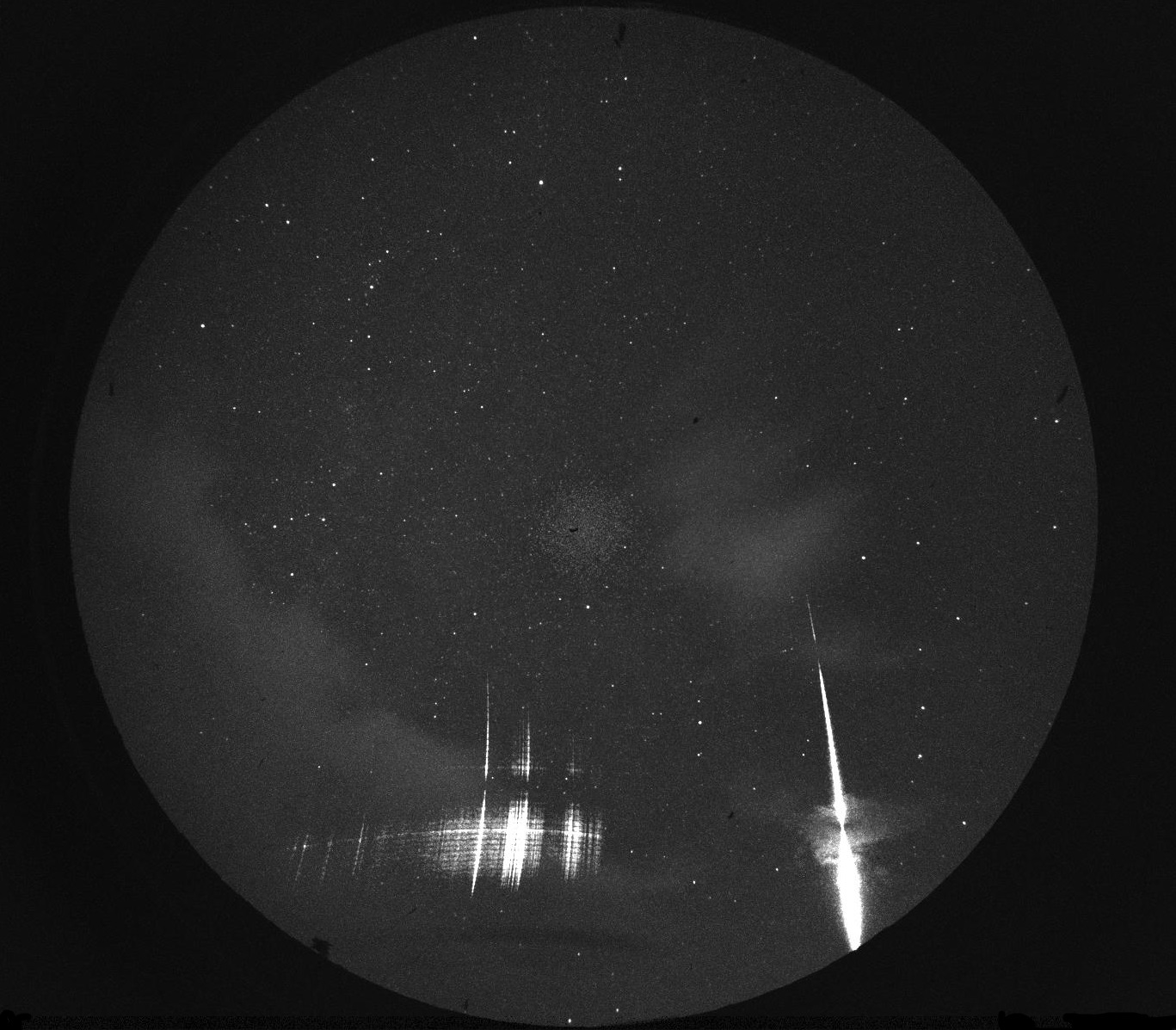}} \caption[f1]{Example of the zero-order image of meteor M20171220\_223458 (trail on the right) with the first-order spectrum (distributed trails on the left) captured by the AMOS-Spec station in AGO Modra.} 
\label{AMOS_ex1}
\end{figure}

The AMOS-Spec is a semi-automatic remotely controlled video system for meteor spectra observations. The display components of the AMOS-Spec consist of a 30 mm f/3.5 fish-eye lens, an image intensifier (Mullard XX1332), a projection lens (Opticon 1.4/19 mm), and a digital camera (Imaging Source DMK 51AU02). This setup yields a 100$^{\circ}$ circular FOV (Fig. \ref{AMOS_ex1}) with a resolution of 1600 x 1200 pixels and a frame rate of 12/s. To our knowledge, AMOS-Spec is one of the largest FOV spectrographs used for meteor observations. Diffraction of the incoming light is provided by a blazed holographic grating with 1000 grooves/mm placed above the lens. The resulting dispersion varies due to the geometry of the all-sky lens with an average value of 1.3 nm/px. This translates to a spectral resolution of approximately R = 200. The limiting magnitude of the AMOS-Spec for meteors (zero order) is similarly to that of the standard AMOS, namely around +4 mag, while the limiting magnitude for a spectral event (first order) is -1 mag. The system covers the whole visual spectrum range of 370–900 nm with a sensitivity peak at 470 nm. An example of an spectral event captured by the AMOS-Spec can be seen in Fig. \ref{AMOS_ex1}.

\subsection{Spectral reduction}

The image processing and reduction of spectra analyzed in this work follow the procedure described in \citet{2017P&SS..143..104M}. In the first step, all spectral recordings are corrected for dark frame, flat-field images and have the star background image subtracted to reduce noise and other sources of illumination. The all-sky geometry of the AMOS-Spec lenses causes slight curvature of events captured near the edge of the FOV. Each spectrum is therefore manually scanned in the ImageJ\footnote{https://imagej.nih.gov/ij/} program on individual frames with spectrum signal. The width of the scanning slit is adapted to the spectrum geometry.

\begin{figure}
\centerline{\includegraphics[width=\columnwidth,angle=0]{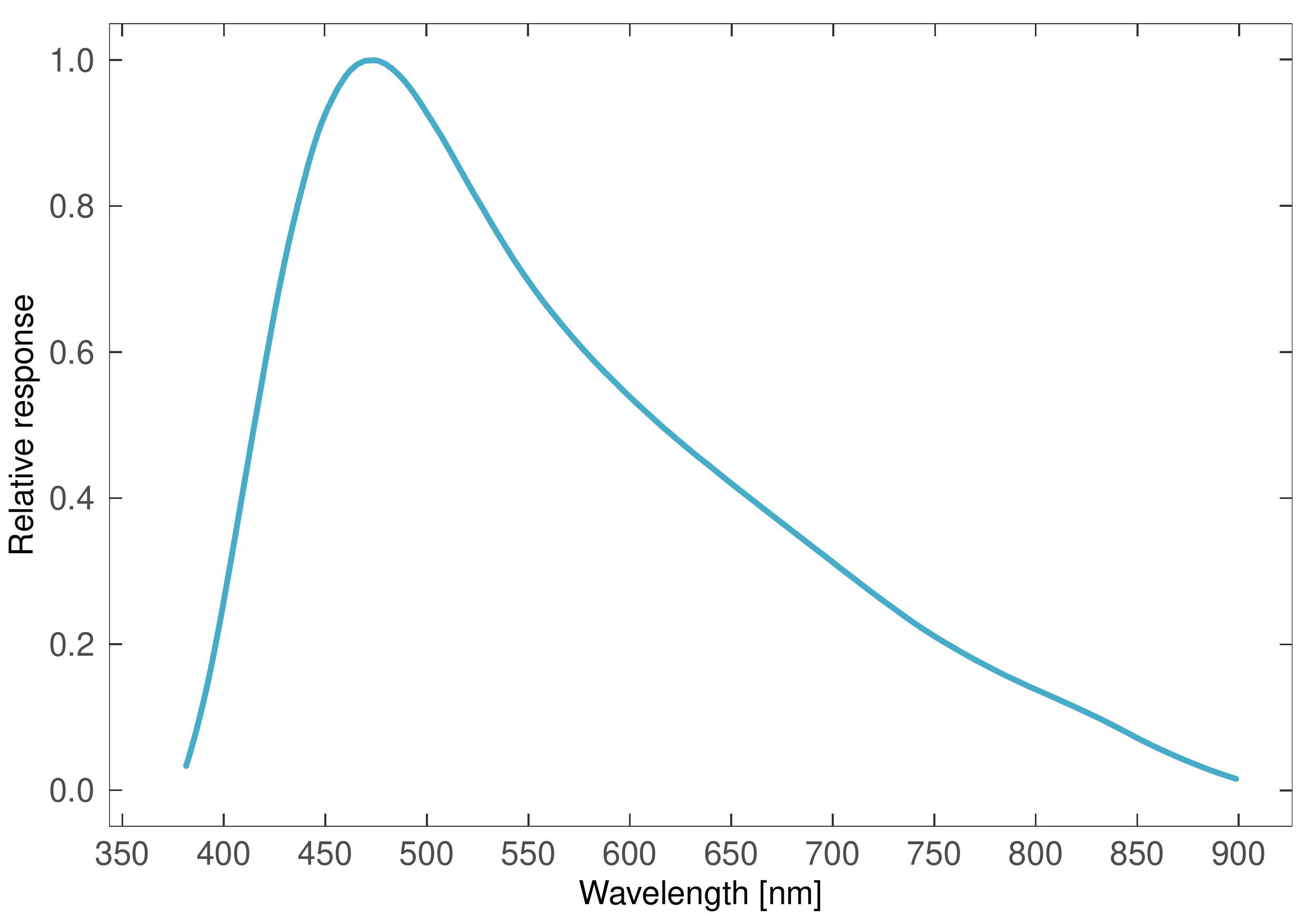}} \caption[f1]{Spectral response curve of the AMOS-Spec system.} 
\label{response}
\end{figure}

All spectra are first scaled based on recognized spectral lines and polynomial fit of second or higher order. The wavelength scale is later fine-tuned during the fitting procedure. The spectrum is corrected for the spectral sensitivity of the system (Fig. \ref{response}). The spectral response curve was determined using a calibration halogen lamp measured in a laboratory environment. The atmospheric transmittance at different stations also affects the specific spectral sensitivity, mainly in the near-UV region. Transmittance differences of < 2\% \citep{2014A&A...564A..46B} can be neglected in the 510-600 nm range of wavelengths under consideration. 

Each spectrum was fitted with the most significant spectral lines (low temperature, high temperature, atmospheric, and wake lines) in meteors using the Fityk software \citep{Wojdyr:ko5121}\footnote{http://fityk.nieto.pl/}. The modeled lines were used to create a synthetic spectrum which was then fit to the measured and calibrated meteor spectrum. The fitting procedure follows the Levenberg–Marquardt algorithm, also known as the damped least-squares method. The modeled lines have Gaussian profiles with a full width at half maximum (FWHM) of typically 3 nm. Moreover, the most notable molecular N\textsubscript{2} bands of the first positive system present in the meteor spectra were fitted using the relative intensities from \citet{2005Icar..174...15B},  adjusted for our spectra. The atmospheric emission lines of O I, N I, and N\textsubscript{2} bands were fitted in the synthetic spectrum depending on the meteor speed and were subtracted before the measurement of meteoroid emission lines.

In this work, we focus on the main meteor multiplets of Mg I – 2, Na I – 1, and Fe I – 15, which form the basis of the spectral classification \citep{2005Icar..174...15B}. The intensities of these multiplets were measured in the fitted synthetic spectrum after subtraction of the continuous radiation and atmospheric emission. The resulting ratios were then applied for the spectral classification. The modeled contributions of all recognized lines of the Fe I - 15 multiplet were summed. These multiplets were selected as they can be accurately measured in most meteor spectra, and are all in the region of high instrumental spectral sensitivity. Nevertheless, the measurement of these lines must be taken with caution, because the Na multiplet overlaps with a sequence of N\textsubscript{2} bands, while the wake lines of Fe can influence the intensity of the Mg line. All these effects were accounted for in the reduction process. The errors of determined Na/Mg and Fe/Mg ratios presented here were determined from the signal-to-noise ratio (S/N) analysis for individual lines.

It must be noted that the spectral classification procedure was originally developed for fainter meteors in the magnitude range +3 to -1, corresponding to meteoroid sizes of 1 - 10 mm. Our system observes meteor spectra of -1 up to -14 magnitude, corresponding to meteoroid sizes of a few mm up to a few decimeters. We expect that the same physical conditions fitted by the thermal equilibrium model applied for bright photographic meteors \citep{1993A&A...279..627B} as well as for fainter meteors \citep{2005Icar..174...15B} can also be assumed for the meteoroid population observed by our system. The effect of self-absorption in brighter meteor spectra was examined in individual cases. Individual frames with saturated meteor spectra were excluded from the analysis.

Figure 4 shows an example of a meteor fit in the 510 - 600 nm region. In specific cases, the positions of atypical lines were searched for in the NIST Atomic Spectra Database \citep{NIST_ASD} and in the Astrophysical Multiplet Table \citep{1945CoPri..20....1M}. Given the large scope of this work, we do not address in detail the minor and atypical emission lines detected during the fitting procedure. The data set analyzed here however can be used for further investigations.

\begin{figure}
\centerline{\includegraphics[width=\columnwidth,angle=0]{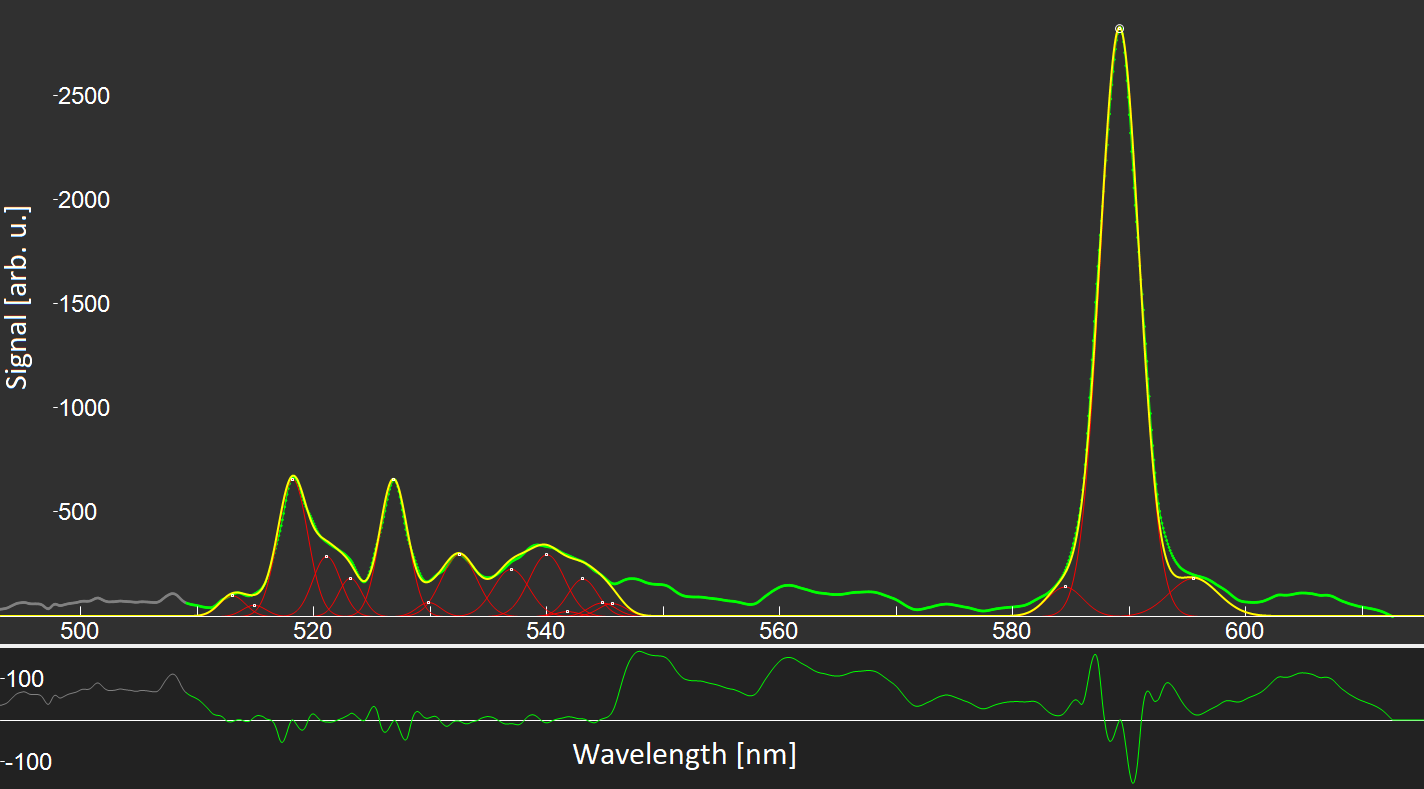}} \caption[f1]{Fit of the synthetic spectrum (yellow) on measured calibrated meteor spectrum (green) as a sum of the main emission contributions (red) in the 510-600 nm region. Residuals of the fit can be seen below the spectrum.} 
\label{fit}
\end{figure}

\subsection{Astrometry and photometry} \label{secAstroPhoto}

Out of the 202 meteors analyzed here, 146 were observed by multiple stations of the AMOS network. Triangulation of meteor paths observed from different stations enable us to accurately determine the atmospheric trajectory and original heliocentric orbit of the incoming meteoroid. Meteors captured by multiple stations, along with their spectra, form the ground base for this work.

The \textit{UFOCapture} detection software \citep{2009JIMO...37...55S} was used during real-time observations to detect and record all of the studied meteors. Star and meteor coordinates from each frame of each event were measured using the original \textit{AMOS} software. In future, the \textit{AMOS} software is also intended to replace the \textit{UFOCapture} software for the real-time detection and positional measurements at all AMOS stations.

Star identification and astrometry was performed using the \textit{RedSky} software \citep[upgraded from an earlier version of][]{KornosIMC17} based on the all-sky procedure of \citet{1995A&AS..112..173B}. \textit{RedSky} uses mathematical transformation for computing rectangular coordinates of catalog stars that are compared with measured stars. The identified stars are used to determine the 13 plate constants from which the zenith distance and azimuth is computed. As a reference, the Sky Catalogue 2000.0 is used up to +6th magnitude including the color index. The precision of the AMOS astrometry is on the order of 0.02 - 0.03$^{\circ}$, which translates to an accuracy of tens of meters for atmospheric meteor trajectory.

Meteor orbits were determined using the \textit{Meteor Trajectory} (\textit{MT}) software based on the procedure of \citet{1987BAICz..38..222C}. The early version of the program has shown promising results in terms of precision \citep{2015pimo.conf..101K, KornosIMC17} and has recently been upgraded for new functionalities. The outputs of the orbital measurements have been, for individual cases, tested and validated on shared observations with the Astronomical Institute of the Czech Academy of Sciences (Dr. Pavel Spurný, private communication). The program computes orbital and geophysical parameters together with their uncertainties based on a Monte Carlo simulation within an uncertainty in the radiant position and geocentric velocity.

Meteor photometry is also performed within the \textit{MT} program, providing apparent and absolute magnitudes and corresponding light curves. The precision of initially obtained magnitudes was further examined individually by visual calibration based on comparison with bright stars, planets, and for bright meteors with the Moon in different phases. Absolute magnitudes were then determined by correction to a standard altitude of 100 km at the observation zenith, and for atmospheric extinction. 

The photometry of the AMOS system is not well adapted to exceptionally bright fireballs ($<$ -8 mag). Based on previous experiences, we suspect that the magnitudes of the brightest fireballs in our sample might be underestimated (Dr. Pavel Spurný, private communication). The underestimation mainly concerns fireballs with bright flares, in which the sharp peaks of brightness are not detected from the integrated video photometry. There are several such cases in our sample. For these, the uncertainty on the determined magnitude and the consequent photometric parameters is higher. The majority of meteors presented here exhibited brightnesses typically between -2 mag and -7 mag (Fig. \ref{magAll}). In these cases, saturation did not significantly affect the photometric calibration. Monochromatic light curves were resolved by our own code based on the relative intensities of the main meteor emission lines (Mg I, Na I, Fe I) at different atmospheric heights.

\section{Spectral classification}

\begin{figure}
\centerline{\includegraphics[width=\columnwidth,angle=0]{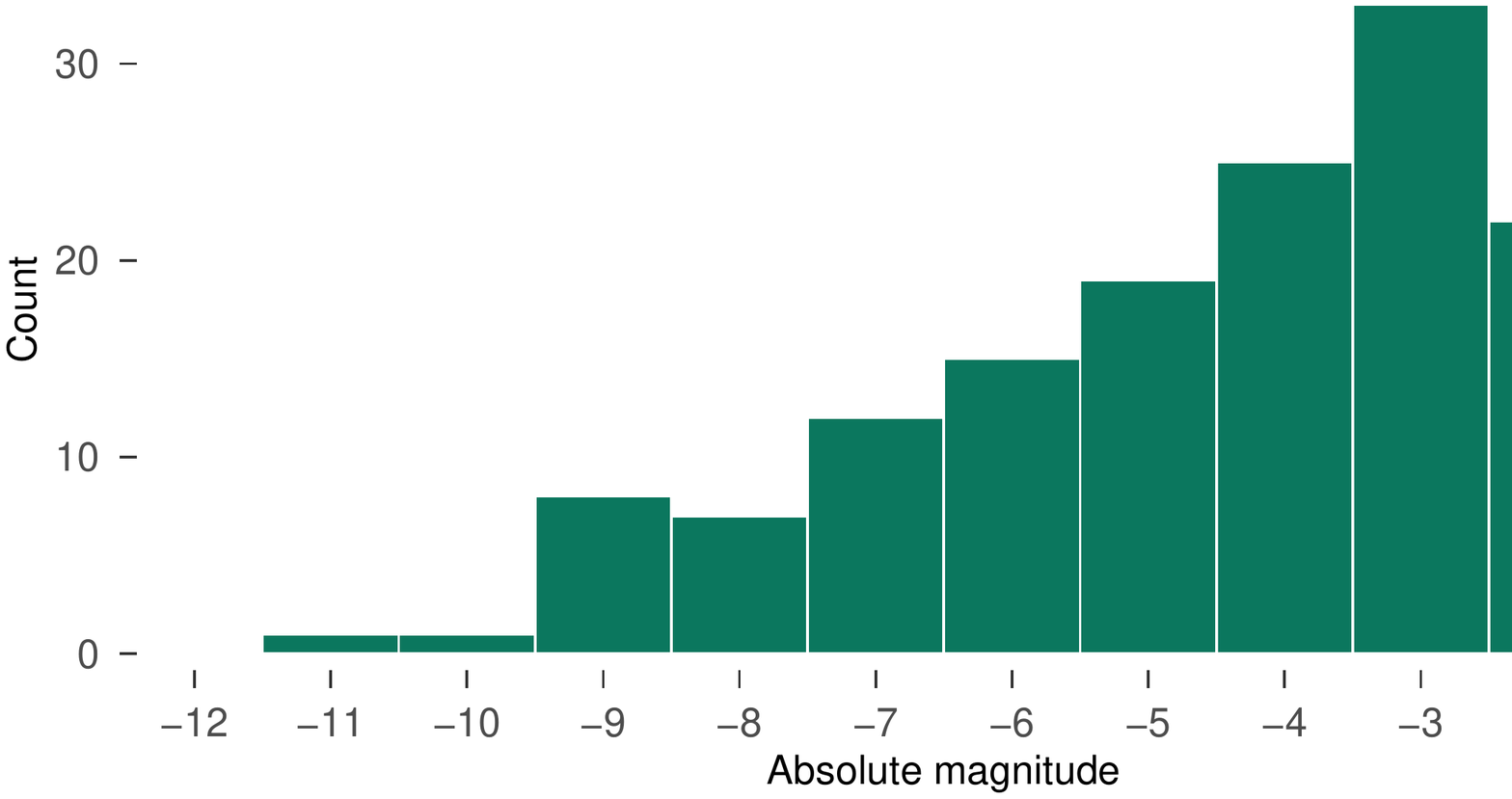}}
\centerline{\includegraphics[width=\columnwidth,angle=0]{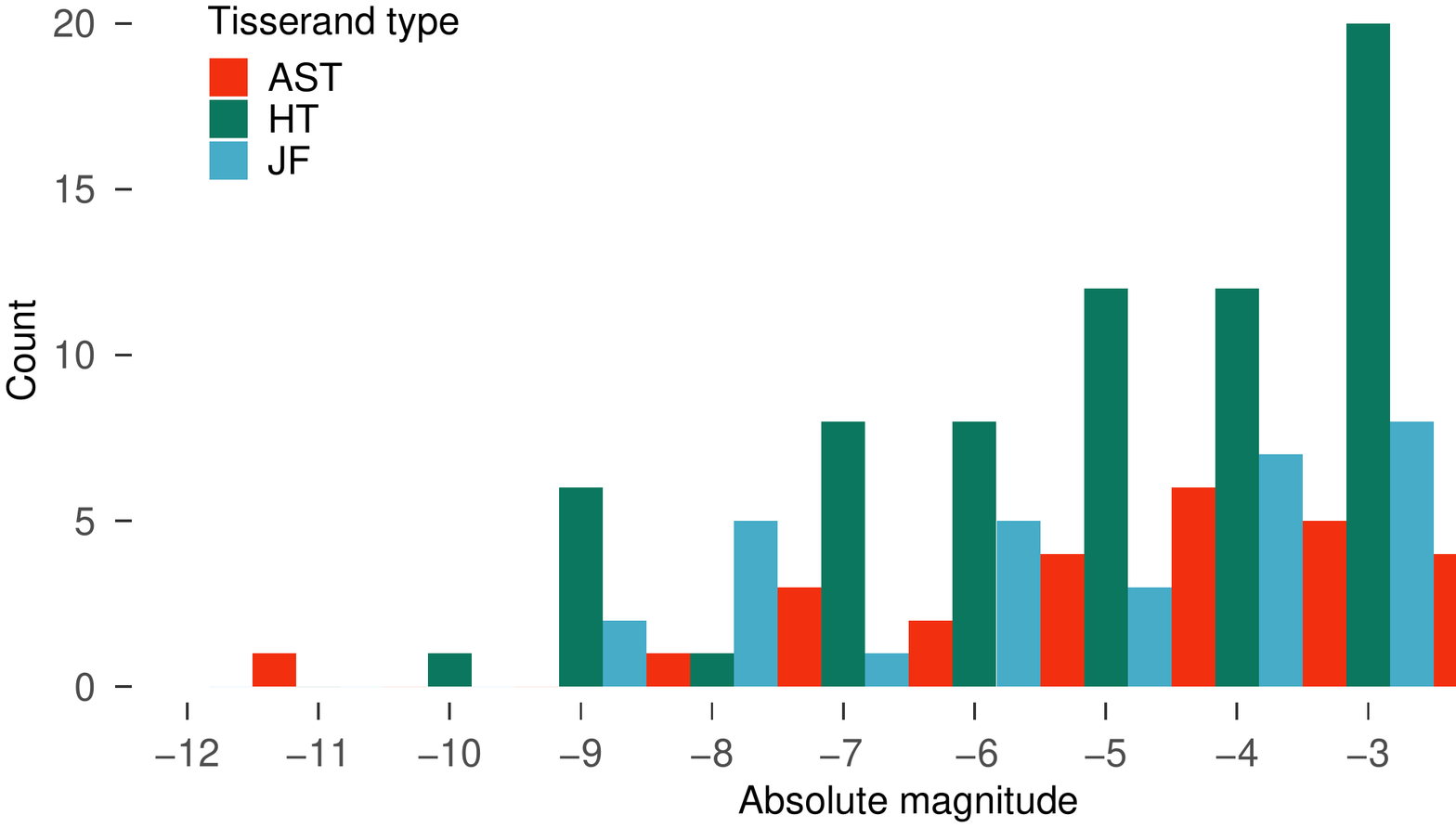}} \caption[f1]{Histogram of absolute magnitudes of meteors in our sample (upper) and magnitude distribution relative to the type of heliocentric orbit based on the Tisserand parameter (lower). The following abbreviations are used: AST for asteroidal orbits, JF for Jupiter-family type orbits, and HT for Halley-type cometary orbits. Only magnitudes of meteors observed by multiple stations are displayed.} 
\label{magAll}
\end{figure}

Overall 202 meteor spectra captured by the AMOS-Spec system during the time period between December 2013 and August 2017 were used for the spectral analysis. The sample includes meteors calculated to be from -1 to -11 mag (Fig. \ref{magAll}). As noted in section \ref{secAstroPhoto}, some of the magnitudes of the brightest fireballs in our sample ($<$ -8 mag) might be underestimated. We estimate that the lower end of the studied magnitude range might reach up to -14 mag. The magnitude range corresponds to meteoroids in the mm to dm size range. 

\begin{figure*}
    \centering
    \begin{subfigure}[b]{0.5\textwidth}
      \centering
      \includegraphics[width=\textwidth]{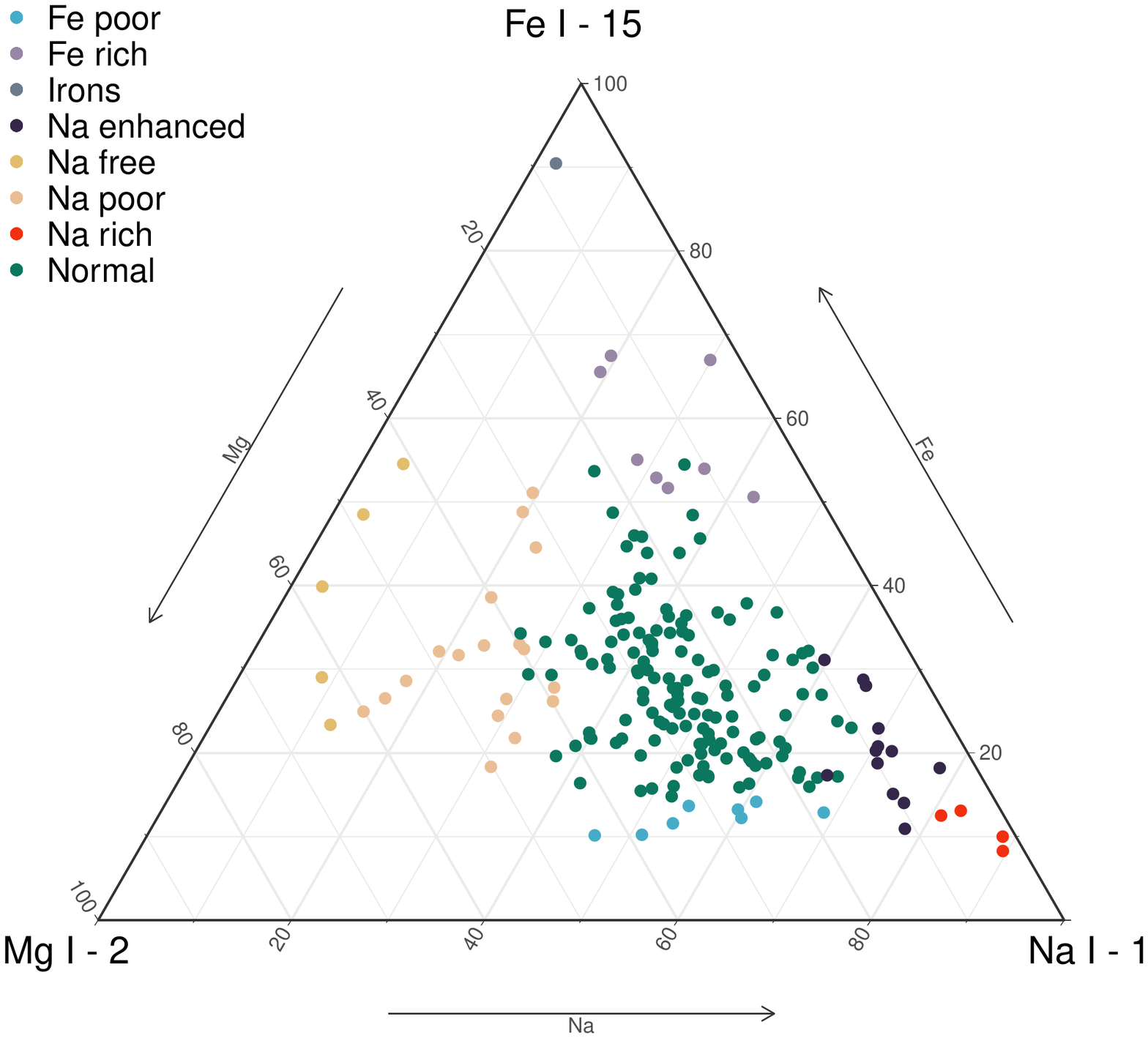}
      \caption{}
      \label{fig:1}
    \end{subfigure}%
    ~
    \begin{subfigure}[b]{0.5\textwidth}
      \centering
      \includegraphics[width=\textwidth]{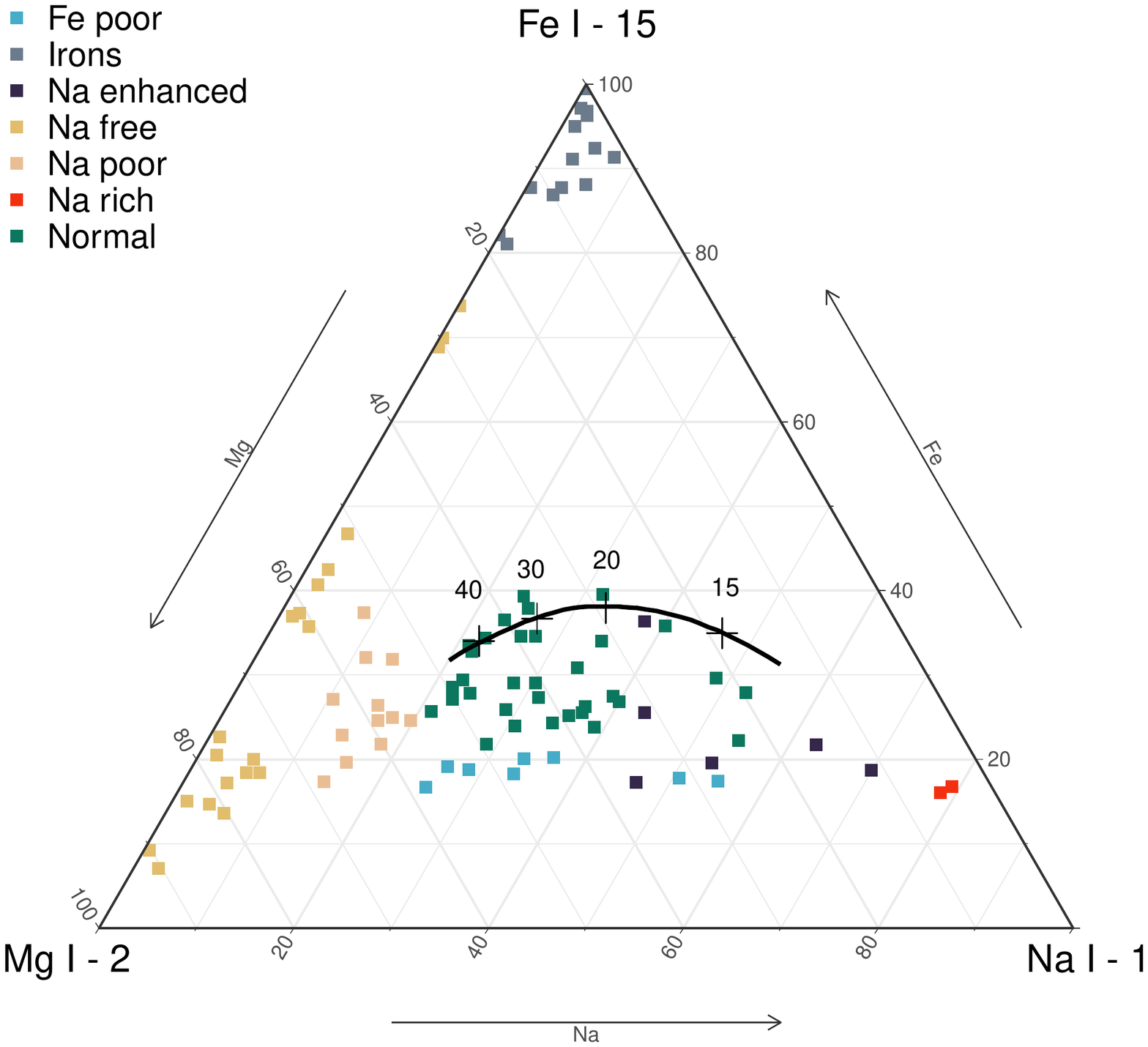}
      \caption{}
      \label{fig:2}
    \end{subfigure}%
    \caption{Ternary diagram displaying the spectral classification of 202 meteors of -1 to -14 mag (corresponding to meteoroids of mm to dm sizes) observed by the AMOS-Spec system during 2013-2017 (a) and (b) comparison to the spectral classification of +3 to -1 mag meteors (corresponding to meteoroid sizes of 1-10 mm) based on data from \citet{2005Icar..174...15B}, including a curve showing the expected range for chondritic composition as a function of meteor speed (in km\,s\textsuperscript{-1}), as modeled by \citet{2005Icar..174...15B}.}
    \label{ternary}
  \end{figure*}

\begin{figure}
\centerline{\includegraphics[width=7.5cm,angle=0]{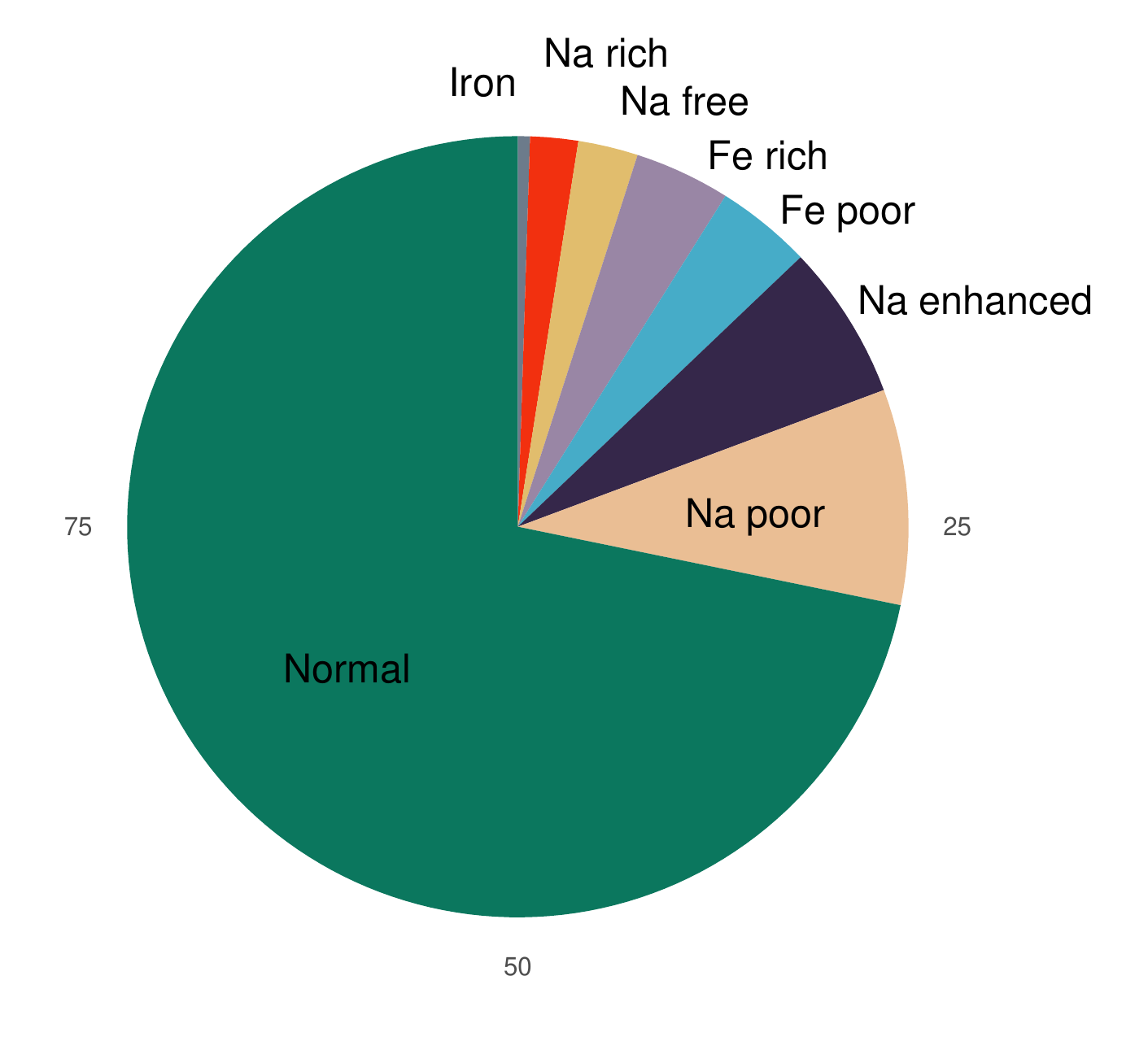}} \caption[f1]{Distribution of spectral classes identified within the sample of 202 meteoroids in the mm to dm size range.} 
\label{pie}
\end{figure}

The spectral classification of our sample based on the relative intensities of Mg I - 2, Na I - 1, and Fe I - 15 is in Fig. \ref{ternary}. Spectra were classified in accordance with the definitions introduced by \citet{2005Icar..174...15B}. Out of the 202 measured meteors, 145 have been identified as normal type, typically positioned in the middle part of the ternary diagram. Normal-type meteors are defined as those lying close to the expected position for chondritic bodies, as modeled by \citet{2005Icar..174...15B} and visible as the chondritic curve in Fig. \ref{ternary}b). These theoretical values and the resulting mean curve were determined assuming chondritic composition and range of temperatures, densities, and sizes of the radiating plasma. This curve might not accurately represent the characteristic chondritic area within our sample of meteors created typically by larger meteoroids. It is likely that the conditions of mean temperature and size of the radiating plasma are shifted in our population. While no theoretical simulation was performed, the characteristic positions of normal-type meteors were simply inferred from the densest part of the distribution in Fig. \ref{ternary}. Normal-type spectra form the largest group of our sample (Fig. \ref{pie}), similarly to previous studies of mm-sized meteoroids \citep{2005Icar..174...15B, 2015A&A...580A..67V, 2019A&A...621A..68V}.

Normal meteoroids are along with Na-poor, Na-enhanced, and Fe-poor meteoroids sometimes referred to as mainstream meteoroids. Na-poor meteors are defined as having a weaker-than-expected (but still
visible) Na line for their given speed. Na-enhanced meteoroids have an obviously stronger-than-expected Na line for their given speed but not so dominant as in Na-rich meteoroids. Fe-poor have the expected Na/Mg ratio but Fe lines that are too faint to be classified as normal. More distinct or so-called nonmainstream classes include three groups: Iron meteors are defined as having spectra dominated by Fe lines. Na-free meteors show no Na line but are not classified as Irons. Finally, Na-rich meteors have spectra dominated by the Na line, with Na/Mg and Na/Fe ratios obviously higher than expected for chondritic bodies \citep{2005Icar..174...15B}.
  
Besides the determined Na/Mg/Fe intensity ratios, the effect of speed dependence (discussed in Section \ref{Sec:orb}) was taken into account for the classification of Na-enhanced and Na-poor meteors. Furthermore, the effects of saturation, self-absorption, and intensities of other Fe multiplets were taken into account for classification of Fe-rich meteors. This is the cause of the partial overlap of spectral classes in Fig. \ref{ternary}.

The shift of the spectral classification among two populations of differently sized meteoroids can be seen by comparison with the ternary diagram in Fig. 6 of \citet{2005Icar..174...15B}. A similar comparison can be made with the ternary diagrams from \citet{2015A&A...580A..67V} or \citet{2019A&A...621A..68V}, which also deal with mm-sized meteoroids and show equivalent distributions. The increase of the Na/Mg ratio in spectra of larger meteoroids is also apparent from Fig. \ref{speedcurve}.

We identified three possible explanations for this effect. The first is instrumental: The spectral sensitivity curve of the AMOS-Spec system shows lower efficiency at 589.2 nm, the peak wavelength of the Na I doublet. We checked the accuracy of the sensitivity curve previously determined from a reference Jupiter spectrum. A new, updated response curve was remeasured and confirmed by several calibration lamps in the laboratory. The response curve error is not sufficient to explain such a shift. 

The second possible explanation is physical. The differences between the magnitude ranges of studied meteor populations are likely to affect the intensity of the Na line. It has been noted that the Na line is often observed earlier (at higher altitudes) than other lines. This relates to the low excitation potential of the Na I line, which can radiate at lower temperatures compared to Mg I. This effect would however favor the increase of Na/Mg ratio in fainter meteors, compared to our sample caused by meteoroids of larger size. The opposite is observed and, as we discuss further below, this effect is much more dependent on meteor speed.

Finally, we suggest that the observed shift does in fact reflect the composition of meteoroids. Sodium is a volatile element, which can be easily depleted from small bodies by space weathering processes. These effects mainly concern the exposure to solar radiation (and specifically thermal desorption when meteoroids are in close proximity to the Sun) and probably the solar wind and cosmic ray irradiation. Depending on the composition and time of exposure, these processes affect mainly the outer layers of interplanetary bodies. For smaller meteoroids, the space weathering processes likely cause more effective volatile depletion within the meteoroid volume. In larger meteoroids studied in our sample, volatiles can be partially depleted from the outer layers but can remain intact below the surface. 

The effect of Na depletion might not only concern the space weathering processes before  entry into Earth's atmosphere. The preheating of the meteoroid in Earth's atmosphere could cause some degree of Na depletion before the meteor spectrum can be observed. The preheating would also more significantly affect smaller meteoroids. Some degree of Na enhancement in 13 larger meteoroids was already noted by \citet{2003M&PS...38.1283T}. The effect is also confirmed by the apparent differences between the number of Na-poor and Na-free meteors observed in our sample compared to that of \citet{2005Icar..174...15B}. Only 11.5\% of our sample (Fig. \ref{pie}) represents these classes, compared to 34\% among mm-sized meteoroids. 

The second most notable difference between the two meteoroid populations is the apparent lack of iron bodies in our sample. Meteor spectra dominated by Fe lines are assumed to come from meteoroids composed mainly of iron--nickel alloy \citep{2005Icar..174...15B, 2017P&SS..143..159C}. Only one iron meteor has been identified in our sample (0.5\%), compared to approximately 14\% among the fainter meteors in the sample of \citet{2005Icar..174...15B}. Similarly, 10\% of the sample detected by \citet{2019A&A...621A..68V} were iron meteoroids. This is an unexpectedly large proportion considering the statistics from known meteorite falls. According to the Meteoritical Bulletin Database\footnote{https://www.lpi.usra.edu/meteor/}, only 3.7\% of all meteorites have been identified as iron meteorites. Meteorite samples only include strong asteroidal bodies, which are able to withstand the atmospheric flight. The proportion of iron bodies in interplanetary space would be expected to be even lower. According to the results of \citet{2005Icar..174...15B}, iron meteoroids are dominant among meteoroids on asteroidal orbits. This is in strong contrast to our results, suggesting that most cm to dm sized asteroidal meteoroids are chondritic. Our results are consistent with the expectations based on what we know about the transport of large main-belt asteroids to NEO space \citep{2017A&A...598A..52G}.

\citet{2017P&SS..143..159C} pointed out the specific ablation process of small (0.7 - 2.1 mm) iron meteoroids. The light curves typically exhibit very steep initial increase of brightness and a gradual decrease. Similar light curves have been detected among small refractory meteoroids by \citet{2015P&SS..118....8C}, likely also composed of iron--nickel alloy or iron sulfide grains. According to \citet{2017P&SS..143..159C} and \citet{2017A&A...606A..63G}, the ablation of these meteoroids is in the form of droplets released from a liquid layer at the meteoroid surface. \citet{2005Icar..174...15B} speculated that if the meteoroid consists of metallic iron with high thermal conductivity, the whole mm-sized meteoroid could be melted completely before the vaporization. The large proportion of iron meteoroids in the mm-sized population could be related to specific formation processes of these bodies and typical sizes of iron--nickel grains. Further investigation is necessary to reveal the processes by which the population of small iron meteoroids is formed.

While the abundance of iron bodies in our sample is very low, we detected several meteors positioned between the normal type and iron class. We introduce a new spectral class of Fe-rich meteors. These are defined as having significantly higher Fe/Mg intensity ratio compared to normal-type meteors, while still showing a notable presence of both Na I and Mg I. These meteors are positioned in the upper part of the ternary diagram (Fig. \ref{ternary}). No meteors were detected in this part of the ternary diagram in the surveys of fainter meteors \citep{2005Icar..174...15B, 2015A&A...580A..67V, 2019A&A...621A..68V}. In some cases, the intensity of Fe can be overestimated in meteors during bright flares. The pixel saturation and optical thickness of the radiating plasma can cause an apparent increase of the Fe/Mg ratio. To reduce this effect, frames with notably saturated spectra were neglected from the line intensity measurements. Furthermore, the Fe/Mg ratio was specifically studied in nonsaturated frames before and after meteor flares, and the relative intensity of the Fe I - 41 multiplet in the 438 - 441 nm region was evaluated to identify Fe-rich meteors.

\section{Meteor orbits}
\label{Sec:orb}

By triangulation of meteors observed by multiple stations, we determined atmospheric trajectories and heliocentric orbits for 146 of the 202 meteors analyzed in this work. The corresponding parameters for all multi-station meteors can be found in Tables 5 and 6 (available at the CDS). In most cases, the meteor entry speed was determined with an uncertainty of $<$ 1 km\,s\textsuperscript{-1}. Hyperbolic solutions were obtained for 15 meteors. In such cases, some orbital elements ($a$, $e$, $T_J$ and $Q$) were neglected from Table 5 (available at the CDS). 

\begin{figure}
\centerline{\includegraphics[width=\columnwidth,angle=0]{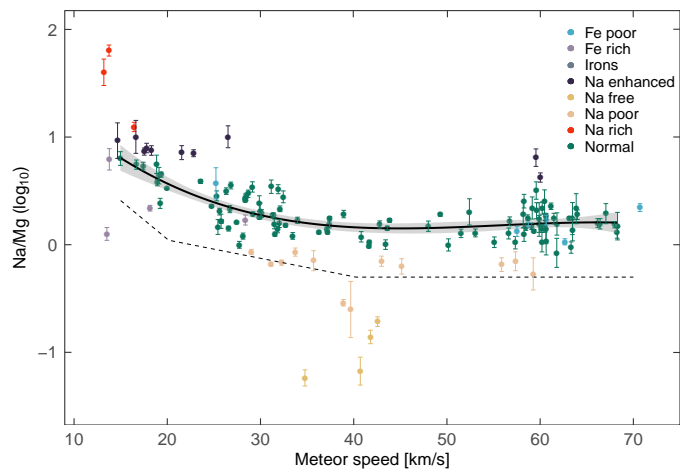}} \caption[f1]{Observed Na/Mg intensity ratio as a function of meteor speed. The solid line represents a cubic fit to the running average with standard error of the mean for meteors classified as normal. The dashed line represents the fit of normal-type meteors from \citet{2005Icar..174...15B} dealing with a mm-sized meteoroid population.} 
\label{speedcurve}
\end{figure}

The dependence of the Na/Mg ratio on meteor speed is shown in Fig. \ref{speedcurve}. This also illustrates the previously noted higher Na/Mg values for larger meteoroids measured by us compared to the ratio in smaller meteoroids studied by \citet{2005Icar..174...15B}. The Na/Mg intensity ratio is dependent on meteor speed (temperature) as a result of the low excitation of Na I line (2.1 eV) compared to Mg I (5.1 eV). Among mm-sized particles \citep{2005Icar..174...15B}, this effect is observed for meteor speeds below 40 km\,s\textsuperscript{-1}. Similar functionality is detected in our sample of larger meteoroids, with the threshold value of meteor speed between 35 and 40 km\,s\textsuperscript{-1} (Fig. \ref{speedcurve}). The breaking point near 35 km\,s\textsuperscript{-1} was recently  found as a good fit for data of \citet{2019A&A...621A..68V}.

\begin{figure}
\centerline{\includegraphics[width=\columnwidth,angle=0]{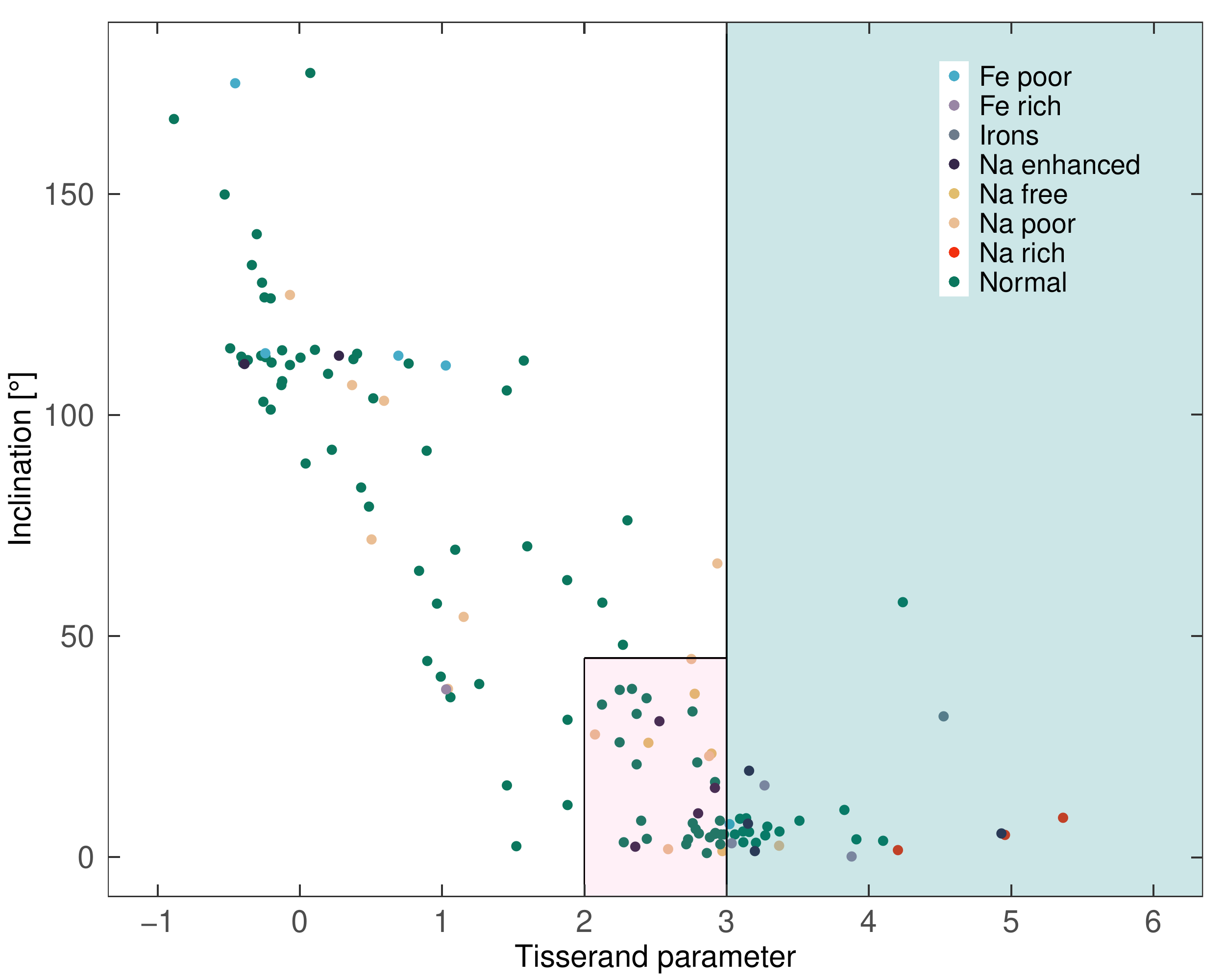}} \caption[f1]{Inclination vs. Tisserand parameter plot for measured meteors of different spectral classes. The colored areas represent asteroidal (green) and Jupiter-family-type (red) orbits.} 
\label{TissPlot}
\end{figure}

The orbits of meteors were classified based on the Tisserand parameter relative to Jupiter. The parameter is often used to define the orbital origin of interplanetary bodies \citep{1979aste.book..289K}, particularly comets, and can be in the same way used for meteoroids. We use it here to distinguish between asteroidal (AST) orbits ($T_J >$ 3), typical for bodies originating in the main asteroid belt, Jupiter-family type (JT) orbits (2 $< T_J <$ 3 and $i <$ 45$^{\circ}$) of bodies from the Kuiper Belt captured by the gravitational influence of Jupiter, and Halley-type (HT) orbits ($T_J <$ 2 and $i >$ 45$^{\circ}$) typical for cometary bodies originating in the Oort cloud. Figure 5 displays a distribution of meteor magnitudes for different types of meteoroid orbits based on the Tisserand parameter.

\begin{figure}[]
\centerline{\includegraphics[width=\columnwidth,angle=0]{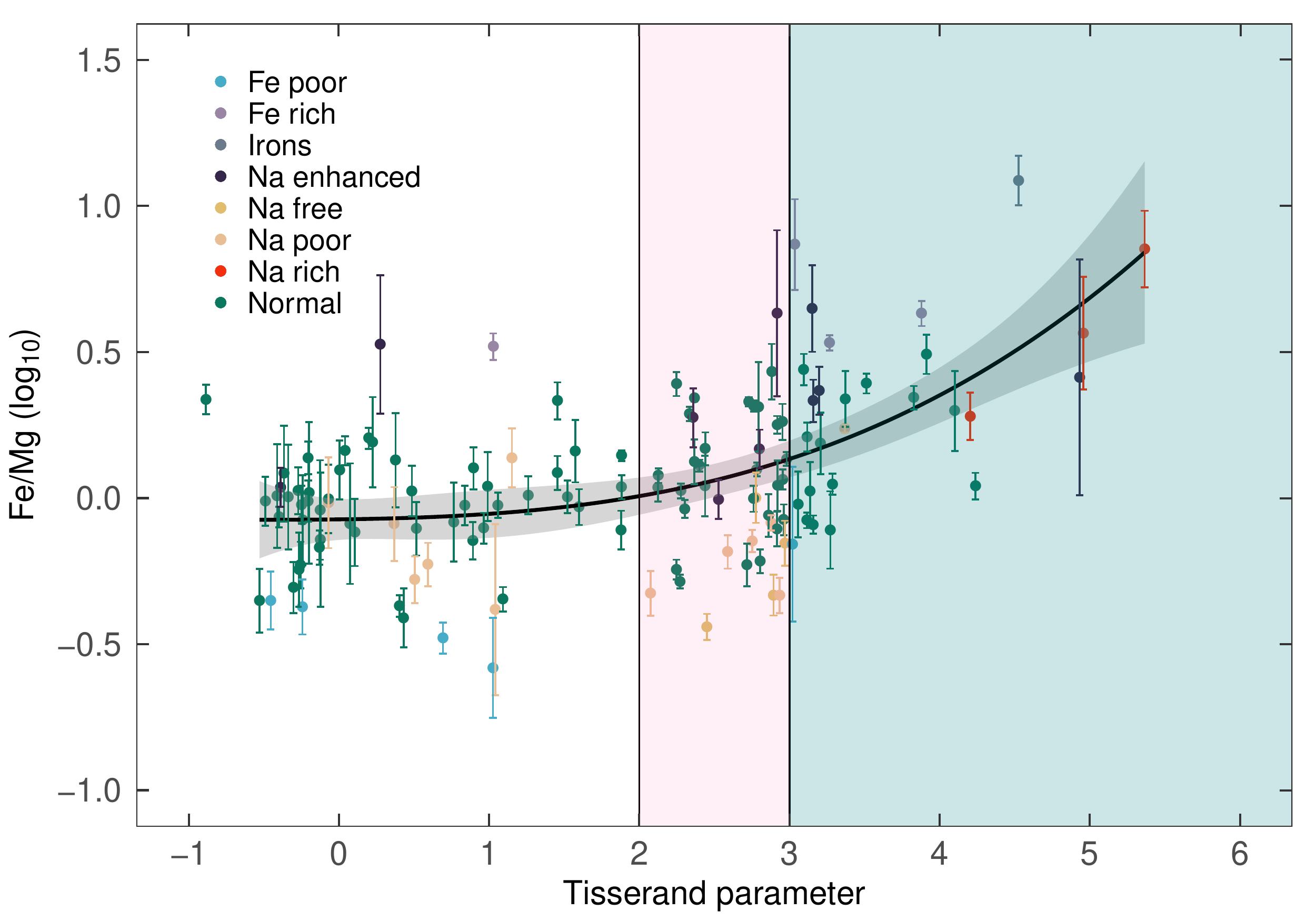}} 
\centerline{\includegraphics[width=\columnwidth,angle=0]{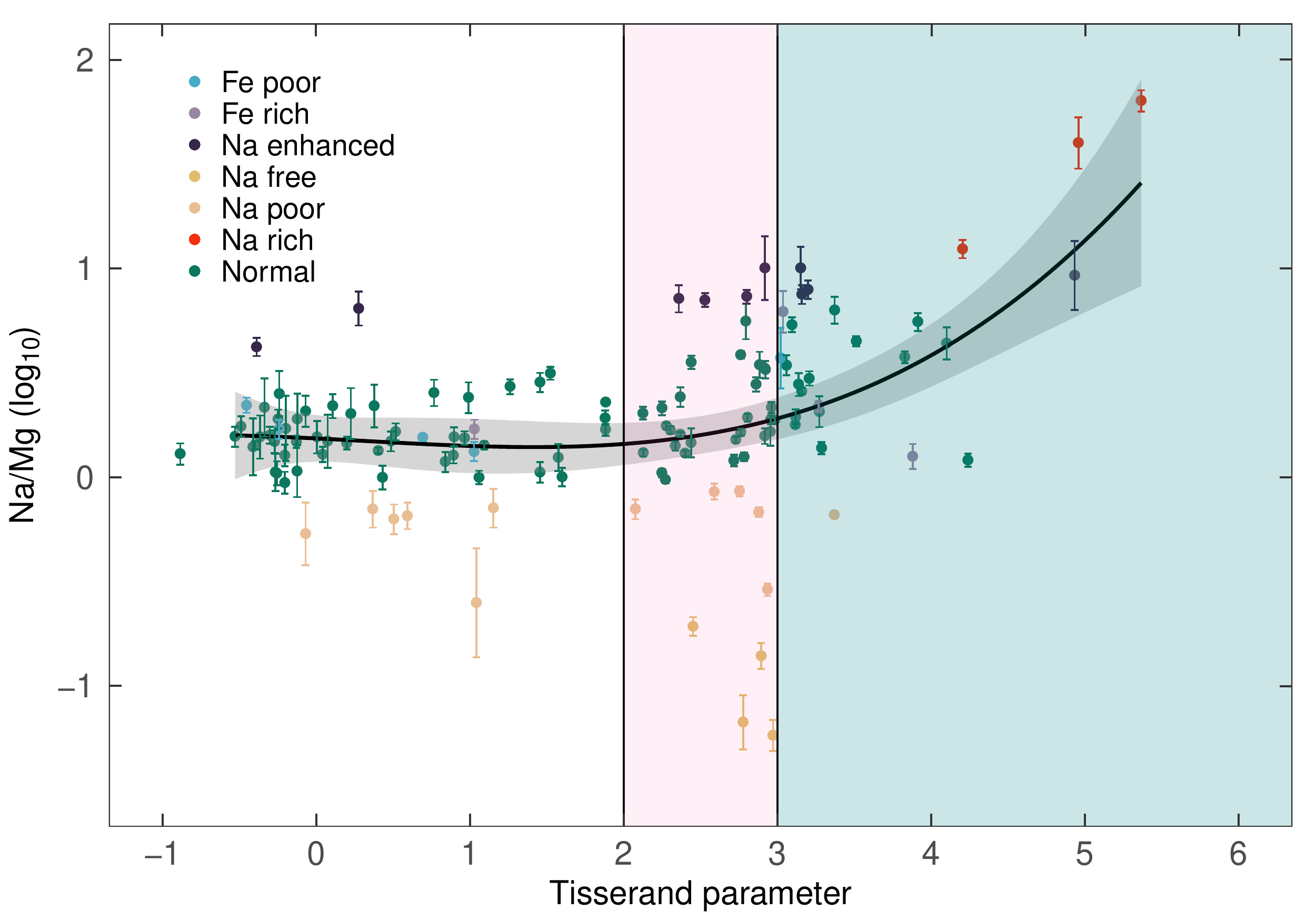}}
\caption[f1]{Observed Fe/Mg intensity ratio (upper) and Na/Mg intensity ratio (lower) as a function of the Tisserand parameter relative to Jupiter. A cubic fit to the running average of all data points and the standard error of the mean are plotted. The colored areas represent asteroidal (green), Jupiter-family type (pink), and Halley-type (white) orbits.} 
\label{TjSpec}
\end{figure}

Overall 61 meteoroids were found on Halley-type orbits, 42 meteoroids on Jupiter-family type orbits, and 28 meteoroids on asteroidal orbits (Fig. \ref{TissPlot}). To some degree, the orbital origin of meteoroids is reflected in the detected spectral properties of meteors. Figure \ref{TjSpec} shows the increase of Fe/Mg intensity ratio for meteoroids originating in the main asteroid belt and decrease of Fe/Mg intensity ratio for meteoroids on Halley-type orbits. We assume that the majority of meteoroids on orbits with $T_J >$ 3 (with the main exception of several meteoroids from the Taurid stream) are of asteroidal origin. 

\begin{figure}
\centerline{\includegraphics[width=\columnwidth,angle=0]{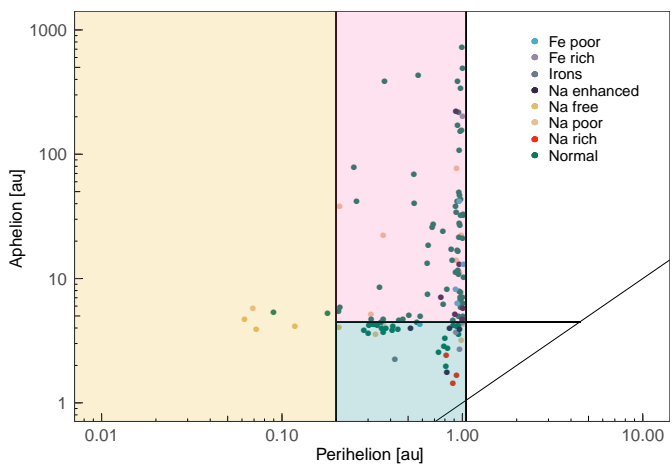}} \caption[f1]{Perihelion vs. aphelion plot in logarithmic scale for all observed meteoroids. The colored areas represent asteroidal (green), cometary (pink), and Sun-approaching orbits (yellow).} 
\label{qQ}
\end{figure}

These chondritic bodies have a larger iron content in general  compared to mostly cometary meteoroids on Halley-type orbits. The distinction between Jupiter-family and Halley-type meteoroids in the characteristic Fe/Mg ratios is less apparent, though on average is also present. This is mainly caused by the presence of Na-free meteoroids in this region and by numerous bodies near the $T_J =$ 3 limit. The mean Fe/Mg intensity ratio ($\pm$ standard error of the mean) for meteoroids on asteroidal orbits ($T_J >$ 3) is $\overline{Fe/Mg}_{AST}$ = 2.81 $\pm$ 0.33, for Jupiter-family meteoroids is $\overline{Fe/Mg}_{JF}$ = 1.28 $\pm$ 0.12, and for Halley-type meteoroids is $\overline{Fe/Mg}_{HT}$ = 1.01 $\pm$ 0.06. The larger standard error of Fe/Mg for asteroidal orbits is mainly caused by the presence of short-period cometary meteoroids, mainly Taurids and $\alpha$-Capricornids.

The Na/Mg ratio apparently increases for meteoroids on asteroidal orbits (Fig. \ref{TjSpec}) with $\overline{Na/Mg}_{AST}$ = 4.29 $\pm$ 1.21 compared to $\overline{Na/Mg}_{JF}$ = 2.31 $\pm$ 0.34 and $\overline{Na/Mg}_{HT}$ = 1.68 $\pm$ 0.10. The larger standard errors emphasize the wide spread of materials on Jupiter-family and asteroidal orbits. Still, the observed effect is mainly caused by an increase of the Na/Mg ratio in slower meteors (Fig. \ref{speedcurve}) characteristic of asteroidal meteoroids. All of the meteors with $T_J >$ 3 have initial velocities $v_i <$ 35 km\,s\textsuperscript{-1}.

Figures \ref{TissPlot} and \ref{qQ} both show that the distinction between asteroidal and cometary bodies near the $T_J =$ 3 boundary is difficult. There is indeed a large overlap between asteroids and Jupiter-family comets near $T_J =$ 3. Therefore, additional structural and spectral data must be considered to distinguish between the two sources. Meteoroid streams positioned near the $T_J =$ 3 boundary are sometimes defined as ecliptical meteor showers. From the showers identified in our sample, we find the Taurid complex showers, the $\alpha$-Capricornids, the $\eta$-Virginids, and the $\kappa$-Cygnids. Meteoroids with $T_J >>$ 3 are likely chondritic. This suggests that all Na-rich meteoroids in our sample are of asteroidal material.

Orbits with $q <$ 0.2 au have been classified separately by some authors as Sun-approaching. This group distinguishes meteoroids with small perihelion distances, in which intense solar radiation affects the spectra. The distinction is displayed on a perihelion versus aphelion plot (Fig. \ref{qQ}). A cluster of meteoroids can be observed near $Q =$ 4.5 au, formed by meteors from ecliptical showers (mainly the Taurids) affected by the the gravitational influence of Jupiter. The aphelion of the orbit of Jupiter is sometimes used as a boundary between comets and asteroids.

\section{Shower meteors}

\begin{figure*}[]
    \centering
    \begin{subfigure}[]{0.5\textwidth}
      \centering
      \includegraphics[width=\textwidth]{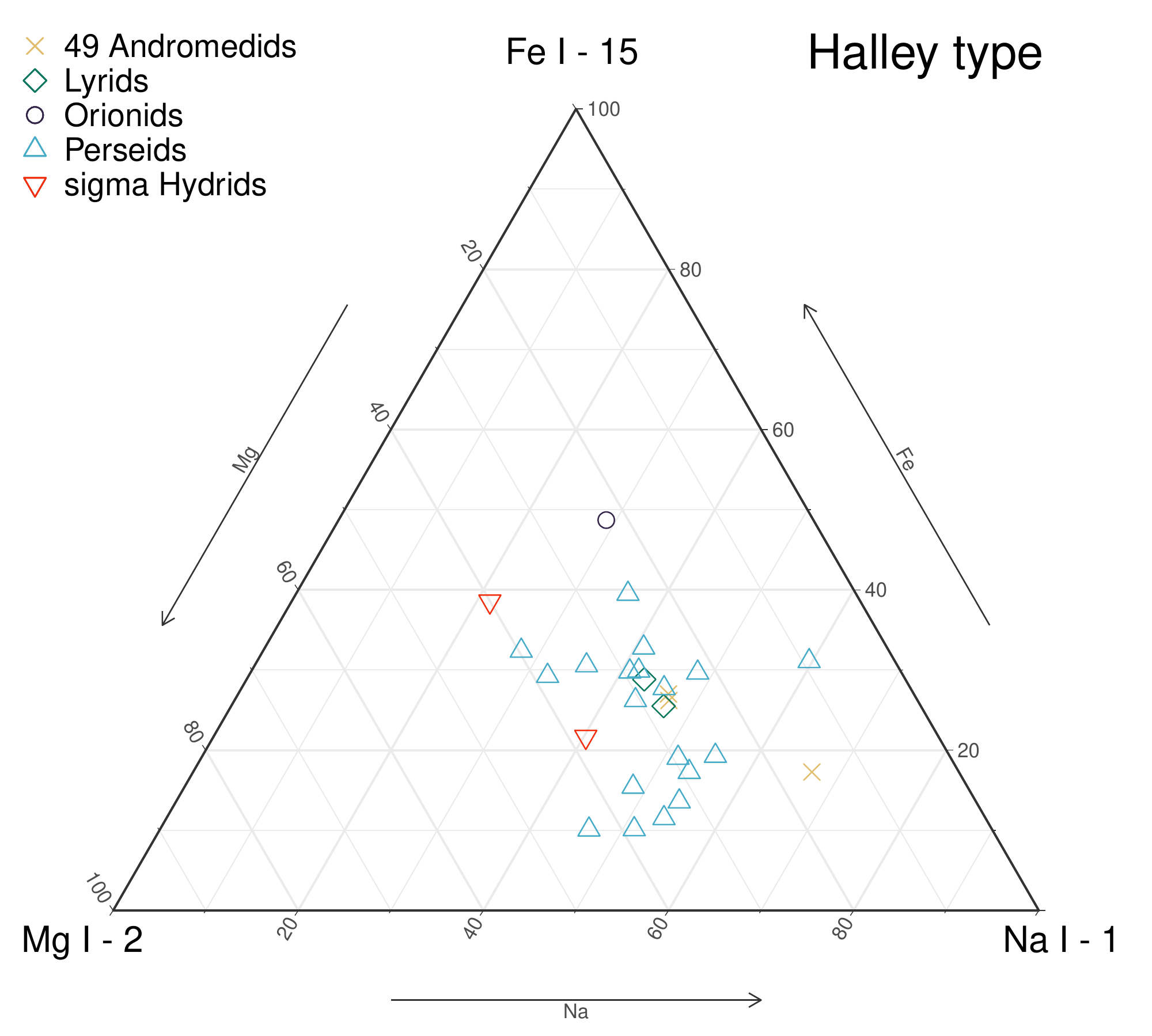}
      \caption{}
      \label{}
    \end{subfigure}%
    ~
    \begin{subfigure}[]{0.5\textwidth}
      \centering
      \includegraphics[width=\textwidth]{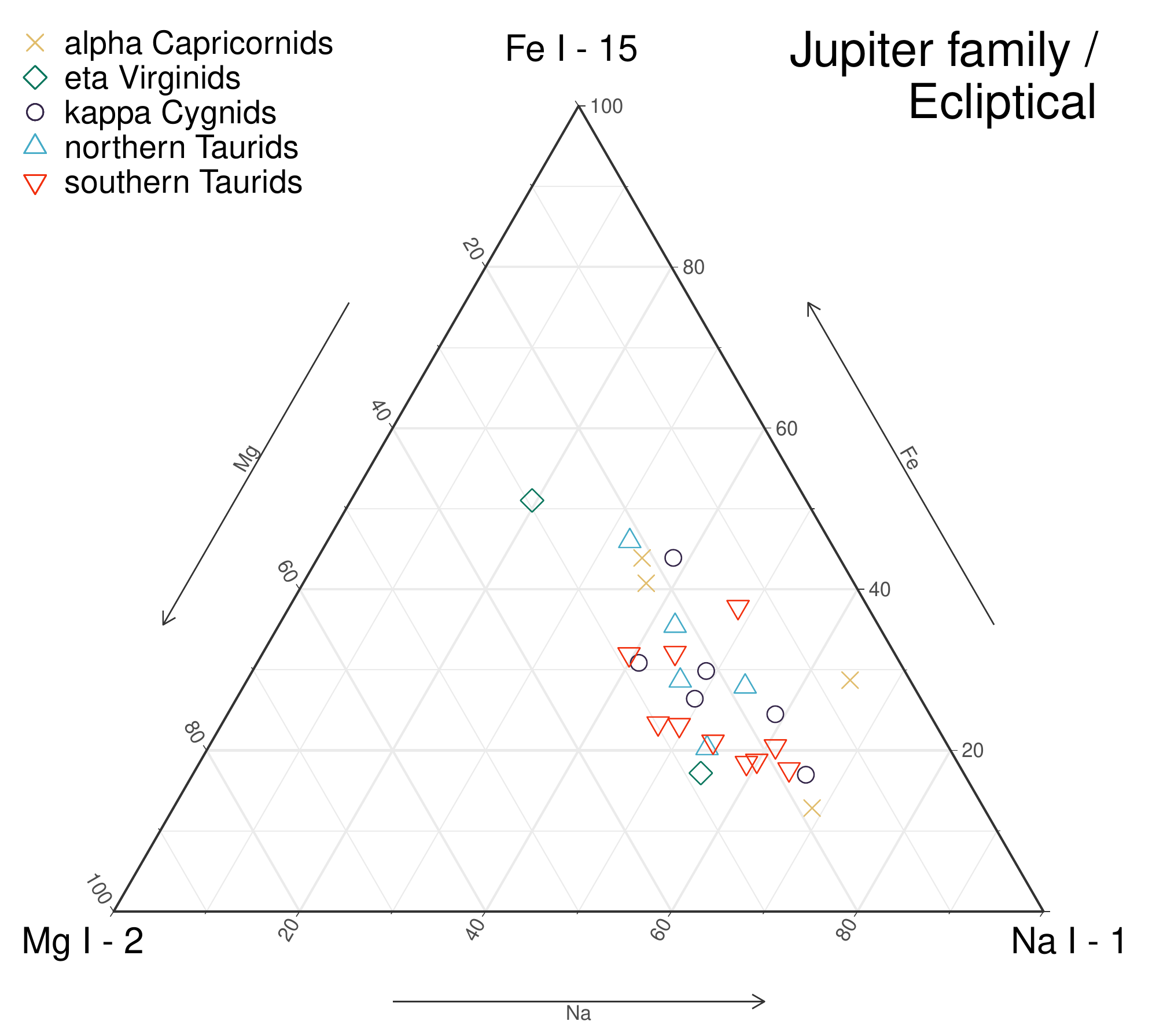}
      \caption{}
      \label{}
    \end{subfigure}%
    \caption{Spectral classification for identified meteors from established meteoroid streams. Separately plotted are meteoroid streams with Halley-type orbits (a) and Jupiter-family type or ecliptical orbits (b).}
    \label{est}
  \end{figure*}%
  
Of the 146 meteoroids for which orbital data is available, 80 were identified as members of known meteor showers and 66 were sporadic. In general, sporadic meteors are thought to have been released from their parent body thousands to millions of years ago and to have moved separately from the recognized parent stream. The shower association was performed within the \textit{MT} software based on radiant position, meteor speed, and solar longitude. The orbital similarity was evaluated using the Southworth-Hawkins criterion \citep{1963SCoA....7..261S} which defines the difference between two meteors represented as a point in a five-dimensional phase space of orbital elements.
  
We considered both established and working list streams of the IAU Meteor Data Center (MDC) \citep{2017P&SS..143....3J} for the identification. In cases where multiple streams were found close to the meteor orbit, we individually estimated the most probable source considering the orbital similarity, radiant position, and shower activity. 
 
Established showers are caused by confirmed meteoroid streams with well-defined meteor activity. The spectral classification of meteors associated with established Jupiter-family-type or ecliptical and Halley-type meteoroid streams is shown in Fig. \ref{est}. Overall, the Jupiter-family and ecliptical showers are shifted slightly to the right in the ternary diagram compared to Halley-type showers. This mainly reflects the generally lower meteor speeds of their meteors. The ternary diagrams in Fig. \ref{est} are dominated by Perseids and Taurids, respectively. All of the Taurid meteoroids analyzed here were observed during the 2015 outburst and were analyzed previously by \citet{2017P&SS..143..104M}.

\subsection{Halley-type}

 \begin{figure}[t]
\centerline{\includegraphics[width=\columnwidth,angle=0]{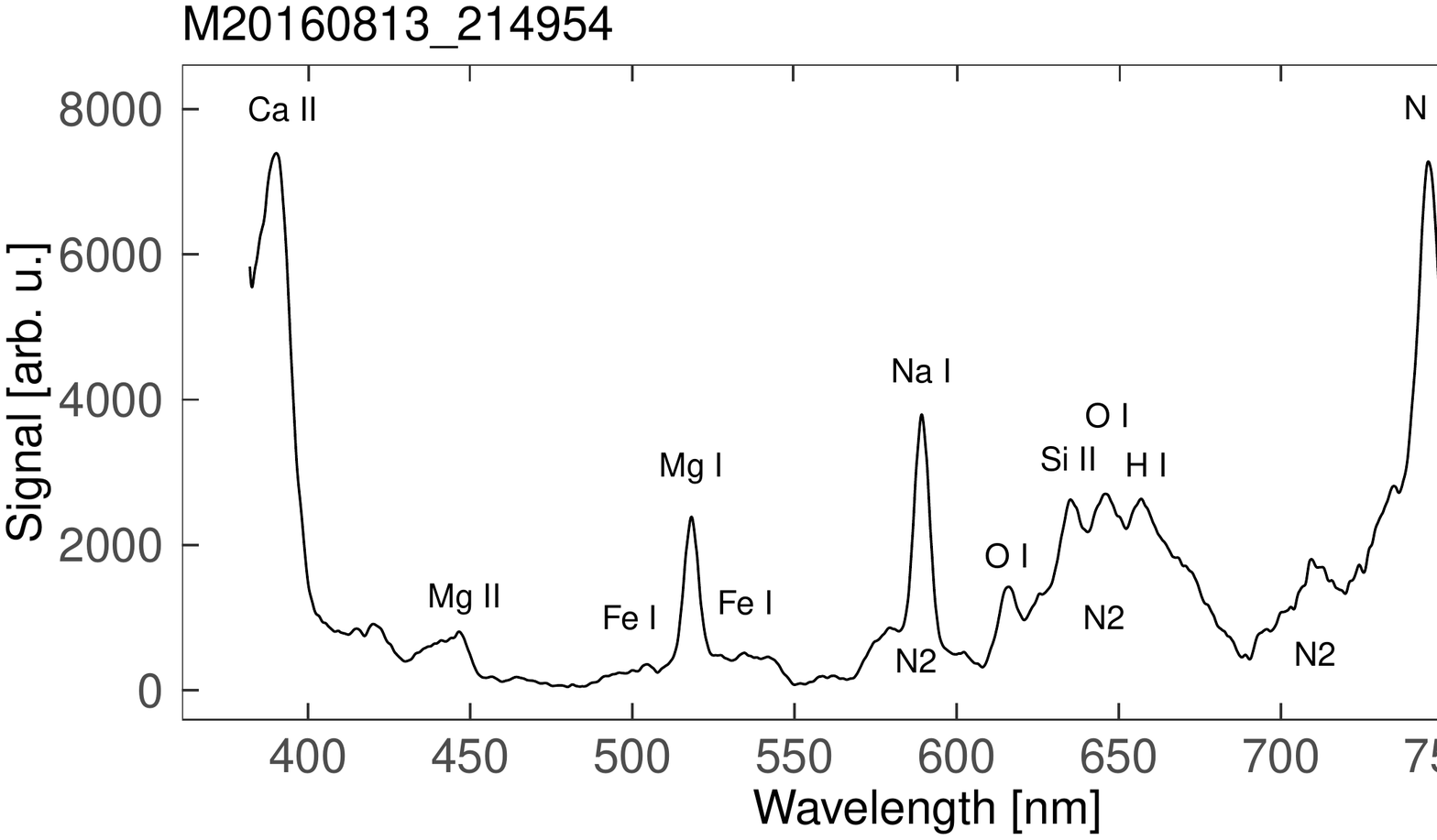}}
\centerline{\includegraphics[width=\columnwidth,angle=0]{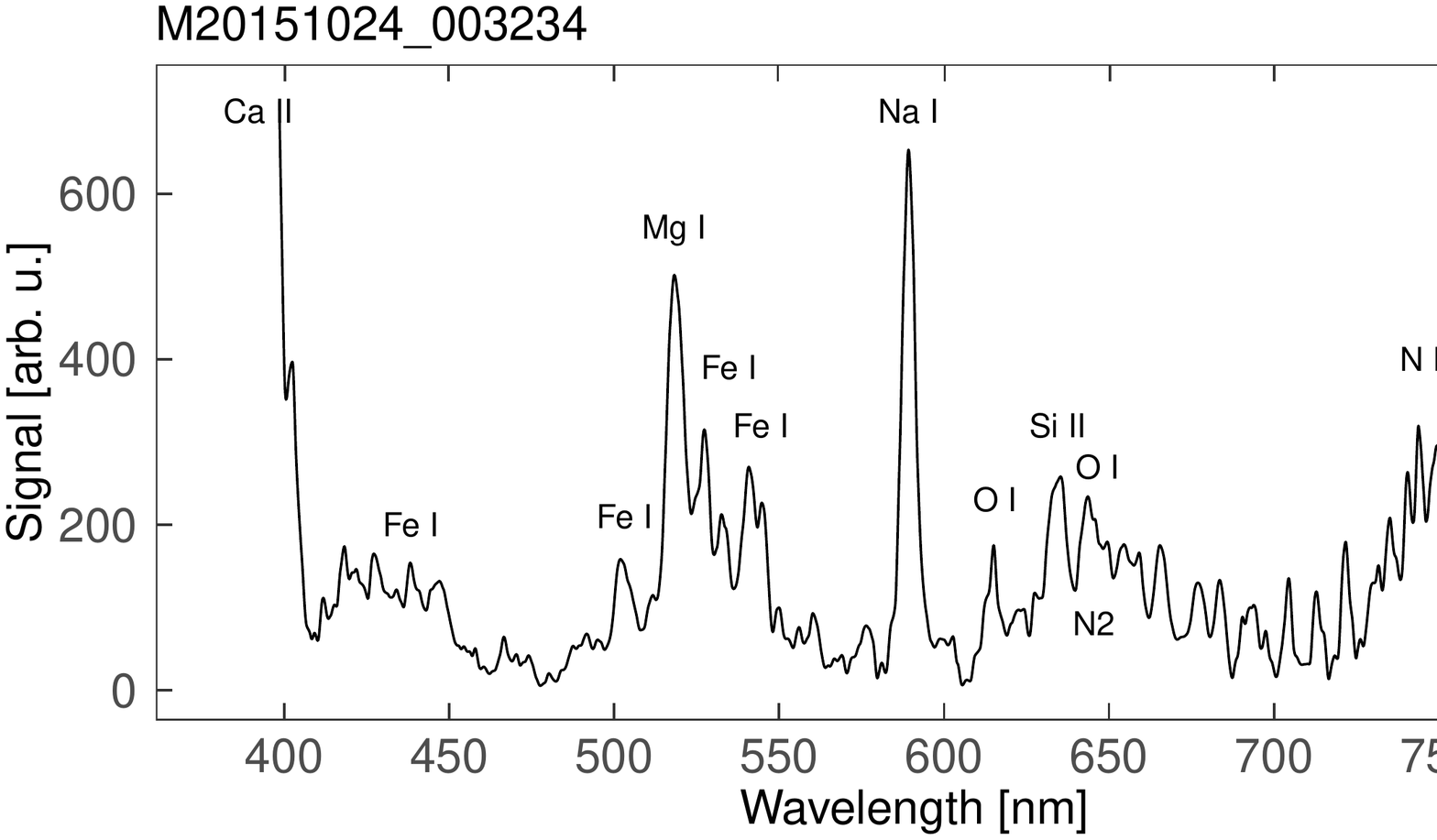}}
\caption[f1]{Spectral profiles of a bright Perseid meteor M20160813\_214954 and a Orionid meteor M20151024\_003234.} 
\label{Perprof}
\end{figure}

The most represented meteoroid stream in our sample are the Perseids originating in comet 109P/Swift–Tuttle. A list of meteors associated with this stream and including their orbits and atmospheric data is provided in Table \ref{tab:Per}. Several more Perseid spectra were observed, but with the only recognizable lines being atmospheric O I and N I. These types of spectra were excluded from our analysis. Most Perseid meteors are positioned in the middle of the ternary diagram (Fig. \ref{est}) with normal-type spectra close to the expected chondritic ratios of Na/Mg/Fe. A relatively large variation of Fe content is observed, between 10 and 40\% in the ternary ratios. Four Fe-poor Perseids are observed. The Na/Mg ratios for Perseids are within the expected normal-type region, with one exception identified as Na-enhanced. One Perseid meteor was classified as Na-poor, pointing out the possible effect of cosmic-ray irradiation causing volatile depletion on Halley-type orbits.
  
\begin{table*}
\centering
\small\begin{center}
\caption {Atmospheric and orbital properties of identified Perseid meteoroids. For references, see Tables 5 and 6 (available at the CDS).}
\vspace{0.1cm}
\resizebox{\textwidth}{!}{\begin{tabular}{llrrrrrrrrrcccccc}
\hline\hline\\
\multicolumn{1}{c}{Code}& %
\multicolumn{1}{l}{Spec. type}& %
\multicolumn{1}{c}{Mag}& %
\multicolumn{1}{c}{$\alpha_g$}& %
\multicolumn{1}{c}{$\delta_g$}& %
\multicolumn{1}{c}{$v_g$}& %
\multicolumn{1}{c}{$H_B$}& %
\multicolumn{1}{c}{$H_E$}& %
\multicolumn{1}{c}{$q$}& %
\multicolumn{1}{c}{$i$}& %
\multicolumn{1}{c}{$\omega$}& %
\multicolumn{1}{c}{$T_J$}& %
\multicolumn{1}{c}{$K_B$}& %
\multicolumn{1}{c}{}& %
\multicolumn{1}{c}{$P_E$}& %
\multicolumn{1}{c}{}\\
\hline\\
M20150812\_232102 & Fe poor     & -7.1 & 46.17 & 57.90 & 61.51 & 126.80 & 86.88 & 0.965 & 114.55 & 155.45 & - & 6.00    & D  & -5.82 & IIIB \\
&  & $\pm$ 1.5 & 0.05 & 0.03 & 0.58 & 0.18 & 0.11 & 0.001 & 0.32 & 0.57 &  &  &  &  &  \vspace{0.1cm} \\
M20150813\_020206 & Fe poor     & -4.2 & 47.99 & 58.03 & 56.34 & 110.87 & 82.60 & 0.937 & 111.13 & 145.38 & 1.03    & 6.75 & C3 & -5.11 & II   \\
&  & $\pm$ 1.0 & 0.09 & 0.01 & 0.41 & 0.17 & 0.02 & 0.002 & 0.30 & 0.88 &  &  &  &  &  \vspace{0.1cm} \\
M20160812\_005618 & Fe poor     & -5.6 & 45.90 & 57.35 & 59.50 & 109.83 & 79.92 & 0.960 & 113.97 & 153.20 & -0.24   & 6.93 & C2 & -5.07 & II   \\
&  & $\pm$ 1.0 & 0.06 & 0.02 & 0.22 & 0.17 & 0.11 & 0.001 & 0.15 & 0.26 &  &  &  &  &  \vspace{0.1cm} \\
M20160813\_214954 & Fe poor     & -9.5 & 53.15 & 57.51 & 57.52 & 119.18 & 82.16 & 0.917 & 113.34 & 141.86 & 0.69    & 6.37 & D  & -5.81 & IIIB \\
&  & $\pm$ 0.6 & 0.08 & 0.05 & 0.66 & 0.13 & 0.07 & 0.004 & 0.44 & 1.30 &  &  &  &  &  \vspace{0.1cm} \\
M20150810\_013034 & Na poor     & -3.4 & 46.10 & 60.72 & 56.16 & 107.64 & 90.28 & 0.925 & 106.69 & 144.38 & 0.37    & 7.00    & C2 & -5.54 & IIIA \\
&  & $\pm$ 1.3 & 0.07 & 0.02 & 0.79 & 0.18 & 0.14 & 0.004 & 0.60 & 1.41 &  &  &  &  &  \vspace{0.1cm} \\
M20150807\_002948 & normal      & -3.0 & 41.59 & 56.60 & 59.46 & 109.99 & 86.79 & 0.940 & 113.19 & 148.41 & -0.41   & 6.92 & C2 & -5.10  & II   \\
&  & $\pm$ 1.4 & 0.04 & 0.02 & 0.25 & 0.03 & 0.01 & 0.001 & 0.12 & 0.22 &  &  &  &  &  \vspace{0.1cm} \\
M20150808\_232655 & normal      & -5.1 & 42.10 & 58.66 & 59.46 & 119.90 & 75.59 & 0.953 & 111.05 & 151.90 & - & 6.30  & D  & -4.69 & II   \\
&  & $\pm$ 1.3 & 0.05 & 0.02 & 0.08 & 0.06 & 0.01 & 0.000 & 0.04 & 0.13 &  &  &  &  &  \vspace{0.1cm} \\
M20150811\_005737 & normal      & -5.0 & 48.45 & 56.63 & 58.16 & 115.95 & 83.84 & 0.919 & 113.81 & 142.88 & 0.40    & 6.48 & D  & -5.28 & IIIA \\
&  & $\pm$ 1.1 & 0.04 & 0.02 & 0.12 & 0.04 & 0.01 & 0.001 & 0.07 & 0.27 &  &  &  &  &  \vspace{0.1cm} \\
M20150813\_011318 & normal      & -3.4 & 47.64 & 58.25 & 60.66 & 112.71 & 83.26 & 0.955 & 113.73 & 152.68 & - & 6.74 & C3 & -4.92 & II   \\
&  & $\pm$ 0.7 & 0.07 & 0.03 & 1.38 & 0.31 & 0.03 & 0.003 & 0.72 & 1.10 &  &  &  &  &  \vspace{0.1cm} \\
M20150814\_020823 & normal      & -8.8 & 48.72 & 59.29 & 59.12 & 117.19 & 83.14 & 0.952 & 111.71 & 151.57 & -0.40   & 6.37 & D  & -5.90  & IIIB \\
&  & $\pm$ 1.3 & 0.04 & 0.11 & 0.66 & 0.17 & 0.16 & 0.002 & 0.46 & 0.80 &  &  &  &  &  \vspace{0.1cm} \\
M20160809\_003127 & normal      & -3.5 & 45.35 & 58.01 & 58.72 & 114.09 & 87.75 & 0.936 & 111.86 & 147.34 & -0.20   & 6.63 & C2 & -5.30  & IIIA \\
&  & $\pm$ 0.8 & 0.06 & 0.03 & 0.37 & 0.23 & 0.04 & 0.002 & 0.24 & 0.59 &  &  &  &  & \vspace{0.1cm}  \\
M20160811\_195200 & normal      & -6.0 & 45.91 & 57.70 & 59.38 & 121.57 & 88.68 & 0.958 & 113.40 & 152.70 & -0.27   & 6.30  & D  & -5.61 & IIIA \\
&  & $\pm$ 0.8 & 0.35 & 0.27 & 1.23 & 1.18 & 0.56 & 0.003 & 0.62 & 1.50 &  &  &  &  &  \vspace{0.1cm} \\
M20160811\_230132 & normal      & -5.0 & 47.66 & 56.70 & 59.02 & 109.51 & 82.46 & 0.946 & 114.68 & 149.24 & 0.11    & 6.99 & C2 & -5.06 & II   \\
&  & $\pm$ 0.5 & 0.03 & 0.09 & 0.56 & 0.17 & 0.10 & 0.002 & 0.29 & 0.87 &  &  &  &  & \vspace{0.1cm}  \\
M20160812\_002835 & normal      & -4.6 & 47.58 & 57.70 & 57.96 & 112.84 & 80.84 & 0.944 & 112.65 & 148.30 & 0.38    & 6.71 & C2 & -5.04 & II   \\
&  & $\pm$ 0.8 & 0.10 & 0.06 & 0.40 & 0.33 & 0.22 & 0.002 & 0.27 & 0.64 &  &  &  &  & \vspace{0.1cm}  \\
M20160812\_011115 & normal      & -1.8 & 44.82 & 58.31 & 62.36 & 101.71 & 85.58 & 0.972 & 114.34 & 157.80 & - & 7.54 & A  & -4.72 & II   \\
&  & $\pm$ 0.6 & 0.04 & 0.03 & 0.67 & 0.07 & 0.42 & 0.001 & 0.36 & 0.56 &  &  &  &  &  \vspace{0.1cm} \\
M20160812\_011418 & normal      & -5.3 & 45.95 & 57.80 & 58.75 & 107.14 & 82.26 & 0.958 & 112.91 & 152.27 & 0.00    & 7.09 & C2 & -5.19 & II   \\
&  & $\pm$ 0.7 & 0.10 & 0.37 & 0.63 & 0.55 & 0.42 & 0.002 & 0.48 & 1.05 &  &  &  &  & \vspace{0.1cm}  \\
M20160812\_014157 & normal      & -7.2 & 48.65 & 59.96 & 57.30 & 101.86 & 75.08 & 0.936 & 109.32 & 147.05 & 0.20    & 7.42 & A  & -5.04 & II   \\
&  & $\pm$ 0.8 & 0.05 & 0.33 & 0.82 & 0.83 & 0.63 & 0.004 & 0.60 & 1.64 &  &  &  &  & \vspace{0.1cm}  \\
M20150811\_012547 & Na enhanced & -3.0 & 43.22 & 56.75 & 58.38 & 103.85 & 80.97 & 0.960 & 113.35 & 152.31 & 0.28    & 7.31 & A  & -4.68 & II   \\
&  & $\pm$ 1.0 & 0.52 & 0.12 & 0.78 & 0.34 & 0.13 & 0.004 & 0.54 & 1.45 &  &  &  &  & \vspace{0.1cm}  \\
M20150812\_011600 & normal      & -2.5 & 43.62 & 56.53 & 59.50 & 114.15 & 86.06 & 0.968 & 114.62 & 155.20 & -0.12   & 6.62 & C2 & -5.02 & II   \\
&  & $\pm$ 1.0 & 0.05 & 0.01 & 0.56 & 0.20 & 0.05 & 0.001 & 0.34 & 0.63 &  &  &  &  &  \vspace{0.1cm} \\
\hline
\end{tabular}}
\label{tab:Per}
\end{center}
\end{table*}

The brightest and most detailed Perseid spectrum in our sample can be seen in Fig. \ref{Perprof}. The corresponding meteor was estimated to be of -9.5 mag, enabling us to observe more detailed spectral features compared to the majority of Perseid spectra dominated by atmospheric emission. The characteristic Ca II ion H and K doublet at 393.4 nm and 396.9 nm is observed even by our system with low sensitivity in the near-UV. Other ion lines typical of the high-temperature component in bright and fast meteors can be also seen: the Mg II line at 448.1 nm, Si II at 634.7 nm, and Ca II at 854.2 nm. The H$\alpha$ line (656.3 nm) of the Balmer series, which can be associated with organic cometary content, is very bright in this spectrum. The spectrum showed very low content of Fe and was classified as Fe-poor. The meteor started to radiate very high in the atmosphere ($H_B$ = 119.2 km) and disintegrated completely at around 82 km altitude, reflecting its very fragile structure. All of the detected features correspond to the soft cometary material of 109P/Swift-Tuttle.

Among other Halley-type showers, we have detected a pair of meteoroids from $\sigma$-Hydrids and Lyrids and three 49 Andromedids showing relatively heterogeneous spectral properties. Some degree of heterogeneity in Fe/Mg and Na/Mg ratios (typically up to 30\% in ternary Na/Mg/Fe content) among meteoroids from one parent object appears to be natural. This is suggested from the wider spectral studies of Leonids, Geminids \citep{2005Icar..174...15B}, Taurids \citep{2017P&SS..143..104M}, and Perseids (this work). Sodium enhancement was confirmed in one 49 Andromedid. One $\sigma$-Hydrid is classified as Na-poor as a result of its small perihelion distance. Our sample also includes one Orionid meteoroid with relatively high Fe intensity (Fig. \ref{Perprof}). The meteor spectrum was observed during a meteor flare, which could cause some degree of Fe intensity overestimation due to saturation and optical thickness of the radiating plasma. Even considering these effects, the observed Fe intensity is unusually high for a cometary body originating in comet 1P/Halley.

\subsection{Ecliptical}
  
Besides Taurids, ecliptical showers (Fig. \ref{est}) were most represented in our sample by the $\kappa$-Cygnids (6 meteoroids) and $\alpha$-Capricornids (4 meteoroids). $\kappa$-Cygnids are spectrally similar to the Taurids, showing relatively similar Na content and higher variations of Fe intensity. The $\alpha$-Capricornids show two distinct types of spectra. We believe that these variations reflect structural differences between the stream meteoroids. The two types of $\alpha$-Capricornids exhibited distinct light curves. The first type shows spectra with increased Fe intensity observed during bright flares associated with sudden disruption. The second type shows Na enhancement and lower Fe content with smooth light curves without flares. The characteristic light curves with bright flares at the end of trajectory were previously observed by \citet{2014Icar..239..273M}, who also analyzed $\alpha$-Capricornid emission spectra and concluded that the meteoroids and their parent body are of nonchondritic composition. The spectrum modeling methodology applied by \citet{2014Icar..239..273M} is however not suitable for the low-resolution spectra and the presented abundances need to be confirmed using higher-resolution data. Overall, the detected light curves and spectra of $\alpha$-Capricornids suggest structural and compositional heterogeneity of the meteoroid stream originating from the inactive comet 169P/NEAT \citep{2010AJ....139.1822J}.

\begin{figure}
\centerline{\includegraphics[width=\columnwidth,angle=0]{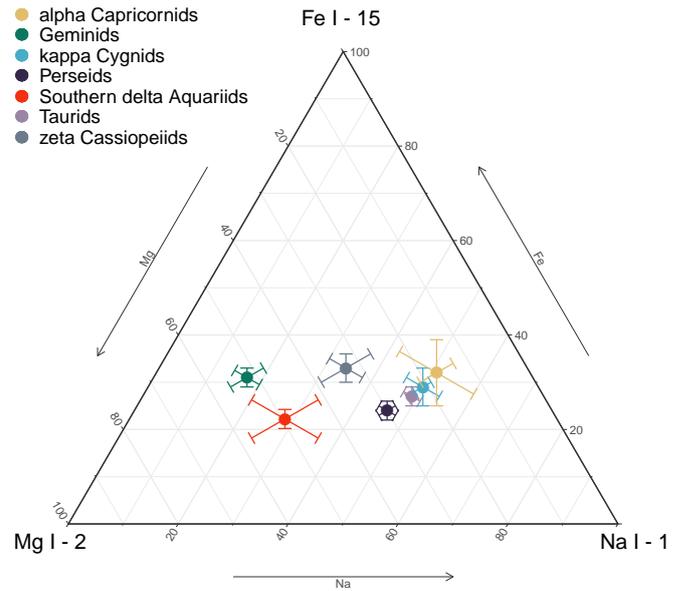}} \caption[f1]{Mean spectral classification (mean Na I/Mg I/Fe I intensity ratios and standard error of the mean) of meteoroid streams with at least three meteoroids in our sample.} 
\label{ternaryavrcomp}
\end{figure}

The spectral differences in mean Na/Mg/Fe ratios between the major meteoroid streams in our sample are displayed on Fig. \ref{ternaryavrcomp}. This plot shows that while the spectral heterogeneity within one stream can be quite substantial (Fig.\ref{est}), the mean Na/Mg/Fe ratios can be used to distinguish between different streams. For accurate characterization of a mean spectral type of a meteoroid stream, sufficient statistical sample is required (over 10 meteors), which is only satisfied for Perseids and Taurids in our sample.

Figure \ref{ternaryavrcomp} also shows the spectral similarity of Taurids and $\kappa$-Cygnids (and potentially also $\alpha$-Capricornids). This might imply that the two streams have parent objects of similar composition. Similarly to Taurids, $\kappa$-Cygnids are thought to be remnants of a previous large comet disruption. The largest remaining object from the break-up associated with $\kappa$-Cygnids is the dormant 2008 ED69 \citep{2008AJ....136..725J}. The spectrum of this body is not known, but the fragile structure and chondritic abundances found in one $\kappa$-Cygnid by \citet{2009MNRAS.392..367T} could suggest a C- or D-type body. On the other hand, comet 2P/Encke associated with the major stream of Taurids is Xe-type but has a spectrum resembling primitive asteroids \citep{2015A&A...584A..97T}. The link between Taurids and primitive carbonaceous materials was also supported by the two associated meteorite falls of Maribo and Sutter's Mill \citep{2015A&A...584A..97T}. Overall, this could imply a similar primitive (C- or D-type) composition of the parent objects of $\kappa$-Cygnids and Taurids.

\subsection{Unconfirmed showers}
  
\begin{figure*}[]
    \centering
    \begin{subfigure}[]{0.5\textwidth}
      \centering
      \includegraphics[width=\textwidth]{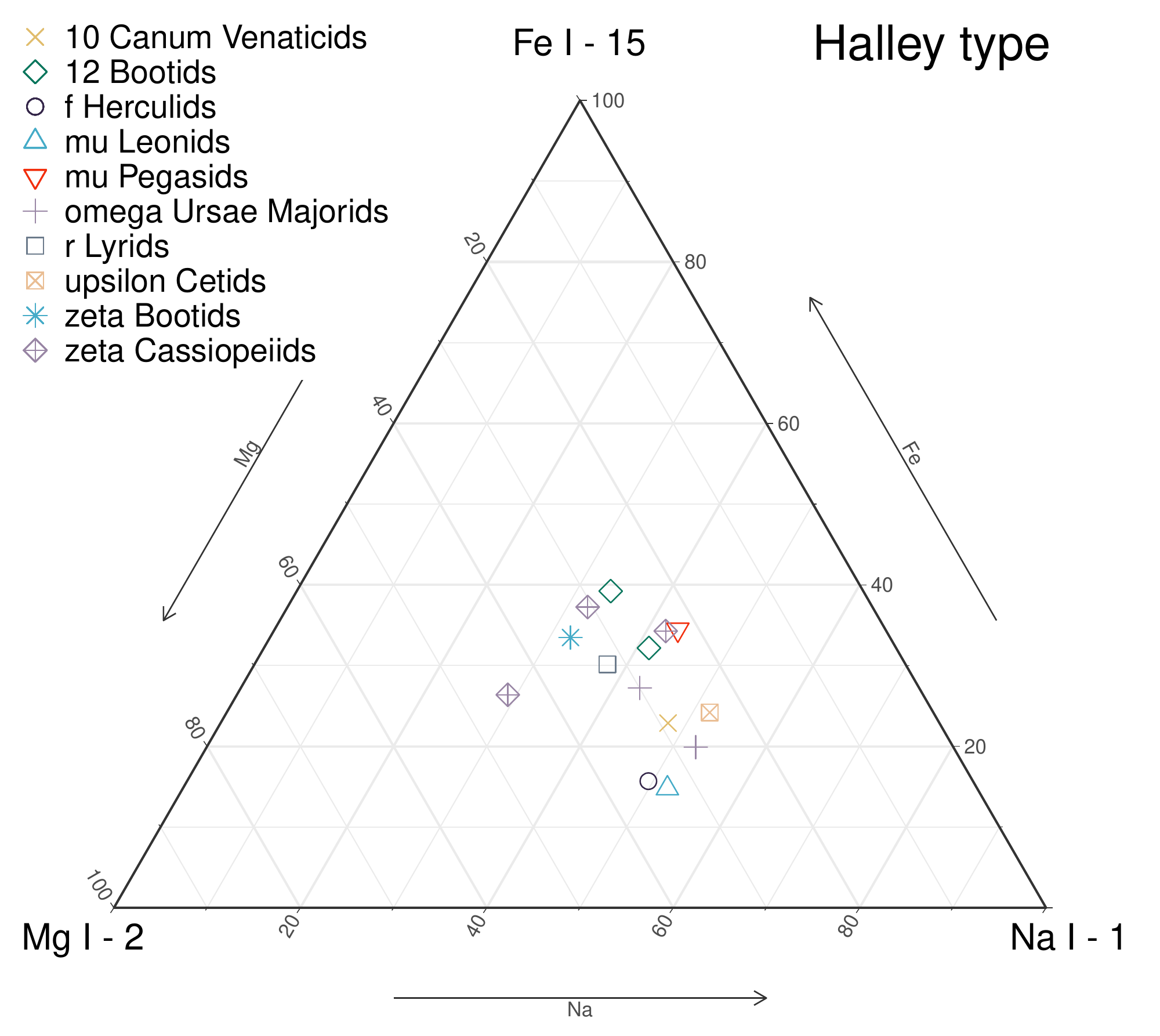}
      \caption{}
      \label{}
    \end{subfigure}%
    ~
    \begin{subfigure}[]{0.5\textwidth}
      \centering
      \includegraphics[width=\textwidth]{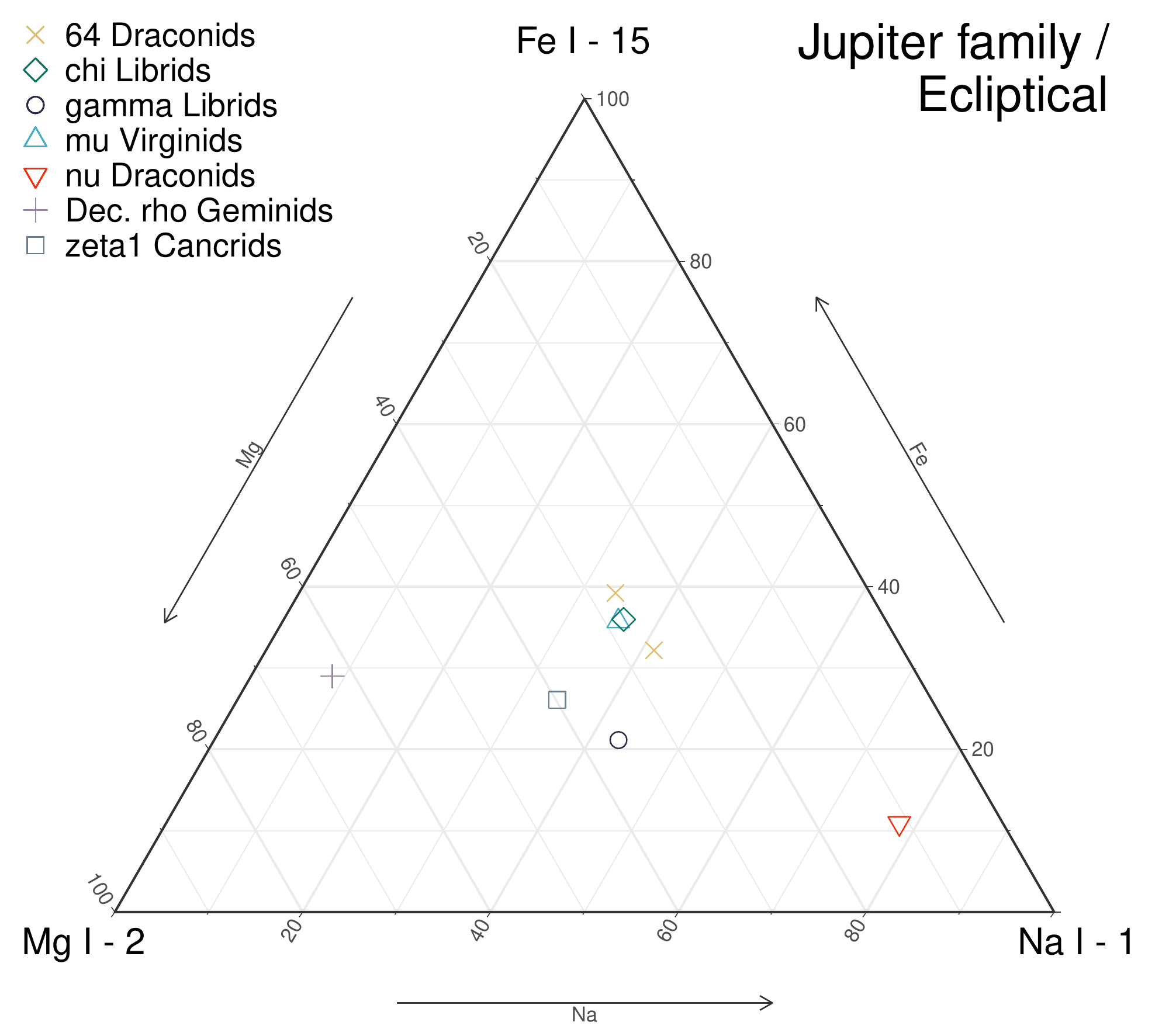}
      \caption{}
      \label{}
    \end{subfigure}%
    \caption{Spectral classification for identified meteors from unconfirmed (working list) meteoroid streams. Separately plotted are meteoroid streams with Halley-type orbits (a) and Jupiter-family type or ecliptical orbits (b).}
    \label{wor}
  \end{figure*}

The showers in the IAU MDC working list are in the process of being confirmed as real individual meteoroid streams. A large survey by \citet{2016Icar..266..355J} recently showed that many of these streams show significant annual activity. Here, we only report on mainly individual meteor spectra associated with working list showers, which if confirmed could provide the first spectral data for these streams. The identified Halley-type and Jupiter-family/ecliptical working list shower meteors are displayed in  Fig. \ref{wor}. 
 
The spectra of Halley-type shower meteors are mostly normal type, with no notable differences from other major stream meteoroids. One of the three $\zeta$-Cassiopeiids showed slight Na depletion and was classified as Na-poor. Overall, their spectral characteristics are similar to Perseids, though on average they exhibit a slight increase in Fe/Na. More distinct spectral classes were detected among Jupiter-family and ecliptical shower meteors (Fig. \ref{wor}). Specifically, one Na-enhanced meteoroid was assigned to the $\nu$-Draconids stream, which was reported as a duplicate of the established August Draconids \citep{2016Icar..266..355J}. The shower is active in August/September with characteristically low-velocity meteors ($v_g$ = 20.1 km\,s\textsuperscript{-1})). The parent body of the stream is not known. As a preliminary estimate, we suspect that the detected Na-enhancement is mainly caused by the low speed of the meteor and does not reflect atypical composition within the stream. This is supported by the detected chondritic Fe/Mg ratio and no identified anomalous lines.

One December $\rho-$Geminid (DRG) meteoroid was detected with a Na-free spectrum, similarly to the typical spectra \citep{2005Icar..174...15B} of the major Geminid stream of 3200 Phaeton. \citet{2016Icar..266..331J} identified DRG meteoroids as members of the Geminid complex. It is assumed that dust particles released from Phaethon, 2005 UD, and potentially another yet-to-be discovered parent body contribute to the streams within the Geminid complex \citep{2016Icar..266..331J}. December $\rho-$Geminid meteoroids have been identified as unusually fast Geminids, which is also confirmed in our sample ($v_g =$ 40.5 km\,s\textsuperscript{-1}). \citet{2016Icar..266..331J} speculated that the observed unusual deceleration of these meteoroids could be linked with atypical physical properties and low density. Our sample shows spectrum depleted in Na as a result of strong solar radiation, and high material strength (type ast/I) of the meteoroid, characteristic for ordinary chondrites. Both spectrally and physically, this meteoroid fits well with other Geminids from the main stream and asteroidal parent body, but the reason behind the high speed of this branch remains unclear.
  
\section{Structure} \label{secSTR}

\begin{figure}
\centerline{\includegraphics[width=\columnwidth,angle=0]{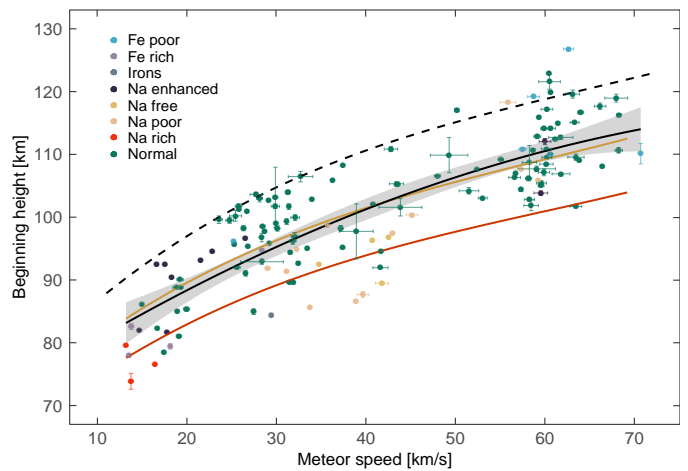}} \caption[f1]{Meteor beginning height as a function of entry speed for different spectral types of meteors. The solid line represents a quadratic fit to the running average of our data and standard error of the mean. The dashed line represents the fit of the data for fainter meteors from \citet{2019A&A...621A..68V}. Orange and red lines show a theoretical model for cometary and asteroidal meteors, respectively, simulated for systems of comparable sensitivity (CAMS-type) by \citet{2018MNRAS.479.4307V}.} 
\label{HbVi}
\end{figure}

The beginning and terminal (end) heights of meteor trajectories can reveal structural properties of the incoming meteoroid. They are used as a basis of the empirical material strength parameters $K_B$ and $P_E$ introduced by \citet{1968SAOSR.279.....C} and \citet{1976JGR....81.6257C}. In general, fragile particles tend to crumble under lower pressures resulting in higher beginning heights of the luminous trajectory. Likewise, only stronger materials can withstand the exerted pressure and reach lower parts of our atmosphere, and only the most robust bodies will eventually fall to the Earth's surface as meteorites. It is observed that depending on the entry velocity and meteoroid mass, the ablation begins at greater heights for cometary bodies compared to asteroidal ones \citep{1976JGR....81.6257C, 2004A&A...428..683K}.

The meteor beginning height is dependent on meteor speed as displayed in Fig. \ref{HbVi}. A similar dependency was observed among fainter meteors by \citet{2019A&A...621A..68V}, but shifted lower by approximately 8 km lower due to the different sensitivity of the system. Our results are also consistent with the simulations of \citet{2018MNRAS.479.4307V}, which suggest that two branches of the plot could approximately recognize asteroidal and cometary bodies. The lowest beginning heights are observed among Na-rich and Fe-rich meteoroids (typically 73 km $ < H_B <$ 83 km compared to most common meteor beginning heights in the range 90 km $ < H_B <$ 115 km). Low beginning heights are also observed for most Na-enhanced meteoroids. In all three groups, the low beginning heights are explained by generally high material strength associated with chondritic-asteroidal bodies. Figure \ref{HbVi} also shows that Na-poor and Na-free meteoroids have lower beginning heights than normal-type meteors of similar speed. On the other hand, most Fe-poor meteors typically show cometary properties with higher beginning heights.

\begin{figure}
\centerline{\includegraphics[width=\columnwidth,angle=0]{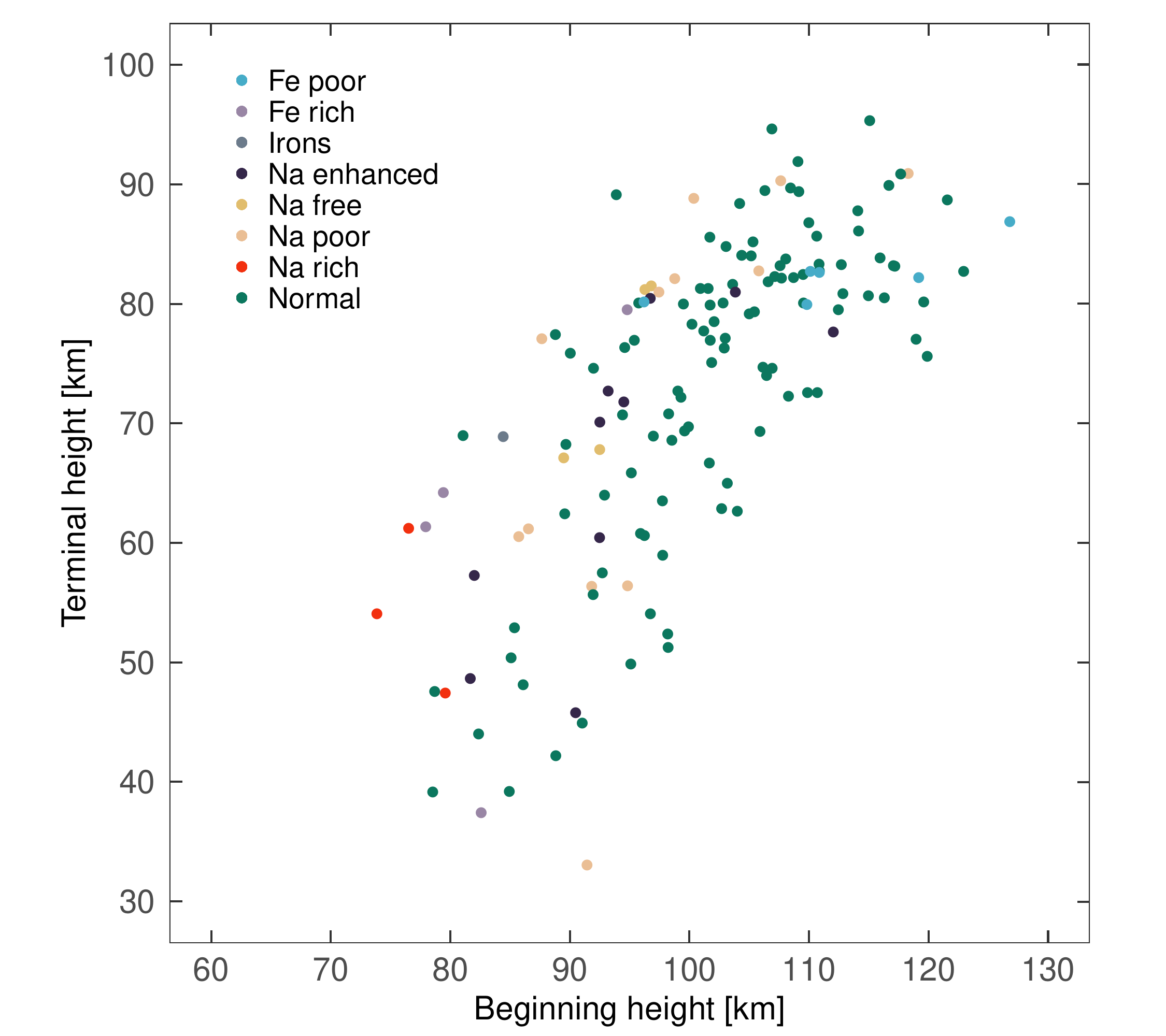}} \caption[f1]{Meteor beginning height vs. meteor terminal height plot for different spectral types of meteors.} 
\label{HbHe}
\end{figure}

\begin{figure}
\centerline{\includegraphics[width=\columnwidth,angle=0]{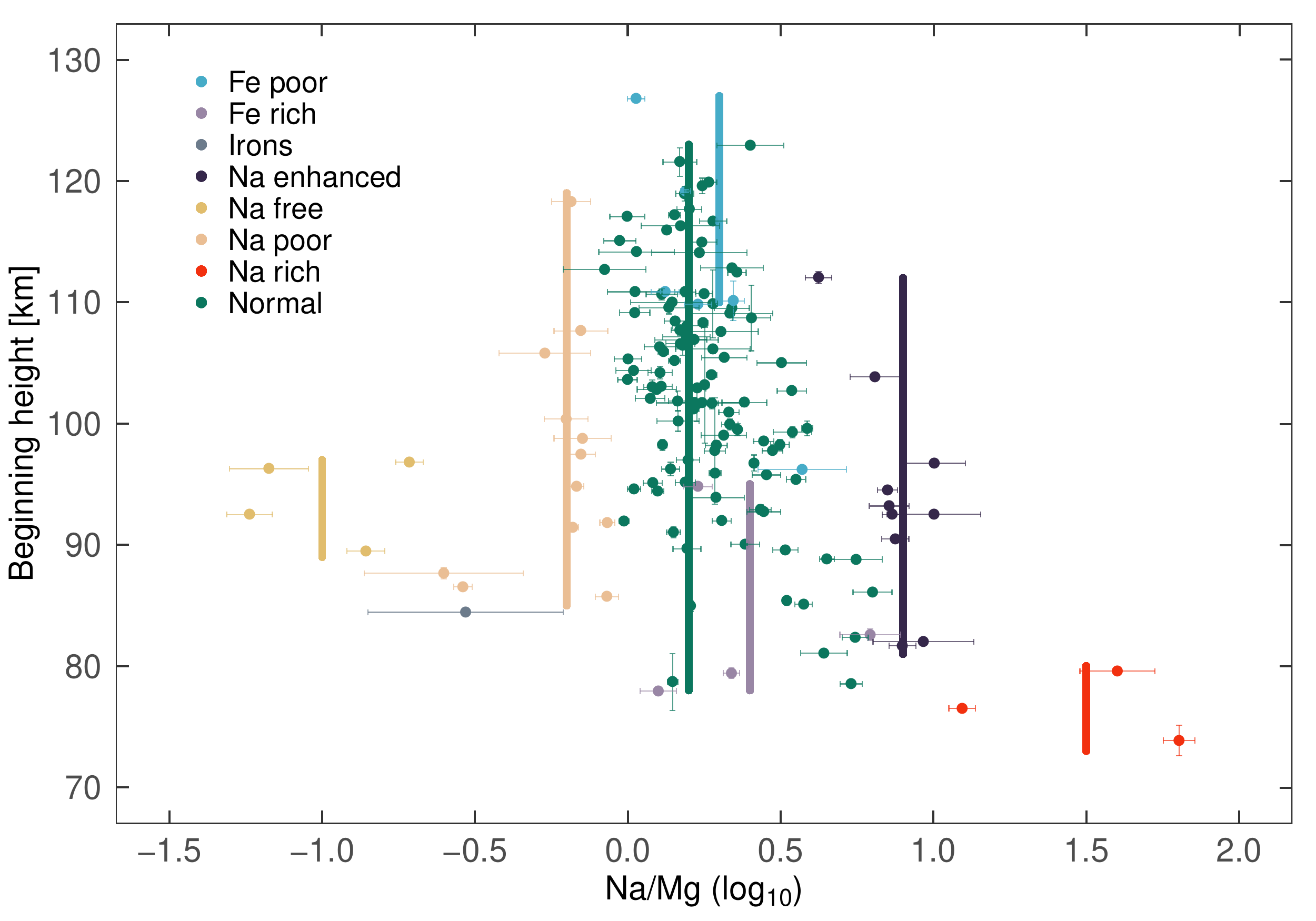}} \caption[f1]{Dependence of Na/Mg intensity ratio on meteor beginning height for different spectral types of meteors. The vertical columns represent the range of detected beginning heights for each spectral class positioned at mean Na/Mg value.} 
\label{HbNaMg}
\end{figure}

A similar functionality is found for meteor terminal heights. The link between beginning and terminal heights is displayed in Fig. \ref{HbHe}. Although terminal heights are more strongly influenced by meteoroid mass, \citet{2004A&A...428..683K} have shown that beginning heights are also mass dependent. The range of meteoroid beginning heights for different spectral types is well distinguished in Fig. \ref{HbNaMg}: 110 - 127 km for Fe-poor, 78 - 123 km for normal type, 85 - 119 km for Na-poor, 89 - 97 km for Na-free, 81 - 112 km for Na-enhanced, 78 - 95 km for Fe-rich, and 73 - 80 km for Na-rich. This plot refers to the characteristic material strengths and structural heterogeneity among each spectral class. The widest range is predictably observed among normal-type spectra, which include both cometary and chondritic meteoroids. Still, the majority of normal-type meteoroids have beginning heights between 100 and 120 km, characteristic of more fragile, cometary bodies. This preference might be affected by the higher number of shower meteors in our sample. Significant heterogeneity is also found among Na-enhanced meteors, this time dominated by stronger meteoroids with lower beginning heights. This distinction also shows the inclusion of both cometary and asteroidal meteoroids in this spectral class, as supported by the determined orbits. 

\begin{figure}
\centerline{\includegraphics[width=\columnwidth,angle=0]{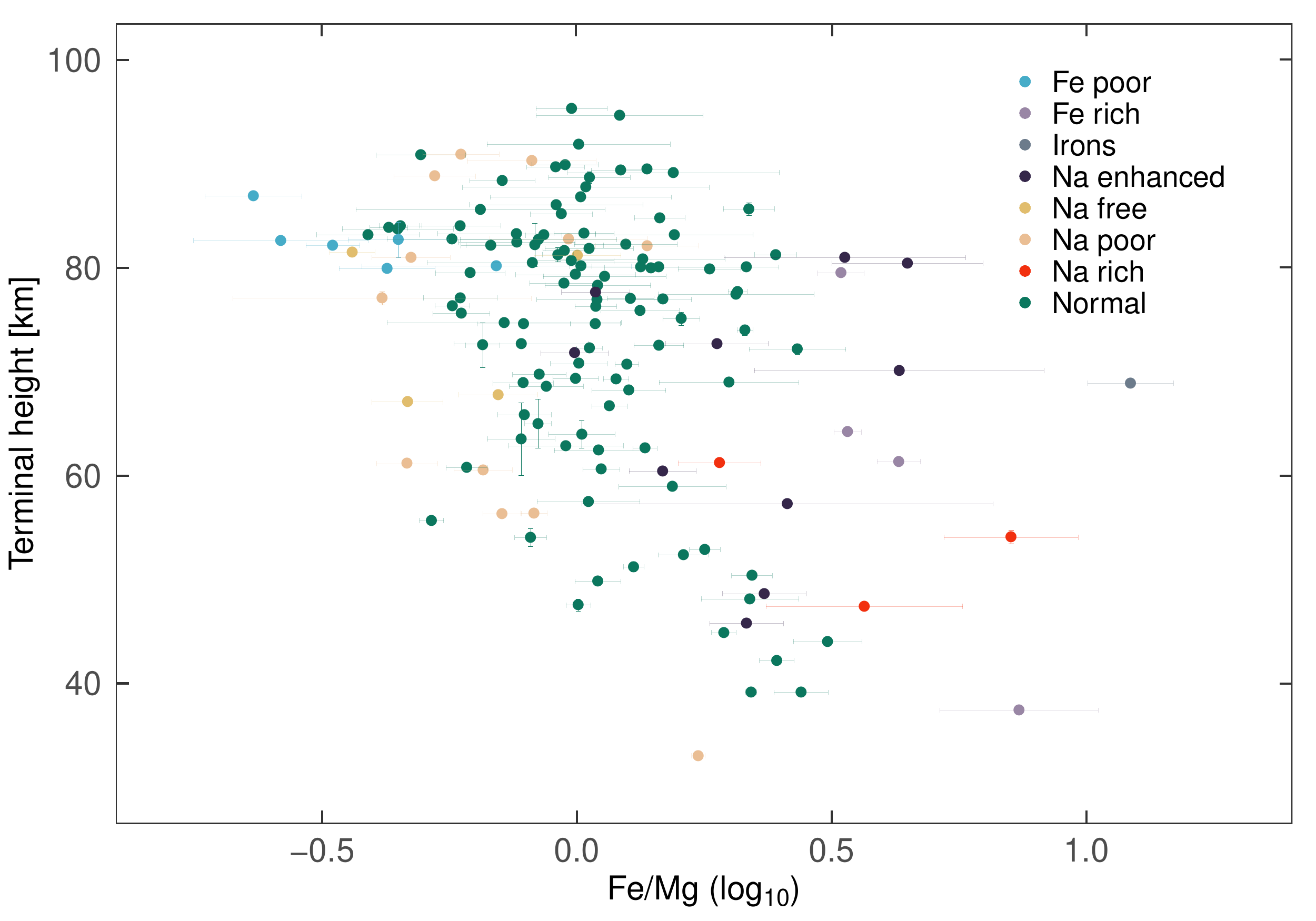}} \caption[f1]{Dependence of Fe/Mg intensity ratio on meteor terminal height for different spectral types of meteors.} 
\label{HeFeMg}
\end{figure}

To some degree, the differences in Fe/Mg ratio can be used to distinguish between cometary and asteroidal bodies. Meteoroids with low iron content exhibit higher terminal heights, characteristic for fragile meteoroids (Fig. \ref{HeFeMg}). The average terminal height for meteors with Fe/Mg $<$ 0.7 is $H_E$ = 79 km, while for meteors with Fe/Mg $>$ 1.0 it is $H_E$ = 69 km. The dependency is not completely straightforward due to the strong effect of entry speed and meteoroid mass. It also only concerns stony meteoroids, since the ablation of iron-rich bodies is atypical and cannot be described by the standard single-body theory. \citet{2017P&SS..143..159C} have shown that iron meteoroids have typically relatively short light curves with low beginning heights but terminal heights usually above 70 km. Similarly, Fig. \ref{HeFeMg} shows that most Fe-rich meteoroids and the only iron meteoroid in our sample had intermediate terminal heights between 80 and 60 km. The lowest terminal height ($H_E$ = 33.0 km) was detected in the brightest meteor in our sample, the Na-poor meteor M20170227\_023123 of estimated -11.2 mag.

\begin{figure}[]
\centerline{\includegraphics[width=\columnwidth,angle=0]{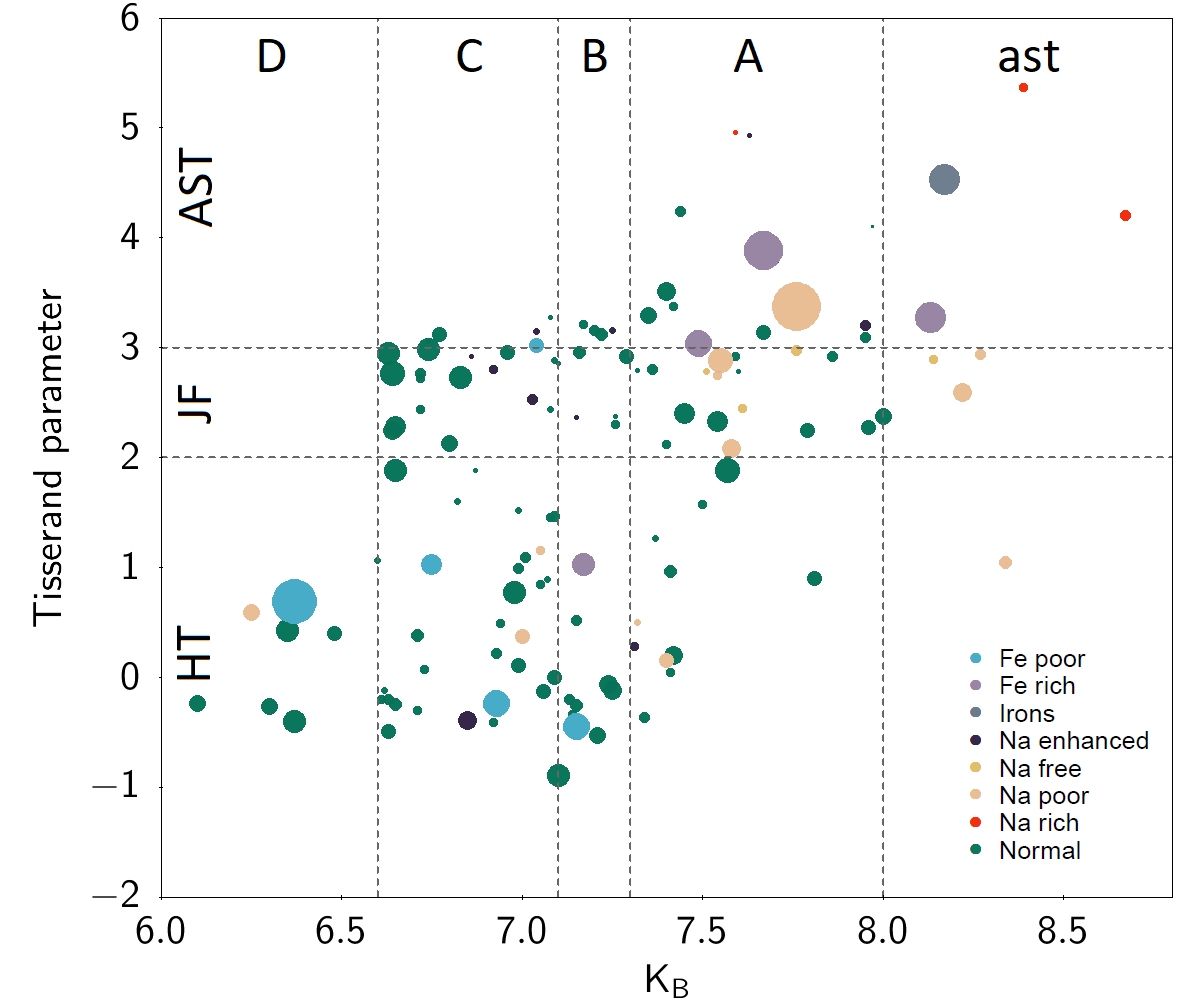}}
\centerline{\includegraphics[width=\columnwidth,angle=0]{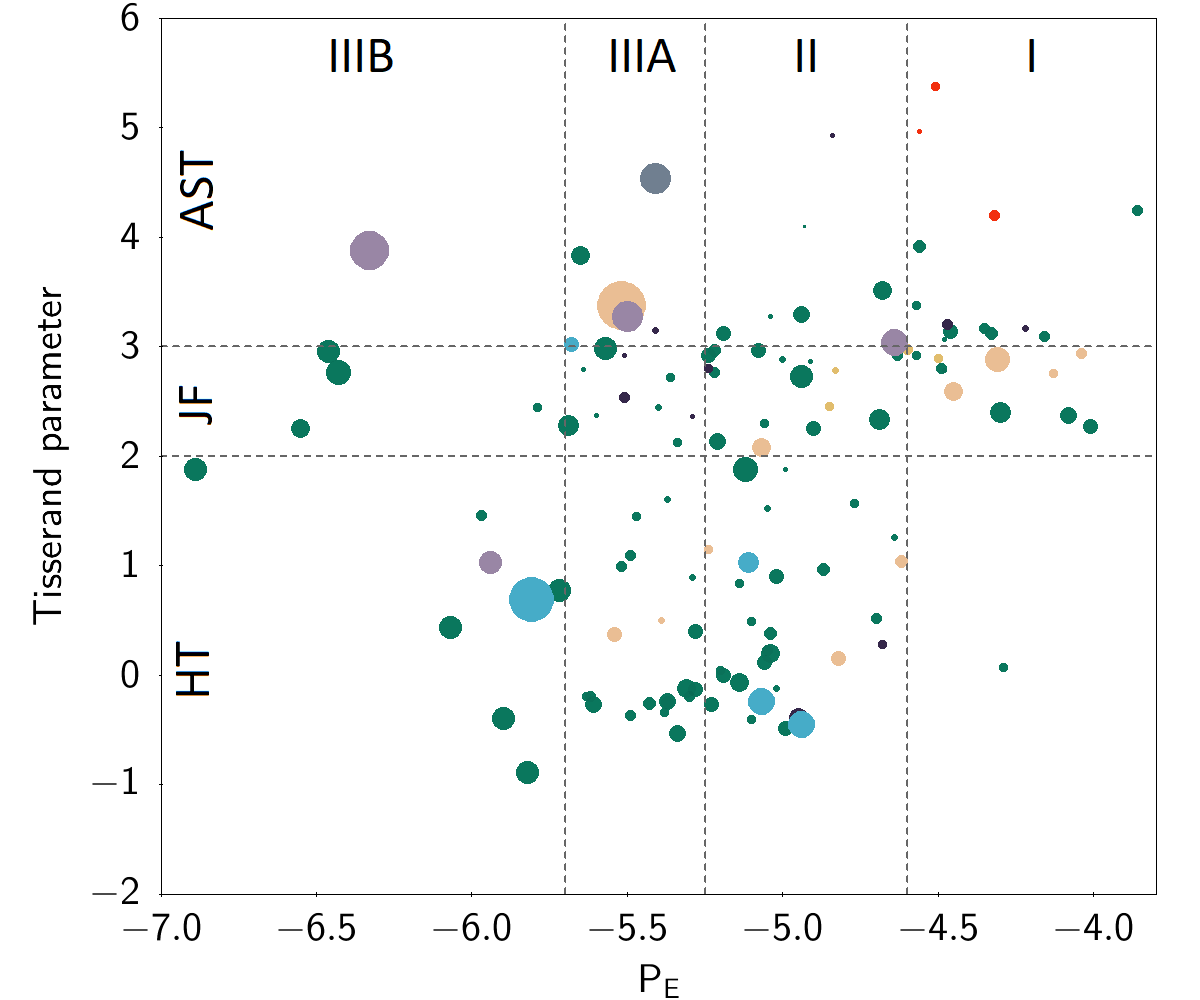}}
\caption[f1]{Material strength classification of all meteoroids observed by multiple stations based on the $K_B$ (upper) and $P_E$ (lower) parameter as a function of the Tisserand parameter. D/IIIB type represents the most fragile cometary material, C/IIIA typical cometary material, B denser cometary material, A/II material similar to carbonaceous bodies, and ast/I material similar to ordinary chondrites. Sizes of meteoroid marks reflect relative meteor magnitudes. Color coding differentiates between the determined meteor spectral classes.} 
\label{KbPeALL}
\end{figure}

The corresponding material strength classification based on $K_B$ and $P_E$ parameters for all meteoroids observed by multiple stations is displayed in Fig. \ref{KbPeALL}. The corresponding values can be found in Table 6 (available at the CDS). Since reliable meteoroid masses were not determined, the dependence of beginning height on original meteoroid mass suggested by \citet{2004A&A...428..683K} cannot be clearly evaluated. Still, no clear preference of meteor magnitude is observed in the $K_B$ classification. The $P_E$ distribution shows that mostly bright fireballs are found among the most fragile IIIB meteoroids, including seven fireballs brighter than -8 mag. The meteor magnitude does not always fully reflect on original meteoroid mass. For example, the strong Na-rich meteoroids are of moderate magnitude, even though we estimated relatively high photometric masses.

Figure \ref{KbPeALL} shows that Na-rich, Na-free, and Na-poor meteoroids are on average composed of the strongest material. Numerous strong, chondritic materials are also identified among normal-type meteors, though none of them are found in the ordinary chondrites group (ast) in the $K_B$ classification. High material strength $K_B$ is also detected in Fe-rich meteoroids, though the $P_E$ values are scattered and generally low. This behavior could reflect the enhanced iron content, as the ablation of iron meteoroids is atypical with relatively short light curves and moderate terminal heights. The same pattern is observed in the only iron meteoroid in our sample which exhibits high material strength at the beginning of ablation (ast), but dissipates quite early (IIIA). The lowest material strength is on average found in Fe-poor meteoroids, similarly to the previous results for smaller meteoroids \citep{2015A&A...580A..67V}.

Our results also demonstrate the correlation between the determined material properties and meteor orbits (Fig. \ref{KbPeALL}). Meteoroids originating in the main asteroid belt or from NEO orbits ($T_J >$ 3) are on average of higher material strength, most often characteristic for carbonaceous and ordinary chondrites. A few fragile meteoroids were also found on asteroidal orbits, mainly including shower meteors of the Taurids and $\alpha$-Capricornids. This population represents fragments of short-period comets, which are deficient compared to asteroidal bodies in this region, but produce a larger amount of dust particles. In case of the Taurids, there are hints that the stream also contains debris from asteroids on similar orbits, though no chondritic meteoroids were identified in the sample from our previous study \citep{2017P&SS..143..104M}.

The range of materials detected on the Jupiter-family orbits is wider and includes both asteroidal and cometary meteoroids (Fig. \ref{KbPeALL}). On average, the majority of the meteoroids in this region are defined by the cometary and carbonaceous material strengths (types C-A/IIIA-II). Several meteoroids with material properties similar to ordinary chondrites are also present, in large part represented by Na-poor and Na-free spectra. In this case, the hardening of material during close approaches to the Sun, associated with the release of volatiles, may be responsible for the detected high material strengths (see section \ref{secNaP}). The $K_B$ values show no meteoroids composed of the softest cometary material (type D) present on Jupiter-family orbits. The presence of the most fragile cometary material in the Jupiter-family zone is however possible, as four samples were identified as $P_E$ type IIIB. Two of these meteoroids are members of the $\kappa$-Cygnids and two are the $\alpha$-Capricornids.

Our results also suggest that meteoroids from Halley-type orbits do not contain ordinary chondrites (type ast/I). The only exception is one Na-poor meteoroid. Another normal-type meteor M20140726\_000917 on a Halley-type orbit was seen to have typically cometary material strength at the ablation beginning (C), but unusually high $P_E$ material strength. The reason behind this discrepancy is not clear. The meteoroid could represent a denser cometary material, possibly formed near a cometary nucleus. Overall, the meteoroids on Halley-type orbits are mainly of fragile structure, including several meteoroids of the softest cometary material (D/IIIB). The most fragile structure was typically found among brighter fireballs and included several stream meteoroids, including the Perseids, Orionid, and $\zeta$-Cassiopeid among Halley-type streams, and $\kappa$-Cygnids and $\alpha$-Capricornids among Jupiter-family streams. This material class D/IIIB was originally defined to include the brittle Giacobinids (now Draconids). Such soft material is apparently not unusual among other meteoroid streams.

Even though Perseid meteoroids were on average of characteristically fragile material strength, Table \ref{tab:Per} shows that some meteoroids showed higher strength comparable to carbonaceous bodies. This is particularly apparent from the $P_E$ values, where over half of our samples were identified as type II. Similarly to the spectra, we observe high heterogeneity of material strengths within Perseids. Partial results for different showers suggest that this is also common among other meteoroid streams. The Perseid spectral and material properties show some degree of correlation. In general, Fe-poor Perseids are more fragile and the Na-enhanced sample is among the strongest (A/II). Unlike Na-depleted meteoroids affected by strong solar radiation, the Na-poor Perseid was of low material strength.

The material strength classification of Halley-type meteoroids (Fig. \ref{KbPeALL}) includes numerous A/II type bodies with strengths similar to carbonaceous materials. The abundance of these meteors is particularly high in the $P_E$ distribution (Fig. \ref{KbPeALL}). This does not imply that all meteoroids characterized as type II are of carbonaceous composition; this group likely also includes dense cometary bodies and fractured chondrites. \citet{1988BAICz..39..221C} noted that this group includes both cometary and asteroidal bodies, as observed in our data. The dominance of cometary materials on Halley-type orbits is apparent in our data. Recently, \citet{2019A&A...621A..68V} reported the existence of iron meteoroids on Halley-type orbits, showing that the evolution of the early solar system could result in a more complicated distribution of materials. The complex mixing and transport of materials was previously revealed from the samples of comet 81P/Wild 2 collected by the Stardust mission \citep{2012M&PS...47..453B}.

\section{Spectral groups}

\subsection{Na-poor and Na-free}
\label{secNaP}

The Na line is one of the most interesting spectral features in meteors. In general, it corresponds to the volatile content in meteoroids. The dependence is however not straightforward, as the Na line intensity in meteors is also influenced by physical conditions, particularly meteor speed. Still, while real Na enhancement can be difficult to reveal, the depletion of Na is directly linked with the loss of volatiles within the meteoroid composition. The detected Na depletion can reveal the thermal evolution of the meteoroid. The main processes of Na loss in meteoroids concern space weathering caused by thermal desorption during close perihelion approaches and cosmic-ray irradiation in the Oort cloud for long-period orbits \citep{2005Icar..174...15B}. Iron meteoroids have naturally low Na content. 

We identified 18 Na-poor and 5 Na-free meteors in our sample, and orbital data is available for 12 and 4 of these meteors, respectively. In terms of spectra, the main difference between the two classes is that the Na line can still be reliably measured in Na-poor meteors. As noted earlier, the detected fraction of Na-depleted meteors is significantly lower compared to the results of studies focused on fainter meteors \citep{2005Icar..174...15B, 2015A&A...580A..67V, 2019A&A...621A..68V}. We argue that this effect reflects the increased preservation of volatiles in larger meteoroids. The space weathering processes more significantly affect smaller bodies. This corresponds to the fact that most Na-poor and Na-free spectra were observed in fainter meteors within our sample. Previously, \citet{2005Icar..174...15B} noted no correlation between the Na content and meteoroid mass within the size range of their sample.

\begin{figure}
\centerline{\includegraphics[width=\columnwidth,angle=0]{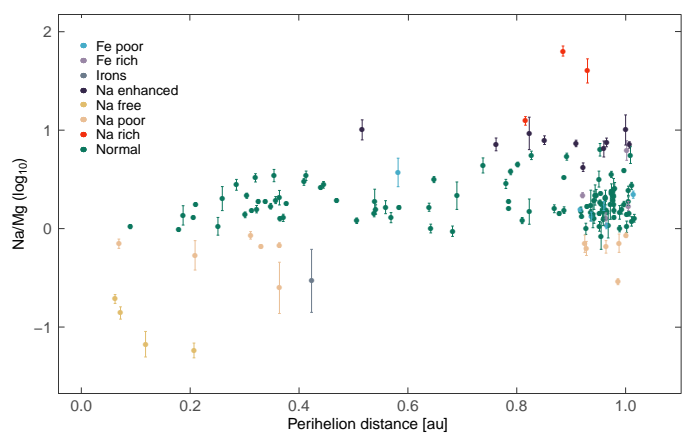}} \caption[f1]{Perihelion distance vs. measured Na/Mg intensity ratio for meteoroids of different spectral classes.} 
\label{qPlot}
\end{figure}

Of the processes causing Na depletion in meteoroids, thermal desorption was identified as the most common. The dependency of Na/Mg ratio on perihelion distance ($q$; Fig. \ref{qPlot}) shows that all four measured Na-free meteors have $q <$ 0.2 au. This correlates with the results of \citet{2005Icar..174...15B} suggesting that Na loss by thermal desorption occurs at $q \leq$ 0.2 au. According to \citet{2006A&A...453L..17K}, sodium abundance in meteoroids does not depend on their perihelion distances at $q \geq$ 0.14 au. This limit is satisfied by three Na-free meteors. Furthermore, Na loss is likely correlated with the dynamic age of a meteoroid. Younger meteoroids have suffered fewer passages in the vicinity of the Sun and can retain more Na. The heating during close perihelion approaches causes loss of volatiles and consequent compaction of the meteoroid material. This effect might be related to the aqueous alteration and is demonstrated by the generally high material strength of Na-free, and to some degree Na-poor meteoroids (Fig. \ref{HbVi} and \ref{KbPeALL}). As seen in Fig. \ref{KbPeALL}, the $K_B$ and $P_E$ classification places Na-free and Na-poor among the strongest spectral classes.

In general, meteors classified as Na-poor have small or intermediate perihelion distances ($q \leq$ 0.37 au) or are on long-period cometary orbits. The process of sodium loss on orbits with 0.2 $< q <$ 0.4 au is not clear. The thermal effects from solar radiation should not be as significant here \citep{2006A&A...453L..17K}. It is possible that some degree of thermal loss in combination with the aging effect is still occurring. We also observed a cluster of Na-poor meteors on Halley-type orbits with 0.9 $< q <$ 1.0 au (Fig. \ref{qPlot}). The Na-depleted material in these meteoroids might come from cometary crust affected by long-term cosmic ray irradiation, as previously suggested by \citet{2005Icar..174...15B}. Our results suggest that this effect is not as significant as solar heating, as no Na-free meteors on such orbits were detected. The atmospheric and orbital properties of Na-free and Na-poor meteoroids are given in Tables \ref{tab:NaF} and \ref{tab:NaP}.

\begin{figure}
\centerline{\includegraphics[width=\columnwidth,angle=0]{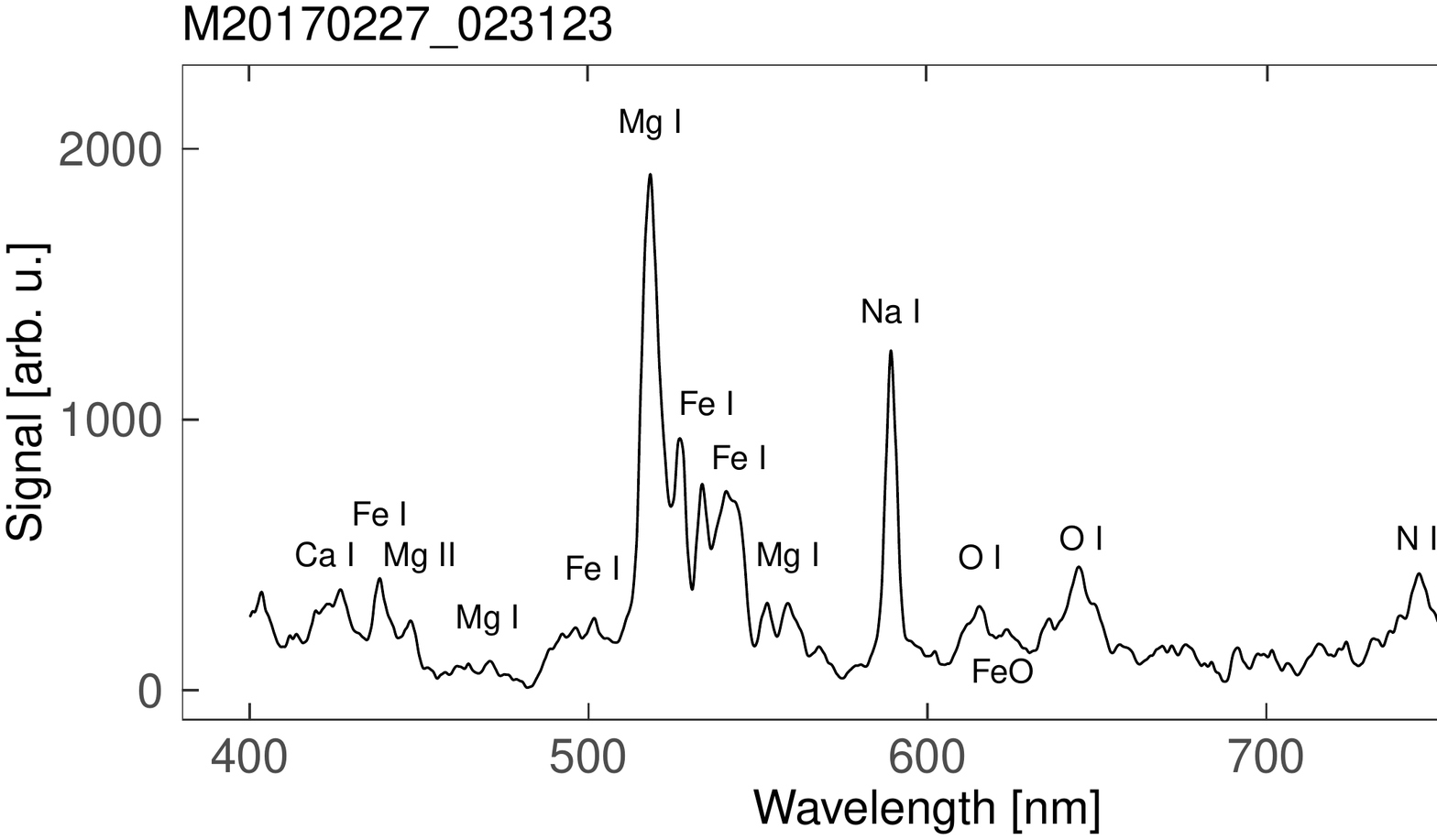}}
\centerline{\includegraphics[width=6cm,angle=0]{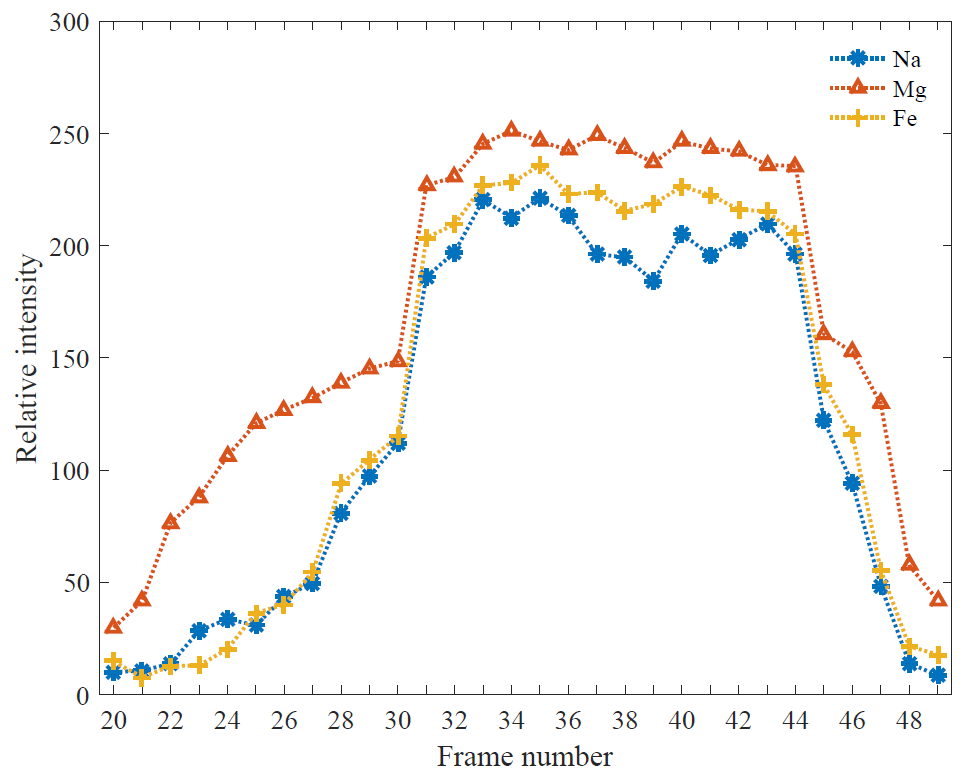}}
\caption[f1]{Spectrum of a bright Na-poor meteor M20170227\_023123 with identified major lines (upper). The monochromatic light curve (lower) shows early emission of Mg, which also remains after Na and Fe lines disappear. Here, the Fe I monochromatic light curve is only represented by the intensity of the 526.9 nm line. One frame corresponds to a time period of 1/12 s.} 
\label{NaP1}
\end{figure}

The most atypical meteor classified as Na-poor is M20170227\_023123 (Fig. \ref{NaP1}). This meteor is by far the brightest of all Na-depleted bodies in our sample, with an estimated -11.2 mag. Among the Na-poor class, it is positioned in the uppermost part of the ternary diagram (Fig. \ref{ternary}), reflecting chondritic values of the Fe/Mg ratio. The meteor had an unusual monochromatic light curve with early release of Mg and delayed emission of Fe and Na lines (Fig. \ref{NaP1}). This behavior is different from the more commonly observed early release of volatile Na. The spectrum of M20170227\_023123 could reflect its larger size with Na depleted from its outer layers. A similar monochromatic light curve was observed in one atypical Geminid by \citet{2019A&A...621A..68V}. The late release of Na was explained by different grain sizes in the two stages of erosion of the meteoroid. Faster depletion of Na is expected for smaller grains contained in the part of the meteoroid which fragmented first.

This meteoroid was identified as a member of the $\eta$-Virginid stream. The orbital data from different authors for the stream show a wide range of mean perihelion distances (0.23 $\leq q \leq$ 0.46 au)\footnote{A collection of mean shower orbits determined by different authors can be found in the IAU MDC: https://www.ta3.sk/IAUC22DB/MDC2007/ \citep{2017P&SS..143....3J}}. Similarly, two $\eta$-Virginids were identified in our sample, M20170227\_023123 with  $q$ = 0.33 au and M20170314\_194345 with $q$ = 0.47 au. The two $\eta$-Virginid have also shown significant spectral differences (Fig. \ref{SunApp}), which could imply a heterogeneous composition within the stream. The parent object of $\eta$-Virginids is not confirmed, but there are hints that the stream could be linked with the short-period comet D/1766 G\textsubscript{1}.

\begin{figure}[!t]
\centerline{\includegraphics[width=\columnwidth,angle=0]{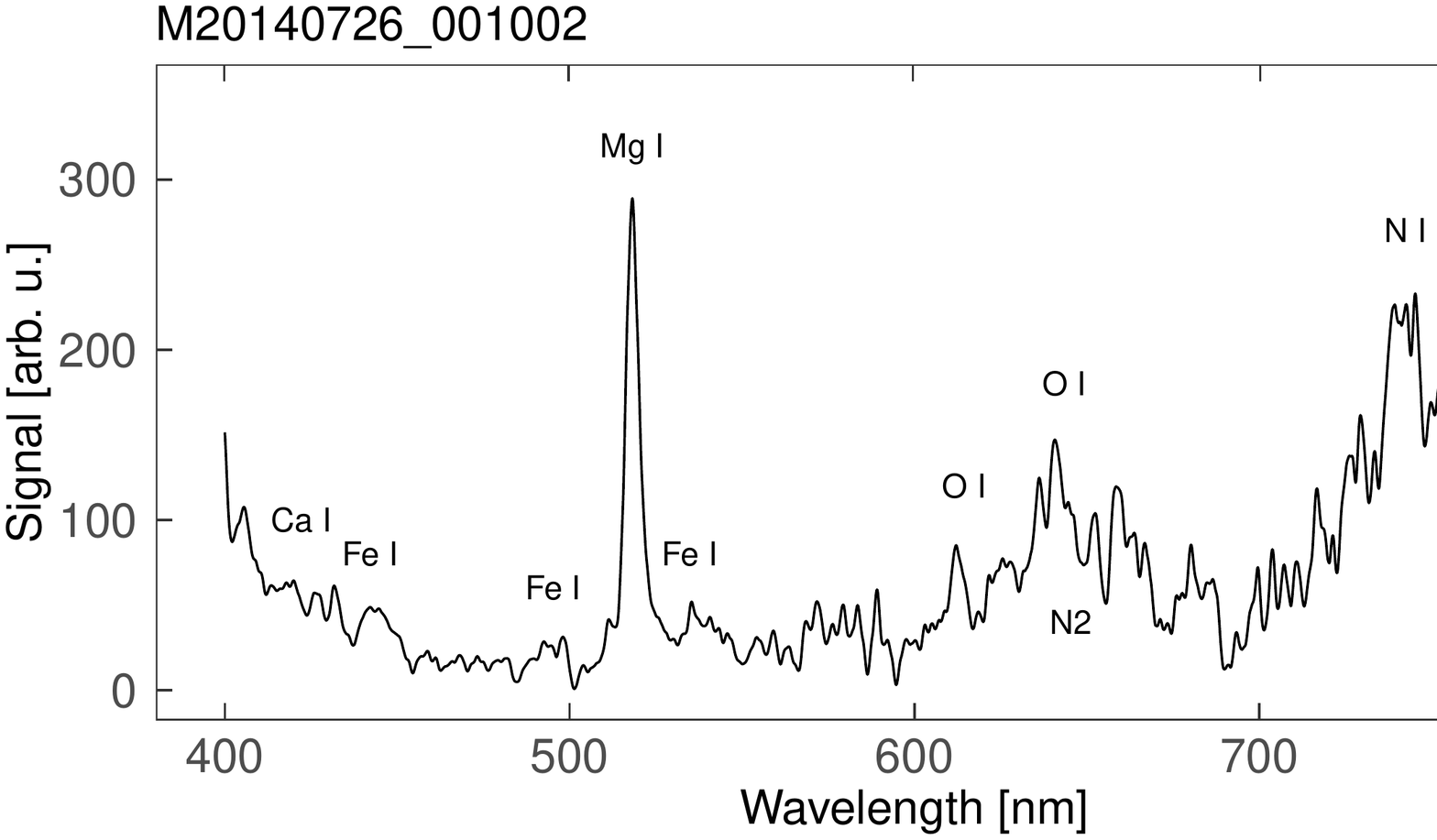}} 
\centerline{\includegraphics[width=\columnwidth,angle=0]{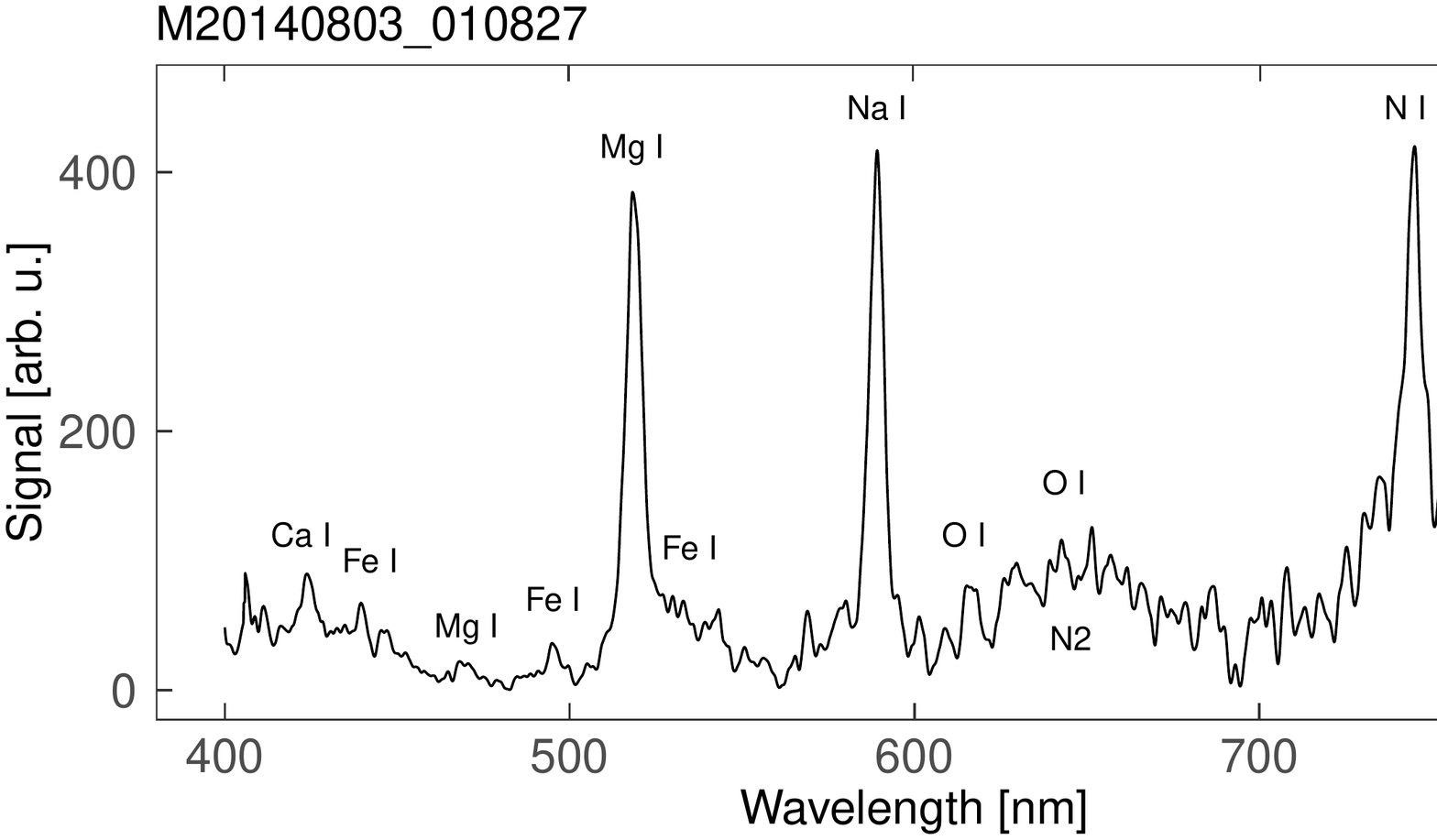}}
\caption[f1]{Spectra of two $\delta$-Aquarid meteors. The difference between Na-free (upper) and normal-type (lower) spectra shows the aging effect among meteoroids in a Sun-approaching stream.} 
\label{SDAprof}
\end{figure}

\begin{figure}[]
\centerline{\includegraphics[width=\columnwidth,angle=0]{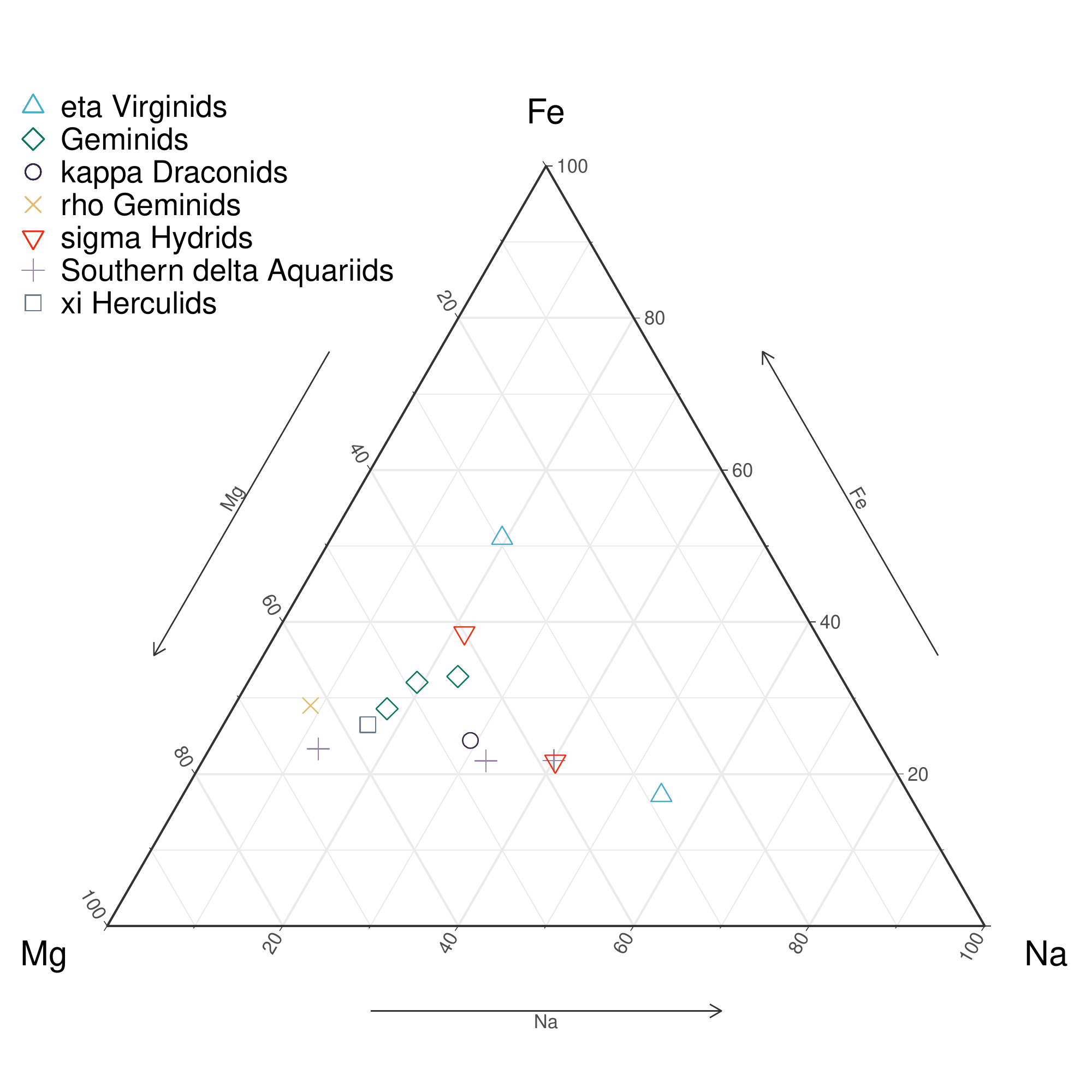}} \caption[f1]{Selected stream meteoroids with detected Na depletion. The displayed Geminids were estimated from single-station data based on radiant position and solar longitude.} 
\label{SunApp}
\end{figure}

Several other meteoroid streams were identified as a source of Na-free and Na-poor bodies. Unsurprisingly, the associated streams have typically small perihelion distances. One Na-free and one Na-poor meteor were found as members of the Southern $\delta$-Aquarids ($q$ = 0.07 au). The $\delta$-Aquarids are a Sun-approaching stream presumably formed from the breakup of a larger body, which also formed the Marsden and Kracht sungrazing comets \citep{2006mspc.book.....J}. As probably the largest remaining fragment of this breakup, 96P/Machholz is assumed to be responsible for most of the dust production to the stream. While the very small $q$ orbits are a natural cause for Na depleted composition, we have also detected one $\delta$-Aquarid meteor with a normal-type spectrum. This spectrum still shows a somewhat lower intensity of Na compared to most normal-type meteors (Fig. \ref{SunApp}). The difference between the Na-free and normal $\delta$-Aquarid spectra is shown in Fig. \ref{SDAprof}. The normal-type sample was a meteor of -5.0 mag compared to -3.1 mag for the Na-free meteor. This could indicate that Na is better preserved in the larger meteoroid, but the size difference is probably not sufficient to explain the spectra. We suggest that the difference between the spectra demonstrates the aging effect of meteoroids on orbits with small $q$. The difference also implies that the meteoroids are being actively supplied to the $\delta$-Aquarid stream. We previously pointed out the diversity of Na/Mg ratios in the two $\delta$-Aquarids in \citet{2016P&SS..123...25R}.

In addition to $\delta$-Aquarids, \citet{2005Icar..174...15B}  identified Geminids ($q$ = 0.14 au) and $\alpha$-Monocerotids ($q$ = 0.19 au) as meteoroid streams containing Na-free and Na-poor meteors. Surprisingly few bright Geminids were collected during the routine observations of the AMOS-Spec in 2013-2017. The observation campaigns for Geminids were in some cases hampered by bad weather, often leading to only single-station data for potential Geminid spectra. Still, several Na-poor spectra were detected near the predicted maximum of the shower activity, as displayed in Fig. \ref{SunApp}. \citet{2005Icar..174...15B} noted a wide spread of Na line intensity in Geminid spectra, also explained by the aging effect. The positions of Geminids identified in Fig. \ref{SunApp} correlate well with the results of \citet{2005Icar..174...15B}. It appears that Geminids are much better represented in spectral surveys of fainter meteors, suggesting that the stream could include fewer larger centimeter- to meter-sized debris. In the future, we intend to carry out a more extensive study of the Geminid composition based on high-resolution spectra from global AMOS-Spec stations.

\begin{table*}[]
\centering
\small\begin{center}
\caption {Atmospheric and orbital properties of detected Na-free meteoroids observed by multiple stations. For references, see Tables 5 and 6 (available at the CDS).} 
\vspace{0.1cm}
\resizebox{\textwidth}{!}{\begin{tabular}{llrrrrrrrrrrrrcc}
\hline\hline\\
\multicolumn{1}{c}{Code}& %
\multicolumn{1}{l}{Shower}& %
\multicolumn{1}{c}{$\alpha_g$}& %
\multicolumn{1}{c}{$\delta_g$}& %
\multicolumn{1}{c}{$v_g$} & %
\multicolumn{1}{c}{$H_B$} & %
\multicolumn{1}{c}{$H_E$} & %
\multicolumn{1}{c}{$a$} & %
\multicolumn{1}{c}{$e$} & %
\multicolumn{1}{c}{$q$} & %
\multicolumn{1}{c}{$Q$} & %
\multicolumn{1}{c}{$i$} & %
\multicolumn{1}{c}{$\omega$}& %
\multicolumn{1}{c}{$\Omega$}& %
\multicolumn{1}{c}{$T_J$} \\
\hline\\
M20140726\_001002 & SDA     & 336.82 & -15.22 & 41.02     & 96.81   & 81.49   & 2.39   & 0.97   & 0.062  & 4.73   & 25.80   & 154.51  & 302.77  & 2.45 \\
                  &         & $\pm$ 0.01  & 0.04  & 0.05  & 0.08   & 0.05  & 0.01  & 0.00   & 0.001  &  0.02      & 0.07 & 0.10 &  &   \vspace{0.1cm}\\
M20150710\_232522 & spo     & 315.16 & -4.01  & 39.07     & 96.28   & 81.19   & 2.13   & 0.94   & 0.118  & 4.15   & 36.84   & 325.00  & 108.17  & 2.78 \\
                  &         & $\pm$ 0.02  & 0.05  & 0.17  & 0.06   & 0.02  & 0.04  & 0.00   & 0.001  &  0.09     & 0.43 & 0.04 &  &  \vspace{0.1cm}\\
M20150829\_015634 & spo     & 354.71 & -1.52  & 33.11     & 92.50   & 67.77   & 2.15   & 0.90   & 0.207  & 4.09   & 1.27    & 312.43  & 155.20  & 2.97 \\
                 &         & $\pm$ 0.03  & 0.03  & 0.09  & 0.08   & 0.14  & 0.02  & 0.00   & 0.001  &   0.04     & 0.04 & 0.09 &  &  \vspace{0.1cm}\\
M20151227\_035406 & DRG     & 128.32 & 25.39  & 40.50     & 89.49   & 67.09   & 2.01   & 0.96   & 0.072  & 3.94   & 23.35   & 332.60  & 274.84  & 2.89 \\
                 &         & $\pm$ 0.10  & 0.07  & 0.75  & 0.18   & 0.16  & 0.25  & 0.00   & 0.004  &  0.50      & 1.35 & 0.43 &  &  \\
\hline
\end{tabular}} 
\label{tab:NaF}
\end{center}

\smallskip

\centering
\small\begin{center}
\caption {Atmospheric and orbital properties of detected Na-poor meteoroids observed by multiple stations. For references, see Tables 5 and 6 (available at the CDS).} 
\vspace{0.1cm}
\resizebox{\textwidth}{!}{\begin{tabular}{llrrrrrrrrrrrrcc}
\hline\hline\\
\multicolumn{1}{c}{Code}& %
\multicolumn{1}{l}{Shower}& %
\multicolumn{1}{c}{$\alpha_g$}& %
\multicolumn{1}{c}{$\delta_g$}& %
\multicolumn{1}{c}{$v_g$} & %
\multicolumn{1}{c}{$H_B$} & %
\multicolumn{1}{c}{$H_E$} & %
\multicolumn{1}{c}{$a$} & %
\multicolumn{1}{c}{$e$} & %
\multicolumn{1}{c}{$q$} & %
\multicolumn{1}{c}{$Q$} & %
\multicolumn{1}{c}{$i$} & %
\multicolumn{1}{c}{$\omega$}& %
\multicolumn{1}{c}{$\Omega$}& %
\multicolumn{1}{c}{$T_J$} \\
\hline\\
M20131203\_050007 & DKD      & 187.13 & 72.22  & 43.72     & 100.37  & 88.80   & 39.17  & 0.98   & 0.928  & 77.40  & 71.77   & 208.27  & 251.00  & 0.50  \\
                 &         & $\pm$ 0.67  & 0.05  & 0.67  & 0.24   & 0.21  & - & 0.04   & 0.002  &   -     & 0.65 & 0.53 &  & \vspace{0.1cm} \\
M20150308\_020714 & XHE      & 256.57 & 42.94  & 37.13     & 86.54   & 61.15   & 2.08   & 0.53   & 0.986  & 3.17   & 66.39   & 191.28  & 346.97  & 2.94  \\
                 &         & $\pm$ 0.04  & 0.03  & 0.07  & 0.05   & 0.03  & 0.02    & 0.00   & 0.000  &   0.04    & 0.08 & 0.06 &  &  \vspace{0.1cm}\\
M20150409\_203107 & spo      & 221.65 & 9.31   & 37.82     & 87.65   & 77.06   & 11.42  & 0.97   & 0.364  & 22.47  & 37.95   & 287.09  & 19.46   & 1.04  \\
                 &         & $\pm$ 0.39  & 0.08  & 0.69  & 0.47   & 0.64  & -  & 0.02   & 0.004  &    -    & 0.91 & 0.80 &  &  \vspace{0.1cm}\\
M20150518\_234943 & spo      & 292.75 & 67.24  & 26.75     & 91.83   & 56.33   & 2.66   & 0.62   & 1.001  & 4.31   & 44.76   & 166.52  & 57.50   & 2.75  \\
                  &         & $\pm$ 0.11  & 0.04  & 0.31  & 0.14   & 0.09  & 0.12    & 0.02   & 0.000  &  0.24    & 0.36 & 0.22 &  & \vspace{0.1cm} \\
M20150803\_202646 & spo      & 301.25 & 61.37  & 33.86     & 98.77   & 82.09   & 11.62  & 0.92   & 0.988  & 22.26  & 54.28   & 198.94  & 130.97  & 1.15  \\
                  &         & $\pm$ 0.06  & 0.04  & 0.26  & 0.12   & 0.10  & 3.21    & 0.02   & 0.000  &   6.39    & 0.25 & 0.17 &  &  \vspace{0.1cm}\\
M20150810\_013034 & PER      & 46.10  & 60.72  & 56.16     & 107.64  & 90.28   & 7.44   & 0.88   & 0.925  & 13.96  & 106.69  & 144.38  & 136.92  & 0.37  \\
                  &         & $\pm$ 0.07  & 0.02  & 0.79  & 0.18   & 0.14  & -  & 0.06   & 0.004  &  -      & 0.60 & 1.41 &  & \vspace{0.1cm} \\
M20151210\_232747 & HYD      & 125.59 & 3.44   & 58.03     & 105.81  & 82.76   & 19.25  & 0.99   & 0.209  & 38.30  & 127.10  & 125.79  & 78.36   & -0.07 \\
                  &         & $\pm$ 0.03  & 0.09  & 0.24  & 0.15   & 0.15  & -   & 0.00   & 0.004  &  -      & 0.35 & 0.72 &  &  \vspace{0.1cm}\\
M20160122\_195954 & ACZ      & 137.80 & 17.57  & 31.60     & 85.72   & 60.52   & 2.71   & 0.89   & 0.311  & 5.12   & 1.74    & 297.07  & 302.06  & 2.59  \\
                  &         & $\pm$ 0.02  & 0.02  & 0.05  & 0.05   & 0.02  & 0.02    & 0.00   & 0.001  &  0.04    & 0.02 & 0.04 &  &  \vspace{0.1cm}\\
M20160727\_013114 & ZCS      & 22.97  & 57.99  & 54.66     & 118.30  & 90.91   & 6.06   & 0.84   & 0.964  & 11.15  & 103.22  & 152.62  & 124.23  & 0.59  \\
                  &         & $\pm$ 0.09  & 0.03  & 0.88  & 0.20   & 0.02  & 8.92    & 0.07   & 0.003  &   -   & 0.68 & 1.25 &  &  \vspace{0.1cm}\\
M20160730\_012149 & SDA      & 340.16 & -15.85 & 41.56     & 97.45   & 80.97   & 2.91   & 0.98   & 0.069  & 5.75   & 27.63   & 152.52  & 307.10  & 2.08  \\
                  &         & $\pm$ 0.04  & 0.07  & 0.10  & 0.06   & 0.07  & 0.02    & 0.00   & 0.000  &   0.03   & 0.25 & 0.08 &  &  \vspace{0.1cm}\\
M20160902\_230214 & spo      & 3.11   & -18.46 & 30.20     & 94.83   & 56.38   & 2.35   & 0.85   & 0.364  & 4.33   & 22.83   & 113.29  & 340.66  & 2.88  \\
                  &         & $\pm$ 0.03  & 0.03  & 0.03  & 0.03   & 0.03  & 0.01    & 0.00   & 0.000  &   0.01     & 0.07 & 0.02 &  &  \vspace{0.1cm}\\
M20170227\_023123 & EVI      & 173.25 & 5.12   & 29.27     & 91.44   & 33.04   & 1.94   & 0.83   & 0.330  & 3.54   & 2.53    & 298.08  & 338.46  & 3.37  \\
                  &         & $\pm$ 0.16  & 0.07  & 0.06  & 0.06   & 0.02  & 0.03    & 0.00   & 0.002  &  0.06      & 0.03 & 0.37 &  &  \\
\hline
\end{tabular}} 
\label{tab:NaP}
\end{center}

\smallskip

\centering
\small\begin{center}
\caption {Atmospheric and orbital properties of detected iron and Fe-rich meteoroids observed by multiple stations. For references, see Tables 5 and 6 (available at the CDS).} 
\vspace{0.1cm}
\resizebox{\textwidth}{!}{\begin{tabular}{llrrrrrrrrrrrrcc}
\hline\hline\\
\multicolumn{1}{c}{Code}& %
\multicolumn{1}{l}{Type}& %
\multicolumn{1}{c}{$\alpha_g$}& %
\multicolumn{1}{c}{$\delta_g$}& %
\multicolumn{1}{c}{$v_g$} & %
\multicolumn{1}{c}{$H_B$} & %
\multicolumn{1}{c}{$H_E$} & %
\multicolumn{1}{c}{$a$} & %
\multicolumn{1}{c}{$e$} & %
\multicolumn{1}{c}{$q$} & %
\multicolumn{1}{c}{$Q$} & %
\multicolumn{1}{c}{$i$} & %
\multicolumn{1}{c}{$\omega$}& %
\multicolumn{1}{c}{$\Omega$}& %
\multicolumn{1}{c}{$T_J$} \\
\hline\\
M20150508\_002949 & Fe-rich & 63.67  & 66.17 & 14.35 & 79.42 & 64.19 & 2.32   & 0.60 & 0.921 & 3.72   & 16.15 & 140.16 & 46.91  & 3.27 \\
 &  & $\pm$ 3.53 & 0.73 & 0.17 & 0.42 & 0.36 & 0.16   & 0.02 & 0.007 & 0.31 & 0.15 & 2.10  &  &  \vspace{0.1cm}\\
M20160416\_211518 & Fe-rich & 113.31 & 35.55 & 8.53  & 82.58 & 37.42 & 2.71   & 0.63 & 1.002 & 4.43   & 3.04  & 173.82 & 27.06  & 3.04 \\
&  & $\pm$ 0.68 & 0.46 & 0.47 & 0.47 & 0.15 & 0.31   & 0.04 & 0.000 & 0.63 & 0.25 & 0.35  &  & \vspace{0.1cm} \\
M20160623\_210113 & Fe-rich & 257.53 & 55.35 & 26.08 & 94.80 & 79.49 & - & 0.99 & 1.005 & - & 37.89 & 192.46 & 92.57  & 1.03 \\
&  & $\pm$ 0.13 & 0.13 & 0.20 & 0.12 & 0.10 & - & 0.01 & 0.000 & - & 0.21 & 0.13  &  &  \vspace{0.1cm}\\
M20170301\_201252 & Fe-rich & 114.83 & 21.81 & 7.75  & 77.94 & 61.33 & 1.84   & 0.47 & 0.964 & 2.71   & 0.07  & 203.36 & 341.43 & 3.88 \\
&  & $\pm$ 0.57 & 0.76 & 0.22 & 0.30 & 0.23 & 0.04   & 0.01 & 0.001 & 0.09 & 0.11 & 0.41 &  &  \vspace{0.1cm}\\
M20160415\_002243 & iron    & 231.60 & 11.20 & 27.24 & 84.43 & 68.89 & 1.34   & 0.68 & 0.423 & 2.25   & 31.78 & 295.19 & 25.25  & 4.53 \\
&  & $\pm$ 0.08 & 0.08 & 0.14 & 0.27 & 0.17 & 0.01   & 0.00 & 0.001 & 0.02 & 0.24 & 0.18  &  &  \\
\hline
\end{tabular}} 
\label{tab:Fe}
\end{center}
\end{table*}

Furthermore, one Na-free meteor (M20151227\_035406) was assigned to the December $\rho$-Geminids \citep{2016Icar..266..331J}. The $\rho$-Geminids are a IAU MDC working-list shower characterized by higher velocities ($v_g$ = 39.5 km\,s\textsuperscript{-1}) and smaller perihelion distances ($q$ = 0.11 au) compared to the Geminids. One $\sigma$-Hydrid meteor ($q$ = 0.22 au) was found to have a Na-poor spectrum. Another $\sigma$-Hydrid was identified as normal-type, though with a lower-than-average Na/Mg ratio (Fig. \ref{SunApp}). Furthermore, the $\kappa$-Draconids, $\xi$-Herculids, and $\eta$-Virginids were also found to produce Na-depleted meteoroids, each with one identified spectrum. Overall, our data on Na-poor meteors suggest that some degree of Na depletion is present also on orbits with intermediate perihelion distances 0.2 $< q <$ 0.4 au, probably also related to the effects of solar radiation. 

\subsection{Fe-rich and irons} \label{secFe}

As noted earlier, only one pure iron meteoroid was identified in our sample. The overall low abundance of iron meteoroids in our sample compared to surveys of smaller meteoroids probably relates to the typical sizes of iron--nickel grains. Iron meteoroids are generally found on asteroidal orbits with a likely origin in the main asteroid belt. However, it appears that the main belt might not be the only source of iron meteors. One iron body was found by \citet{2015A&A...580A..67V} on a borderline Jupiter-family orbit, and \citet{2019A&A...621A..68V} even pointed out the surprising existence of iron meteoroids on Halley-type orbits. Common features are observed among irons, including similar trajectory heights, lengths, and light curves, which cannot be explained by simple single-body ablation theory.

\begin{figure}[]
\centerline{\includegraphics[width=\columnwidth,angle=0]{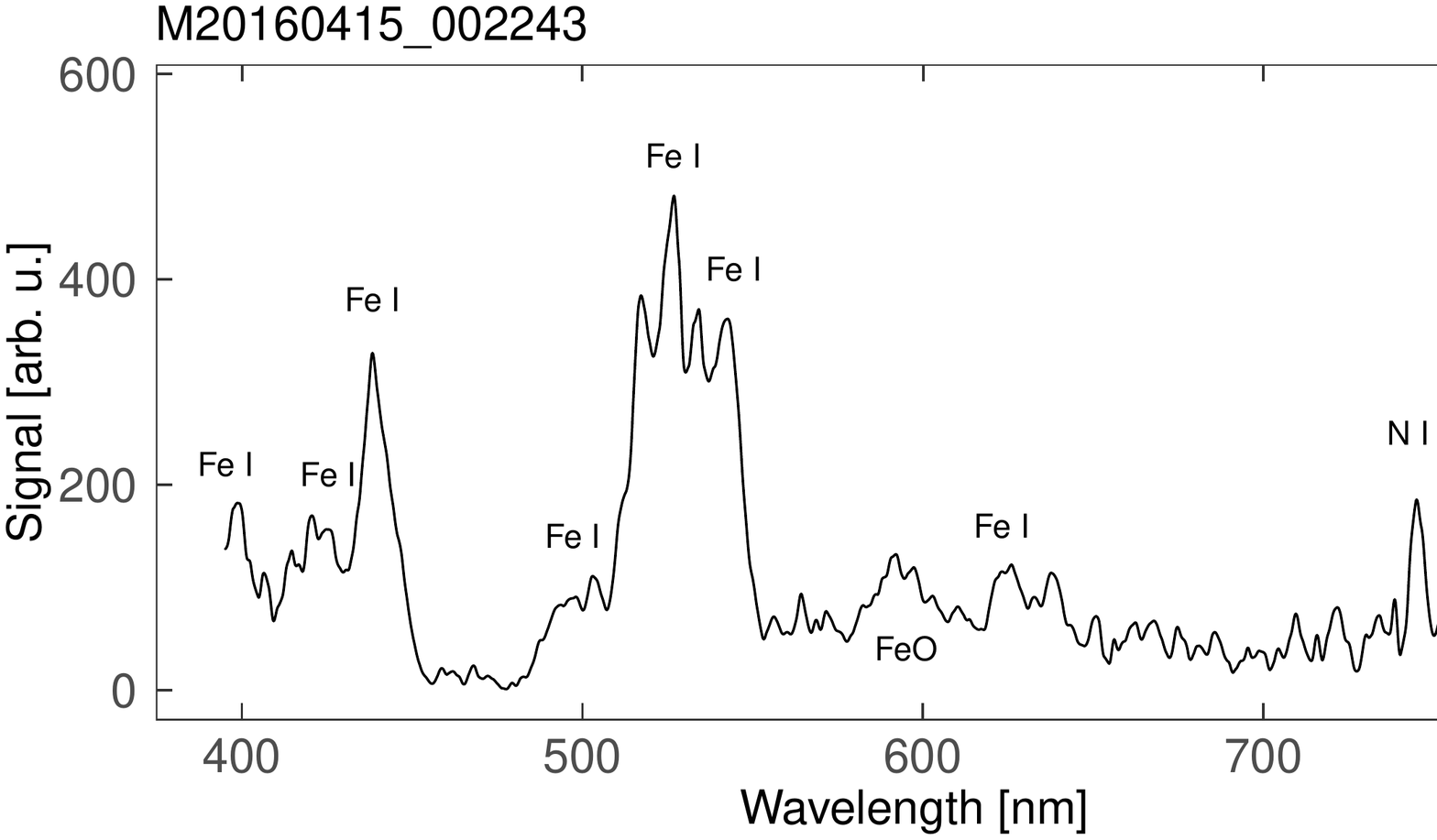}}
\centerline{\includegraphics[width=\columnwidth,angle=0]{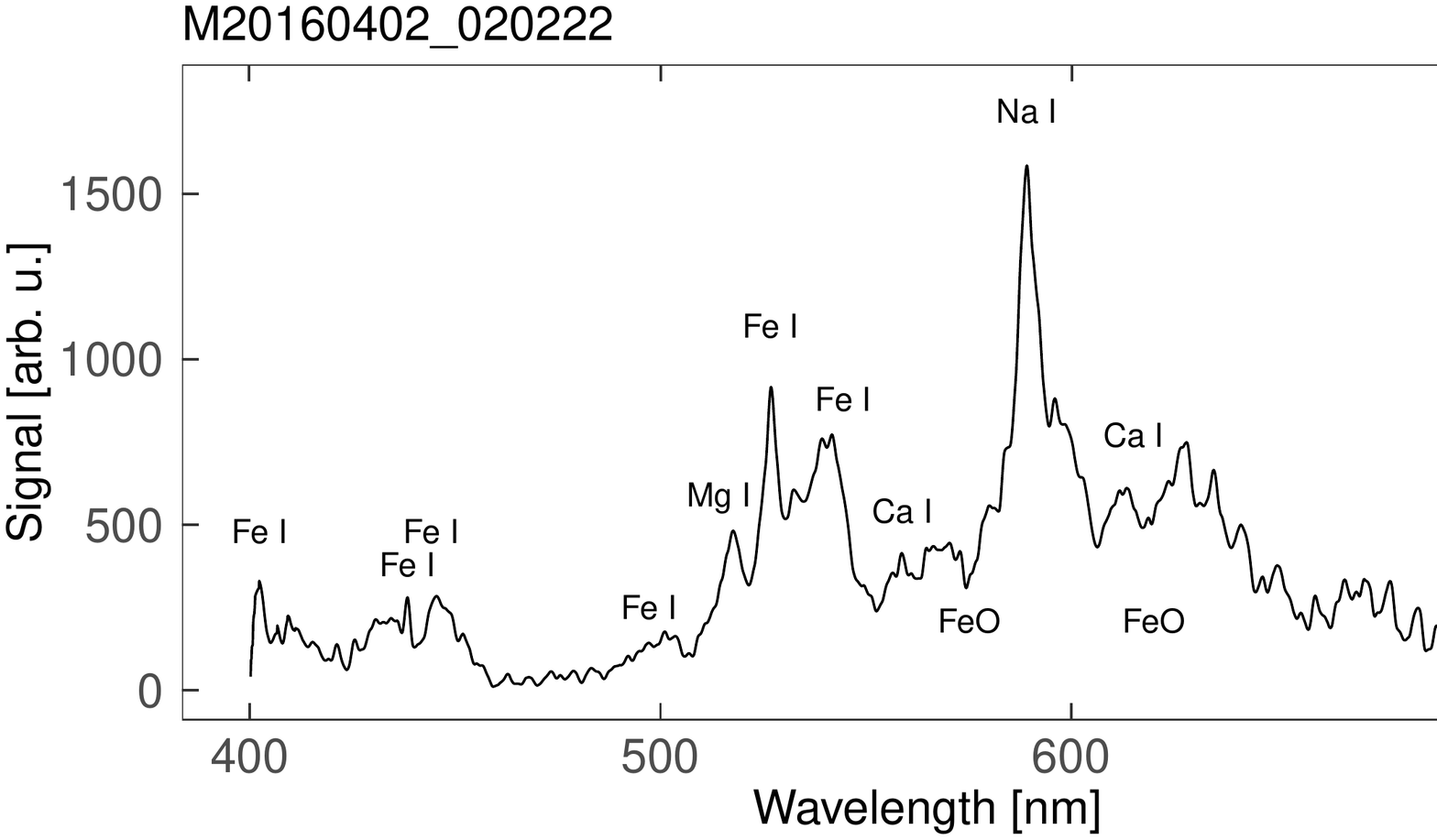}} \caption[f1]{Spectral profile of an iron meteor M20160415\_002243 (upper) and a Fe-rich meteor M20160402\_020222 (lower).} 
\label{FeR}
\end{figure}

The iron meteor M20160415\_002243 (Fig. \ref{FeR}) from our sample shows orbital and atmospheric parameters typical for an asteroidal body (Table \ref{tab:Fe}). The speed of the meteor ($v_i$ = 29.4 km\,s\textsuperscript{-1}) is higher than for most iron meteors (typically below 20 km\,s\textsuperscript{-1}). A similar speed has however already been detected in one meteor by \citet{2005Icar..174...15B} and an even higher speed was found among the iron meteors on Halley-type orbits by \citet{2019A&A...621A..68V}.  

\begin{figure}[]
    \centering
    \begin{subfigure}[b]{0.5\columnwidth}
      \centering
      \includegraphics[width=\textwidth]{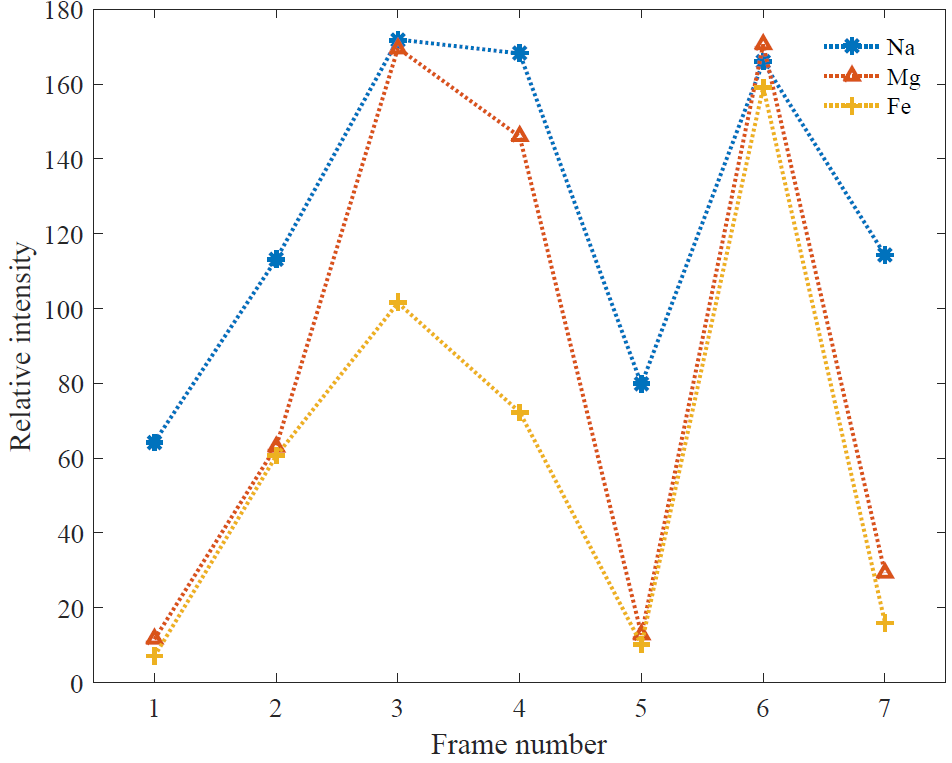}
      \caption{}
      \label{fig:1}
    \end{subfigure}%
    ~
    \begin{subfigure}[b]{0.5\columnwidth}
      \centering
      \includegraphics[width=\textwidth]{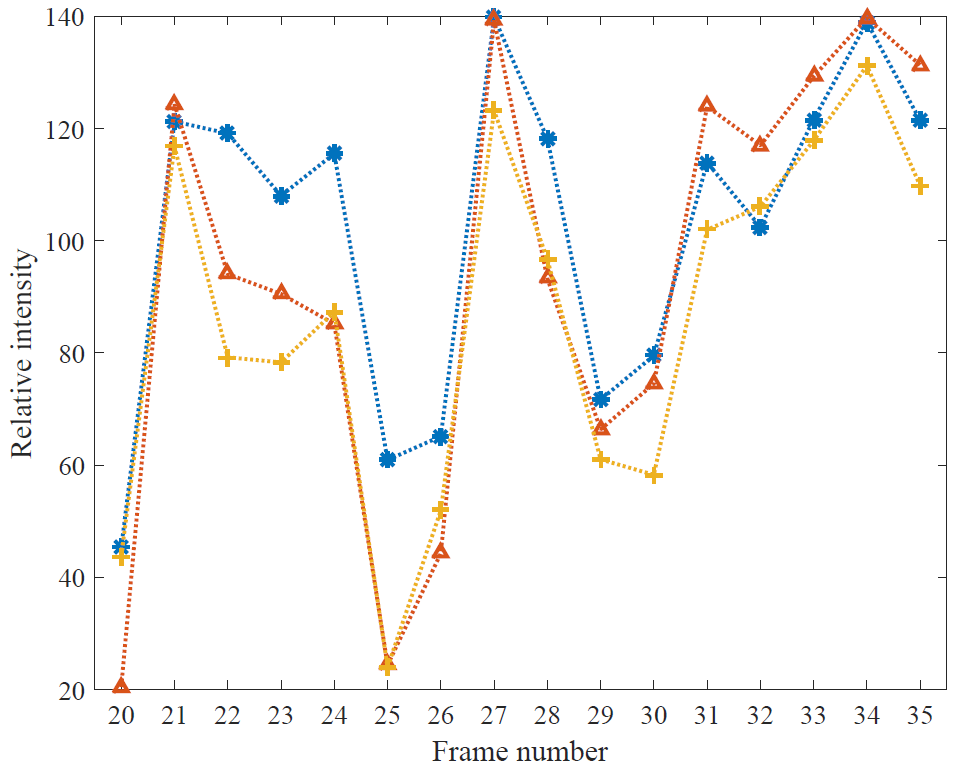}
      \caption{}
      \label{fig:2}
    \end{subfigure}%
    \caption{Monochromatic light curves of Fe-rich meteors M20150508\_002949 (a) and M20170301\_201252 (b) showing multiple flares. Here, the Fe I monochromatic light curve is only represented by the intensity of the 526.9 nm line. One frame corresponds to a time period of 1/12 s.}
    \label{FeRmono}
  \end{figure}

We identified two types of meteoroids observed as Fe-rich meteors. The majority of the increase in Fe/Mg ratio can be caused by the enhanced presence of iron--nickel or iron sulfide grains in smaller meteoroids. The most probable composition of these meteoroids is similar to H-type chondrites, with detected Fe enhancement either caused by heterogeneous content of iron grains or unresolved overestimation of Fe intensity due to the optical thickness of the radiating plasma. The effects of saturation and optical thickness were examined in each case. The spectrum of these meteors is often obtained from a few frames during sudden meteoroid disruption (exhibited as a meteor flare). One such example is meteor M20150508\_002949 which showed a light curve with numerous continuous flares, suggesting a fractured structure of a relatively strong material. A similar light curve and spectral type was detected in meteor M20170301\_201252. Monochromatic light curves of the two meteors are shown in Fig. \ref{FeRmono}. The orbital data for both these meteoroids (Table \ref{tab:Fe}) indicate an origin in the main asteroid belt ($T_J =$ 3.27 and $T_J =$ 3.88 respectively). None of the iron or Fe-rich meteors in our sample were assigned to a known meteoroid stream. Similarly, all of the iron meteoroids detected by \citet{2005Icar..174...15B} and \citet{2019A&A...621A..68V} were sporadic.

We also assume the possibility that Fe-rich spectra represent meteoroids similar to stony iron meteorites. These types of bodies formed either in the transition region between the core and the mantle of differentiated asteroids (pallasites) \citep{2002cem..book.....N}, or by low-velocity collisions of large differentiated asteroids (mesosiderites) \citep{1985Natur.318..168W}. The best candidate for a stony iron meteoroid is meteor M20160402\_020222. The spectral profile of this meteor can be seen in Fig. \ref{FeR}. The meteor exhibited similar spectral features from all 15 frames of the recording. The trajectory of the meteor started low and was short, with a smooth light curve. These characteristics are not unlike iron meteors. The main spectral difference is the presence of the Na I line and a possible contribution of Mg I, which is however likely blended with surrounding Fe I lines. The lines at 558.9 nm and 616.3 nm are well fitted by Ca I. This could be explained by the Ca-rich plagioclase grains commonly found in mesosiderites \citep{papike1998planetary}. All other unmarked lines are most probably caused by Fe I and FeO with a possible contribution of CaO. A detailed study of this spectral type captured in higher resolution is needed to confirm the corresponding composition.

Meteor M20160402\_020222 was not captured by multiple stations of the AMOS network, so the trajectory and orbit could not be determined from our data. To obtain some information about the source of this body, the meteor was searched for in the EDMOND (European viDeo MeteOr Network Database) \citep{2014me13.conf..225K}. According to EDMOND, the meteor had a low velocity ($v_i$ = 14 km\,s\textsuperscript{-1}), low beginning height ($H_B$ = 74.5 km), and low terminal height ($H_E$ = 54.4 km). This corresponds to an Apollo-type asteroidal orbit ($a$ = 1.8 au, $i$ = 10.4 deg) with medium eccentricity ($e$ = 0.46). Since numerous networks contribute to EDMOND, the precision of the data can vary from case to case.

\subsection{Na-enhanced and Na-rich} \label{secNaE}

Na-enhanced and Na-rich meteors constitute a relatively large part of our sample (Fig. \ref{pie}). Above, we mentioned the detected overall enhancement of the Na/Mg intensity ratio compared to surveys of fainter meteors. The overall results of this work show that sodium is better preserved in larger meteoroids (Fig. \ref{speedcurve}). The study of \citet{2019A&A...621A..68V} suggests that the preservation of Na is particularly efficient for meteoroids consisting of larger grains. Considering the differences between the size populations, we individually evaluated each meteor before it was classified as Na-enhanced or Na-rich, depending on meteor speed and position within the ternary diagram. The contribution of Na-enhanced and Na-rich meteors is still notably higher than found among smaller meteoroids studied by \citet{2005Icar..174...15B}. Our sample includes 15 Na-enhanced and 4 Na-rich meteors, of which 10 and 3 were observed by multiple stations, respectively.

As apparent from Fig. \ref{speedcurve} and similarly to the results of \citet{2005Icar..174...15B}, we found a strong connection between the increase of Na/Mg intensity and meteor speed. While the effect is taken into account for the spectral classification, the distinction between normal-type and Na-enhanced (Na-rich) meteors is particularly difficult for very slow meteors. Various degrees of Na enhancement are already apparent for all meteors with entry speed $v_i <$ 27 km\,s\textsuperscript{-1}. Meteors with $v_i <$ 15 km\,s\textsuperscript{-1} were identified mainly as Na-rich, Na-enhanced, or Fe-rich.

The distinct spectra of Na-rich meteors raised the possibility of atypical meteoroid composition, possibly related to inhomogeneous structures in comet interiors. Previous surveys of fainter meteors however consisted of only limited samples of Na-rich meteoroids, specifically lacking orbital data. Based on our results, the Na enhancement detected in slow meteors is mainly, if not exclusively, caused by the speed preference of the low excitation Na I line. Essentially, the effect reflects the lower temperature of the radiating plasma \citep{2005Icar..174...15B} in combination with typically moderate magnitudes of these meteors. Still, two Na-enhanced meteors were detected with high meteor speeds of $\approx$ 60 km\,s\textsuperscript{-1}. The meteoroids were members of the Perseid and 49 Andromedid stream. The two samples confirm that real compositional Na enhancement can be present in meteoroids.

Based on the determined orbital and structural properties, we report that the Na-enhanced class is a heterogeneous group of both cometary and asteroidal meteoroids, which are not linked by composition. Na-rich meteors have
all been found to be asteroidal, moving on heliocentric orbits very similar to several Apollo asteroids. The asteroidal orbits correspond well with the determined high material strengths characteristic of ordinary or carbonaceous chondrites (Fig. \ref{KbPeALL}). We believe this spectral group consists of chondritic fragments of S-type or C-type Apollo asteroids. The full analysis including higher-resolution data collected from global AMOS stations will be presented separately in \citet[][in preparation]{MatInPrep}.

\section*{Conclusions}

We present the first wide survey of spectral and physical properties of mm- to dm-sized meteoroids. While the specific material types (e.g., types of chondrites) cannot be distinguished from the low-resolution emission spectra \citep{2015aste.book..257B,2018A&A...613A..54D}, the combination of spectral, orbital, and atmospheric meteor data allows us to reveal the rough composition, structure, and source of meteoroids.

The obtained results reveal the compositional differences between cm- to dm-sized and mm-sized meteoroids. The preservation of volatiles in larger meteoroids is directly observed. We find an overall increase of Na/Mg ratio in cm- to dm-sized meteoroids compared to the population of mm-sized bodies. We suggest that this distinction is caused by the weaker effects of space weathering (solar radiation, solar wind and cosmic rays), which less efficiently alter the bulk composition of larger meteoroids. 

We detect a very low abundance of iron meteoroids in our sample. This is in strong contrast to mm-sized bodies, among which irons dominate the meteoroids on asteroidal orbits. Our results suggest that most cm- to dm-sized meteoroids on asteroidal orbits are chondritic. The large ratio of iron meteoroids in the mm-sized population could relate to the specific formation process of these bodies and typical sizes of iron--nickel grains. Moreover, a new spectral group was discovered - Fe-rich meteors, which show a higher intensity of Fe compared to most chondritic meteoroids. We suggest that this group may contain iron-enhanced H-type chondrites and bodies similar to stony iron meteorites. 

Thermal desorption during close perihelion approaches was confirmed as the main cause of Na depletion in medium-sized meteoroids, similarly to what is observed among smaller bodies. Our results demonstrate that the alteration of meteoroids and loss of volatiles during this process leads to higher material strength. A significant contribution of Na-enhanced and Na-rich meteoroids was detected, with spectra in most cases affected by low meteor speed and moderate magnitudes. Some Na-enhanced and all Na-rich meteoroids were identified as asteroidal. A definite compositional Na enhancement was detected in two cometary meteors of Perseids and 49 Andromedids, which had surprisingly high material strengths. 

Several major meteor showers were identified within this work, revealing significant heterogeneity in their meteoroid streams. We have found that Perseids originating in comet 109P/Swift-Tuttle contain a wide range of materials from Na-poor to Na-enhanced meteoroids. Significant structural and spectral differences were also found among the ecliptical $\alpha$-Capricornids of comet 169P/NEAT and sungrazing $\delta$-Aquarids of comet 96P/Machholz. The detected heterogeneities are affected by environmental factors and thermal history, but likely also reflect real inhomogeneities in comet interiors. Furthermore, we find spectral similarity between $\kappa$-Cygnids and Taurids, which could imply a similar composition of the parent objects of the two streams.

The heterogeneity of materials originating in one parent object can also be significant for asteroidal bodies, as was suggested from the Příbram and Neuschwanstein meteorite pair \citep{2003Natur.423..151S} and observed in the variety of samples found in the Almahata Sitta \citep{2009Natur.458..485J} or Benešov falls \citep{2014A&A...570A..39S}. Regarding sungrazing meteoroids, the dynamic age of these bodies is presented by the degree of Na depletion in their spectra. For the $\alpha$-Capricornids, diverse light curves and emission spectra reflect a structurally heterogeneous stream with a high volatile content. We also present the first spectral data for individual meteors originating in several minor meteoroid streams.

\begin{acknowledgements}
          We are thankful to Dr. Robert Jedicke for his valuable review. The first author would like to thank Dr. Jiří Borovička, Dr. Pavel Spurný and Dr. Mária Hajduková for their comments and suggestions, which helped improve this work.

          This work was supported by the Slovak Research and Development Agency under the contract APVV-16-0148 and by the Slovak Grant Agency for Science grant VEGA 01/0596/18. 
      
      This work was also supported by the Comenius University grant UK/179/2018 and special scholarship of the Faculty of Mathematics, Physics and Informatics, Comenius University in Bratislava.

\end{acknowledgements}

\bibliographystyle{aa}
\bibliography{references}

\end{document}